\documentclass[11pt,a4paper]{article}
\usepackage{verbatim}

\usepackage{dsfont}
\usepackage{sidecap}
\sidecaptionvpos{figure}{t}
\usepackage[utf8]{inputenc}
\usepackage[english]{babel}
\usepackage{empheq}
\usepackage{amssymb,amsmath}
\usepackage{graphicx, xcolor, varwidth}
\usepackage{setspace}
\usepackage[permil]{overpic}
\usepackage{graphicx}
\usepackage{color}
\usepackage{braket}
\usepackage{simplewick}
\usepackage{jheppub}
\usepackage[T1]{fontenc}
\usepackage[autostyle]{csquotes}
\usepackage{ulem}
\usepackage{float}
\usepackage{subcaption}
\usepackage{mathtools}

\usepackage[bottom]{footmisc}

\usepackage{simpler-wick}

\usepackage{feynmp}
\colorlet{darkblue}{blue!70!black}
\colorlet{darkgreen}{green!70!black}

\numberwithin{equation}{section}

\numberwithin{equation}{section}

\newcommand{\tr}{\mathrm{tr}}
\newcommand{\Tr}{\mathrm{Tr}}



\title{	Large $N$ algebras and generalized entropy}

\author{Venkatesa Chandrasekaran,$^1$  Geoff Penington,$^{1,2}$ and  Edward Witten$^1$}
\affiliation{$^1$School of Natural Sciences, Institute for Advanced Study,\\ 1 Einstein Drive, Princeton, NJ 08540 USA}
\affiliation{$^2$Center for Theoretical Physics and Department of Physics, University of California,\\  Berkeley, CA 94720 USA}
\emailAdd{venchandrasekaran@ias.edu}
\emailAdd{geoffp@berkeley.edu}
\emailAdd{witten@ias.edu}

\abstract{We construct a Type II$_\infty$ von Neumann algebra that describes the large $N$ physics of single-trace operators in AdS/CFT in the microcanonical ensemble, where there is no need to include perturbative $1/N$ corrections. Using only the extrapolate dictionary, we show that the entropy of semiclassical states on this algebra is holographically dual to the generalized entropy of the black hole bifurcation surface. From a boundary perspective, this constitutes a derivation of a special case of the QES prescription without any use of Euclidean gravity or replicas; from a purely bulk perspective, it is a derivation of the quantum-corrected Bekenstein-Hawking formula as the entropy of an explicit algebra in the $G \to 0$ limit of Lorentzian effective field theory quantum gravity. In a limit where a black hole is first allowed to  equilibrate and then is later potentially re-excited, we show that the generalized second law is a direct consequence of the monotonicity of the entropy of algebras under trace-preserving inclusions. Finally, by considering excitations that are separated by more than a scrambling time we construct a ``free product'' von Neumann algebra that describes the semiclassical physics of long wormholes supported by shocks. We compute R\'{e}nyi entropies for this algebra and show that they are equal to a sum over saddles associated to  quantum extremal surfaces in the wormhole. Surprisingly, however, the saddles associated to ``bulge'' quantum extremal surfaces contribute with a negative sign.}

\begin{document}
\maketitle 
\section{Introduction}
\begin{figure}[t]
\begin{center}
  \includegraphics[width = 0.45\linewidth]{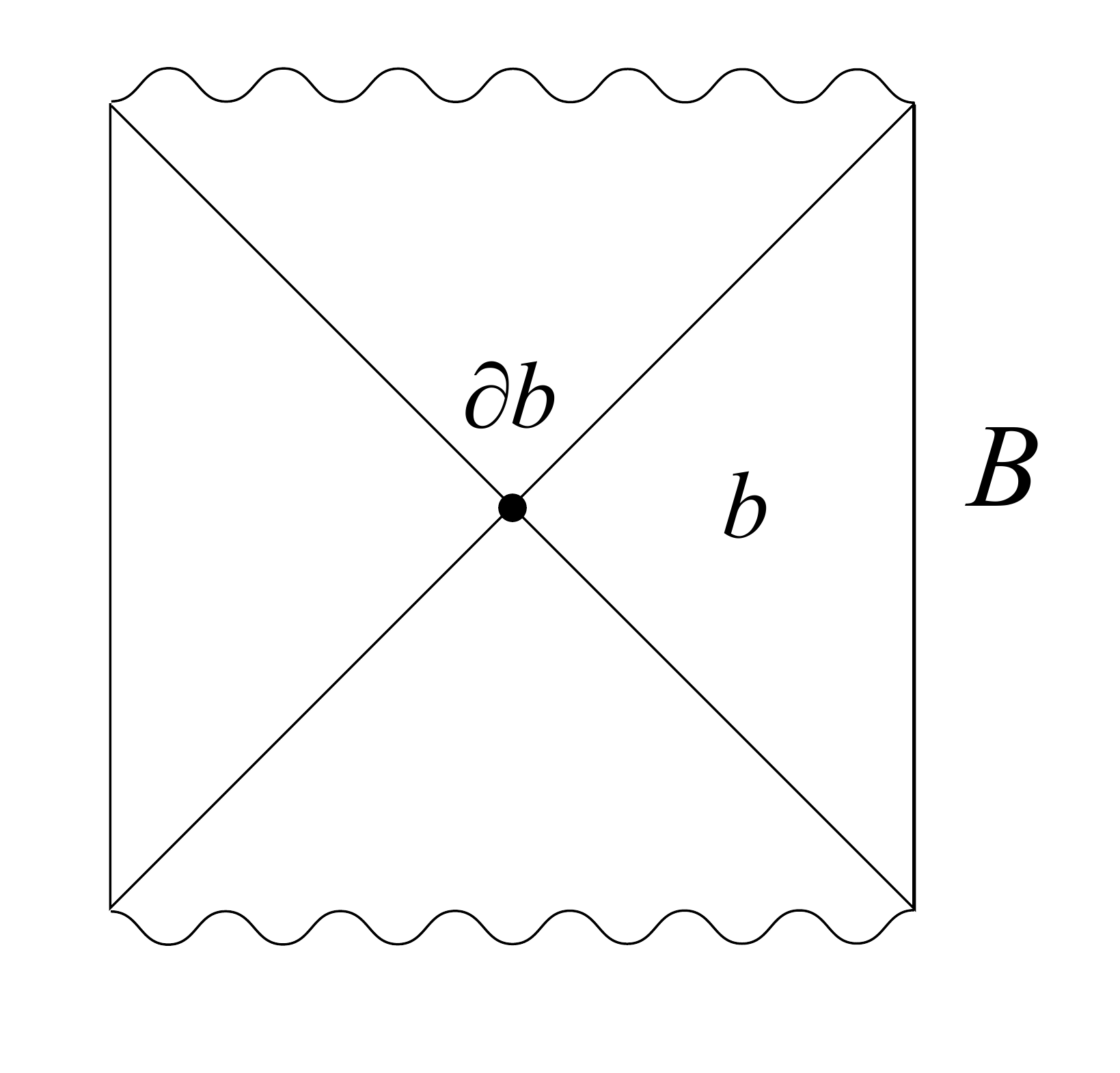} 
\end{center}
\caption{When the boundary region $B$ is the right asymptotic boundary of a two-sided black hole, the entanglement wedge $b$ is simply the right exterior of the black hole. The entanglement entropy of $B$ is given by the generalized entropy $S_\mathrm{gen}(b) = A(\partial b)/4G + S_\mathrm{bulk}(b)$ where $A(\partial b)$ is the area of the black hole bifurcation surface and $S_\mathrm{bulk}(b)$ is the entanglement entropy of bulk quantum fields in the right exterior.}
\label{fig:BH_EW}
\end{figure}

Over two decades after its discovery, the AdS/CFT correspondence \cite{Maldacena:1997re,Gubser,Witten:1998qj} remains our most powerful insight into microscopic quantum gravity -- and simultaneously a deep mystery whose underlying mechanisms we do not remotely understand. Both of these features are nicely exhibited by the quantum extremal surface (QES) prescription \cite{Engelhardt:2014gca} for holographic entanglement entropy, which says that the entanglement entropy $S(B)$ of a CFT subregion $B$ is equal to the \emph{generalized entropy}
\begin{align} \label{eq:sgen}
    S_\mathrm{gen}(b) = \frac{A(\partial b)}{4G} + S_\mathrm{bulk}(b)~,
\end{align}
of a dual bulk quantum gravity region $b$ known as the \emph{entanglement wedge} of $B$.\footnote{The QES prescription is a generalization of the earlier Ryu-Takayanagi formula \cite{Ryu:2006bv} that accommodates time-dependent spacetimes \cite{Hubeny:2013gba, maximin} and quantum corrections $S_\mathrm{bulk}$ \cite{Faulkner:2013ana}. For simplicity, we will use the broad heading of the QES prescription to refer to all of these important developments. In an abuse of terminology, it is also common to refer to $S_\mathrm{gen}(b)$ as the generalized entropy of the bounding surface $\partial b$, and we will sometimes do so.} Here $A(\partial b)$ is the area of the codimension-two surface $\partial b$ (called the quantum extremal surface) that bounds the wedge $b$, $G$ is Newton's constant, and $S_\mathrm{bulk}(b)$ is the entanglement entropy of bulk quantum fields in $b$. In general, the region $b$ is defined as the smallest generalised entropy region with conformal boundary $B$ that is an extremum (or, more precisely, a critical point) of \eqref{eq:sgen} under local perturbations of the quantum extremal surface $\partial b$. In this paper we will be primarily interested in states that are small perturbations of a two-sided black hole, with $B$ the right (or sometimes left) asymptotic boundary; for such states the region $b$ is simply the right (or left) exterior of the black hole, as shown in Figure \ref{fig:BH_EW}. 

The QES prescription has had a profound impact in our conceptual understanding of AdS/CFT, leading to the idea that the bulk spacetime itself emerges from quantum entanglement in the boundary theory \cite{Raamsdonk:2010aa, Maldacena:2013aa}. It has also evolved into a crucial technical tool for many important quantum gravity computations; perhaps most famously it was the key ingredient in the derivation of the Page curve \cite{Penington:2019npb, Almheiri:2019vm, Penington:2019kki, Almheiri:2019qdq}. However we have only a minimal understanding of why the QES prescription should be true in the first place.\footnote{Previous hints at possible answers to this question include tensor network toy models \cite{Swingle:2009bg,HaPPY,Hayden:2016cfa,Bao:2018pvs} and a close relationship between the QES prescription and quantum error correction \cite{Almheiri:2014lwa,Harlow:2016vwg,Akers:2021fut}.} What is the origin of the area term $A(\partial b)/4G$? Why isn't the entropy $S(B)$ just equal to the bulk entropy $S\mathrm(b)$? Why is the region $b$ dual to $B$ determined by extremising $S_\mathrm{gen}(b)$? 

To be clear: it is possible to derive the QES prescription using a beautiful argument first introduced by Lewkowycz and Maldacena \cite{Lewkowycz:2013nqa, Faulkner:2013ana, Dong:2017aa, Penington:2019kki, Almheiri:2019qdq}. However that derivation relies on the AdS/CFT correspondence to relate the boundary entanglement entropy to a bulk Euclidean gravity path integral on a replicated manifold, followed by a bulk calculation that is essentially a more sophisticated version of the Gibbons-Hawking Euclidean gravity derivation of the Bekenstein-Hawking entropy. As such, the argument inherits all the interpretational  and conceptual questions of that earlier derivation. In effect, Euclidean gravity acts as a magical black box that reliably gives us the right answer, but doesn't give the slightest hint about how it was able to obtain that answer. A principal goal of this paper will be to begin to remove some of this mystery, and to understand the formula \eqref{eq:sgen} as an emergent description of the entanglement entropy of a certain class of CFT states in the large $N$ limit. 

Recently, Liu and Leutheusser \cite{Leutheusser:2021ab,Leutheusser:2021aa} studied the large $N$ limit of thermal correlation functions of single-trace CFT operators. They argued that, above the Hawking-Page transition, these correlation functions were described by a Type III$_1$ von Neumann algebra $\mathcal{A}_{R,0}
$. For detailed reviews of von Neumann algebras aimed at physics audience, we refer readers to \cite{Witten:2018wc, Witten:2021jzq}. However heuristically Type III von Neumann algebras describe degrees of freedom that have both divergent entanglement and divergent entanglement fluctuations (Type III$_1$ means additionally that the  entanglement spectrum takes on all positive real values). Crucially, the algebra of observables associated to any subregion\footnote{By subregion we mean a domain of dependence with a nonempty boundary, i.e. one that is not both open and closed.} in quantum field theory is Type III$_1$. Liu and Leutheusser identified the boundary Type III$_1$ algebra $\mathcal{A}_{R,0}$ that they had constructed with the bulk algebra $\mathcal{A}_{r,0}$ of right exterior QFT observables in a black hole background in the $G \to 0$ limit. Any representation of a Type III von Neumann algebra $\mathcal{A}$ has commutant algebra $\mathcal{A}'$ -- the algebra of operators that commute with all operators in $\mathcal{A}$ -- that is also Type III. In the bulk, the commutant $\mathcal{A}_{r,0}'$ of the right exterior algebra is simply the algebra $\mathcal{A}_{\ell,0}$ of operators in the left exterior of a two-sided black hole. On the boundary, the Hartle-Hawking state of a two-sided black hole is holographically dual to the thermofield double state
\begin{align} \label{eq:TFDdef}
    \ket{\mathrm{TFD}} \propto \sum_i e^{-\beta E_i/2} \ket{\overline{E_i}}_L \ket{E_i}_R.
\end{align}
This is a canonical purification of the thermal ensemble on two copies of the CFT Hilbert space that correspond to the left and right asymptotic boundaries of the two-sided black hole; the sum in \eqref{eq:TFDdef} is over all energy eigenstates $\ket{E_i}$, with $\ket{\overline{E_i}}$ their CPT duals. The Liu-Leutheusser algebra $\mathcal{A}_{R,0}$ and its commutant $\mathcal{A}_{L,0} = \mathcal{A}_{R,0}'$ describe the large $N$ limit of single-trace CFT operators acting on the thermofield double state $\ket{\mathrm{TFD}}$ at the right and left boundaries respectively.

Further progress was made in \cite{wittengcp} where it was shown that, if we introduce a rescaled version of the boundary Hamiltonian in the large $N$ algebra, and include perturbative $1/N$ corrections in a formal power series, then the resulting large $N$ algebra describes the so-called \emph{crossed product} of the algebra $\mathcal{A}_{R,0}$ by a group of modular automorphisms. This crossed product construction is a central tool in the mathematical theory of Type III von Neumann algebras because it converts a Type III algebra into a Type II$_\infty$ algebra.

Type II von Neumann algebras feature infinite entanglement, just like Type III algebras. However, unlike Type III algebras, Type II algebras have finite entanglement fluctuations. As a result, on a Type II algebra $\mathcal{A}$ we can define a \emph{trace} $\tr$. This is a linear functional on $\mathcal{A}$ satisfying
\begin{align} \label{eq:deftrace}
    \tr[a b] = \tr[b a]~,
\end{align}
for all operators $a,b \in \mathcal{A}$ and positive in the sense that $\tr[a a^\dagger]>0$ for all nonzero $a$. The trace $\tr$ on a Type II algebra $\mathcal{A}$ should not be confused with the usual trace $\Tr$ on the Hilbert space $\mathcal{H}$ on which the algebra $\mathcal{A}$ acts. In fact, the trace $\Tr(a)$ will be infinite for any operator in $a \in\mathcal{A}$. However the algebraic condition \eqref{eq:deftrace} encourages us to think of the trace $\tr$ as an infinitely rescaled, renormalized version of $\Tr$. The difference between the two subtypes of Type II von Neumann algebras is that in a Type II$_1$ algebra $\mathcal{A}$ the trace $\tr[a]$ of any bounded operator $a \in \mathcal{A}$ is finite, whereas in a Type II$_\infty$ algebra only ``trace-class'' observables have a finite trace, as for the usual trace $\Tr$ on an infinite-dimensional separable Hilbert space. If the Type II algebra is a factor, meaning that its center consists only of c-numbers, then the trace $\tr$ is unique up to rescaling. However for Type II$_\infty$ algebras there is generally no canonical choice of scaling.

The existence of the trace $\tr$ allows us to define a density matrix $\rho_\Phi$ for any state $\ket{\Phi}$ by the condition
\begin{align}
    \tr[\rho_\Phi a] = \braket{\Phi|a|\Phi} \,\,\,\,\forall a \in \mathcal{A}.
\end{align}
The existence and uniqueness of $\rho_\Phi$ follows from the nondegeneracy of the trace $\tr$. In turn, we can define an entanglement entropy
\begin{align} \label{eq:entdef}
    S(\Phi)_{\mathcal{A}} = -\tr[\rho_\Phi \log \rho_\Phi].
\end{align}
For a Type II$_\infty$ algebra, this entropy may take any value in the range $[-\infty,\infty]$. If we interpret the trace $\tr$ as an infinitely rescaled version of a standard Hilbert space trace, then the entropy \eqref{eq:entdef} agrees with the usual definition of entanglement entropy, except for the subtraction of a state-independent infinite constant. The freedom to rescale the trace $\tr$ by any finite factor allows us to shift the entropy \eqref{eq:entdef} by any finite state-independent constant.

In contrast, for Type III algebras, the divergent entanglement fluctuations mean that one cannot construct density matrices or entropies (even renormalized ones!) in a mathematically rigorous way. Of course, such issues have never stopped physicists before, and indeed the study of entanglement entropy of Type III algebras in quantum field theory is a rich and important subject. These entropy computations can sometimes be given a rigorous interpretation in terms of relative entropies, which can be defined for Type III algebras using Tomita-Takesaki theory.\footnote{Tomita-Takesaki theory will play an important role in this paper; see for example \cite{Witten:2018wc} for a review.} For example, the mutual information $I(A:B)$ between two regions $A$ and $B$ -- which is traditionally defined as a finite combination $I(A:B) = S(A) + S(B) - S(AB)$ of entropies that are individually divergent -- can instead be defined as the relative entropy on those regions of the state itself, relative to a state that is indistinguishable from the original state on each individual region, but that has no correlation between the regions.

An unfortunate feature of the algebra constructed in \cite{wittengcp} is that the formal factors of $1/N$ that appear in the crossed product algebra lead to factors of $N$ appearing in exponents when one tries to define a trace $\tr$. As a result, one cannot define a density matrix as a formal power series, or even as a formal Laurent series, in $1/N$. Happily, in Section \ref{sec:alg}, we show that a similar crossed product algebra can be obtained without any need to consider a formal power series in $1/N$, by taking a slightly different $N \to \infty$ limit. Rather than starting with the usual thermofield double state $\ket{\mathrm{TFD}}$, which has divergent energy fluctuations at large $N$, we start with a microcanonical version of the thermofield double state where the energy fluctuations are $O(1)$. Because the energy fluctuations are finite, shifting the boundary Hamiltonian $H_R$ by a divergent additive constant $E_0$ leads to an operator $h_R = H_R - E_0$ that has a sensible large $N$ limit. (In particular, unlike in the canonical ensemble, there is no need to additionally rescale $h_R$ by a factor of $1/N$.) The large $N$ algebra $\mathcal{A}_R$ generated by the Leutheusser-Liu algebra $\mathcal{A}_{R,0}$, together with the renormalized boundary Hamiltonian $h_R$, is a Type II$_\infty$ factor. The large $N$ Hilbert space $\mathcal{H}$ on which the algebra $\mathcal{A}_R$ naturally acts 
describes the $G \to 0$ limit of perturbative excitations around a two-sided black hole, along with quantum fluctuations in the relative  timeshift between the two boundaries. The bulk dual $\mathcal{A}_r$ of the algebra $\mathcal{A}_R$ is generated by the right exterior bulk QFT algebra $\mathcal{A}_{r,0}$ (gravitationally dressed to the right asymptotic boundary), together with the renormalized right ADM Hamiltonian.

In Section \ref{sec:sgen} we turn our attention to density matrices and entropies on the Type II$_\infty$ algebra $\mathcal{A}_R$. We focus on a particular class of states, which we call semiclassical states, where the fluctuations in the timeshift between the two boundaries is small, and hence the bulk geometry is approximately fixed. Our central result is that for such states the entropy
\begin{align} \label{eq:ent=genent}
    S(\mathcal{A}_R) \approx \braket{\beta h_R} - \braket{\log p(h_R)} - S_\mathrm{rel}(\Phi||\Psi)~.
\end{align}
Here $p(h_R)$ describes the probability distribution of the renormalized energy $h_R$ while $S_\mathrm{rel}(\Phi||\Psi)$ is the relative entropy (on the right exterior) of the state $\ket{\Phi}$ of the bulk quantum fields relative to the Hartle-Hawking state $\ket{\Psi}$. Applying a previous argument of Wall \cite{Wall:2011hj}, we use Raychaudhuri's equation to show that the right hand side of \eqref{eq:ent=genent} is exactly equal to the generalized entropy of the bifurcation surface of the black hole, up to the state-independent constant inherent in the definition of $S(\mathcal{A}_R)$:
\begin{align} \label{eq:ent=genent2}
    S(\mathcal{A}_R) \approx \left\langle\frac{A}{4G}\right\rangle + S_\mathrm{bulk}(\mathcal{A}_{r,0})_\Phi + \mathrm{const}~.
\end{align}
All three terms on the right-hand side of \eqref{eq:ent=genent2} are individually divergent\footnote{Even at finite $G$, the second term is infinite because $\mathcal{A}_{r,0}$ is a Type III algebra. This divergence is cancelled by the perturbative renormalization of $G$ in the first term. The combination of the first and second terms is therefore finite at finite $G$, but has a state-independent divergence as $G \to 0$. This last divergence is cancelled by the divergent constant third term.}; however their sum is finite, as can be seen from the formula \eqref{eq:ent=genent} for the left-hand side of \eqref{eq:ent=genent2} in terms of purely finite quantities.

To reiterate the logic above: we started with a large $N$ algebra $\mathcal{A}_R$ of CFT operators generated by the algebra $\mathcal{A}_{R,0}$ of single-trace operators together with the renormalized Hamiltonian $h_R$ acting on a microcanonical version of the thermofield double. The bulk dual of the algebra $\mathcal{A}_{R,0}$ is the algebra $\mathcal{A}_{R,0}$ of QFT operators in the right exterior, while $h_R$ is dual to the renormalized right ADM energy. The latter can be thought of as an extra mode (conjugate to the timeshift between the two boundaries) that is present in quantum gravity but not in quantum field theory in curved spacetime. Counterintuitively, including this extra mode makes the entanglement entropy in quantum gravity better defined than it would be in quantum field theory because the algebra becomes Type II rather than Type III. Finally, we obtained a formula \eqref{eq:ent=genent2} for this entropy that included (the divergent physicist's definition of) the QFT entanglement entropy for the Type III algebra $\mathcal{A}_{R,0}$ -- plus an additional term that was equal to the area of the black hole horizon. All of this followed purely from the large $N$ Lorentzian physics of the boundary CFT. The only input from holography was the identification of $\mathcal{A}_{R,0}$ with $\mathcal{A}_{r,0}$ and of $h_R$ with the renormalized ADM mass. Both follow directly from the extrapolate dictionary of AdS/CFT.

We could alternatively have started directly with the bulk algebra $\mathcal{A}_r$ in an effective field theory of gravity together with other fields. The derivation in Section \ref{sec:sgen} then gives us a relationship, for small $G$, between generalized entropy in this effective field theory and the entropy of a Type II von Neumann algebra $\mathcal{A}_r$.
 We used this approach 
in a companion paper \cite{Chandrasekaran:2022cip} to understand the entropy of the cosmological horizon in de Sitter space, where no dual theory analogous to the boundary CFT is known. In Section \ref{sec:BHentropy}, we therefore reinterpret the derivation of \eqref{eq:ent=genent2} as a derivation of the Bekenstein-Hawking entropy in Lorentzian effective field theory of gravity, and argue that it has significant advantages over previous derivations.

An important property of generalized entropy is that it satisfies the so-called generalized second law (GSL). Originally motivated by Bekenstein \cite{Bekenstein:1973ur}, and later proven by Wall \cite{Wall:2010cj, Wall:2011hj}, this states that the generalized entropy of a cut through a black hole event horizon monotonically increases with time. Although the close analogy with the ordinary second law of thermodynamics has been clear since its original conception, a precise microscopic principle behind the generalized second law has long since been lacking. In recent work \cite{Engelhardt:2018aa, Bousso:2019dxk}, it has been argued that the generalized entropy of a horizon cut\footnote{Technically the horizon here is an apparent horizon rather than the event horizon of the black hole.  Perturbative backreaction from semiclassical matter on the eternal black hole background only causes the event horizon to differ from the apparent horizon at $O(G)$, but, because of the factor of $1/G$ in \eqref{eq:sgen}, their generalized entropies then differ by an $O(1)$ amount.} is related to a particular coarse-grained boundary entropy; however those arguments rely on the QES prescription, and hence on Euclidean gravity calculations, in an essential way.

In Section \ref{sec:gsl}, we consider a simple case of the generalized second law, where the black hole is temporarily allowed to come to equilibrium (for a time much longer than the thermalization time but shorter than the scrambling time) before again being disturbed. The generalized second law says that the generalized entropy of the equilibrated horizon should be larger than that of the bifurcation surface, but smaller than that of the horizon at future infinity. In the large $N$ limit, we obtain a Type II algebra of operators accessible outside the black hole at late times (after the equilibration). This algebra forms a von Neumann subalgebra of the larger algebra of operators accessible at any time. A general fact about trace-preserving inclusions of algebras says that the entropy of the late time subalgebra must always be larger than the entropy of the full boundary algebra. We show that this inequality underlies the generalized second law in this situation. We also comment on an analogous construction for one-sided black holes formed from collapse. 

Unfortunately we have not been able to extend this line of argument to horizon cuts that are out of equilibrium, because we were not able to find a Type II algebra whose entropy is equal to the generalized entropy of such a cut. It seems likely that some new ingredient beyond Type II von Neumann algebras is necessary to understand this; one possibility that we discuss briefly is that these generalized entropies describe the entropies of subspaces of observables, rather than  actual algebras.

Finally, in Section \ref{sec:scrambling}, we study the 
case of perturbations of a black hole that are made, possibly out of time order, at times that differ
by more than a scrambling time.
In this situation, we show that the appropriate large $N$ algebra is naturally constructed as a so-called ``free product'' of von Neumann algebras, and acts on states that describe long wormholes supported by shocks. We compute R\'{e}nyi 2-entropies for the large $N$ algebra and show that they reduce to a sum over terms associated to quantum extremal surfaces. Interestingly, bulge quantum extremal surfaces appear to contribute to the R\'{e}nyi entropy computation with a negative sign; this conclusion has important implications for the gravitational interpretation of subleading Weingarten terms in random unitary integrals \cite{Stanford:2021bhl}. 

In Appendix \ref{app:corrections}, we return to the canonical ensemble, and argue that one can indeed define entropies of states as a Laurent series in $1/N$ in a satisfactory manner,  even though the definition of traces and density matrices is problematic. In Appendix \ref{app:otherscharges}, we discuss the inclusion of other symmetries and charges (beyond the Hamiltonian) in the large $N$ algebra. Finally, in Appendix \ref{app:isotrace} we prove an important technical result needed in Section \ref{sec:sgen} relating the definitions of the trace in different constructions of the crossed product algebra.

\section{Constructing the algebra \label{sec:alg}}

\subsection{Canonical ensemble \label{sec:canon}}
We start by briefly reviewing the construction in \cite{Leutheusser:2021aa, Leutheusser:2021ab} of an algebra of single-trace operators from the large $N$ limit of thermal CFT correlation functions. In $\mathcal{N} = 4$ super Yang-Mills, appropriately normalized, single-trace operators of the general form $t = \text{Tr}T - \langle \text{Tr}T\rangle_\beta$ (where $T$ is a polynomial in matrix-valued fields and their derivatives) have vanishing thermal one-point functions $\braket{t}_\beta$, while their connected higher $k$-point functions scale as $1/N^{k-2}$. In the strict large $N$ limit, only the two-point functions survive; the large $N$ limit is therefore referred to as a generalized free field theory. Following \cite{wittengcp}, we initially restrict to single-trace operators that are noncentral at large $N$, i.e. that have commutators with other single-trace operators whose expectation value is nonzero in the large $N$ limit.\footnote{More precisely, we restrict to operators $t$ such that $\braket{t c}_\beta = \braket{c t}_\beta = 0$ for all single-trace operators $c$ that are central at large $N$. In combination with the set of central single-trace operators, such operators span the space of all single-trace operators.}

From this data we can construct a Hilbert space $\mathcal{H}_0$ and an algebra $\mathcal{A}_{R,0}$ that describe the large $N$ thermal physics as follows. We first construct a vector space $\mathcal{V}_0$ that contains a distinguished state $\ket{\Psi}$ (which we will shortly identify with the thermofield double state) and is spanned by states of the form
\begin{align}
   t_1 t_2 \dots t_n \ket{\Psi}~, 
\end{align}
for any finite $n$ and product of single-trace operators $t_1 t_2 \dots t_n$. The algebra of finite products of single-trace operators naturally acts on $\mathcal{V}_0$ by left multiplication. We can  use the thermal
correlation functions $ \braket{t_1 t_2 \dots t_n}_\beta$ to define an inner product on $\mathcal{V}_0$:
\begin{align} \label{eq:innerproductdef}
    \braket{\Psi| t_1 t_2 \dots t_n |\Psi} = \lim_{N\to\infty} \braket{t_1 t_2 \dots t_n}_\beta~.
\end{align}
Taking the completion of $\mathcal{V}_0$ with respect to this inner product leads to a separable Hilbert space $\mathcal{H}_0$.\footnote{This procedure is commonly known as the GNS construction \cite{GelfandNaimark, bams/1183510397}.} The double commutant of the algebra generated by (bounded functions of) single-trace operators acting on $\mathcal{H}_0$, or equivalently the closure of that algebra in either the strong or weak operator topologies, is a von Neumann algebra $\mathcal{A}_{R,0}$. When the temperature $T = 1/\beta$ of the thermal correlation functions is above the Hawking-Page transition, the algebra $\mathcal{A}_{R,0}$ is believed to be a Type III$_1$ von Neumann factor \cite{Leutheusser:2021ab, Leutheusser:2021aa}.

As we claimed above, there is a natural identification between the state $\ket{\Psi} \in \mathcal{H}_0$ and the large $N$ limit of the thermofield double (TFD) state: 
\begin{align}
    |\text{TFD}\rangle = \sum_{i}e^{-\beta E_i/2}|E_i\rangle_L |E_i\rangle_R. \label{eq:tfd}
\end{align}
The thermofield double state is a canonical purification of the thermal density matrix $\rho_{\beta} = Z[\beta]^{-1}e^{-\beta H}$ on two copies of the boundary Hilbert space (known as the left and right boundary). It is therefore a pure state with exactly thermal correlation functions; the same is true of $\ket{\Psi}$ because of the definition of the inner product \eqref{eq:innerproductdef}. Unlike the finite $N$ boundary Hilbert space, however, the Hilbert space $\mathcal{H}_0$ does not factorize into a product of Hilbert spaces associated to the left and right boundary because the entanglement between the two boundaries diverges at large $N$. Instead, observables on the right and left boundary are described respectively by the Type III algebra $\mathcal{A}_{R,0}$ and its commutant $\mathcal{A}_{L,0} = \mathcal{A}_{R,0}'$.\footnote{In particular, the antiunitary modular conjugation operator $J_\Psi$ that exchanges $\mathcal{A}_{R,0}$ and $\mathcal{A}_{L,0}$ is identified with the operator in the finite $N$ theory that time reverses and then exchanges the left and right boundaries.}

\begin{figure}[t]
\begin{center}
  \includegraphics[width = 0.55\linewidth]{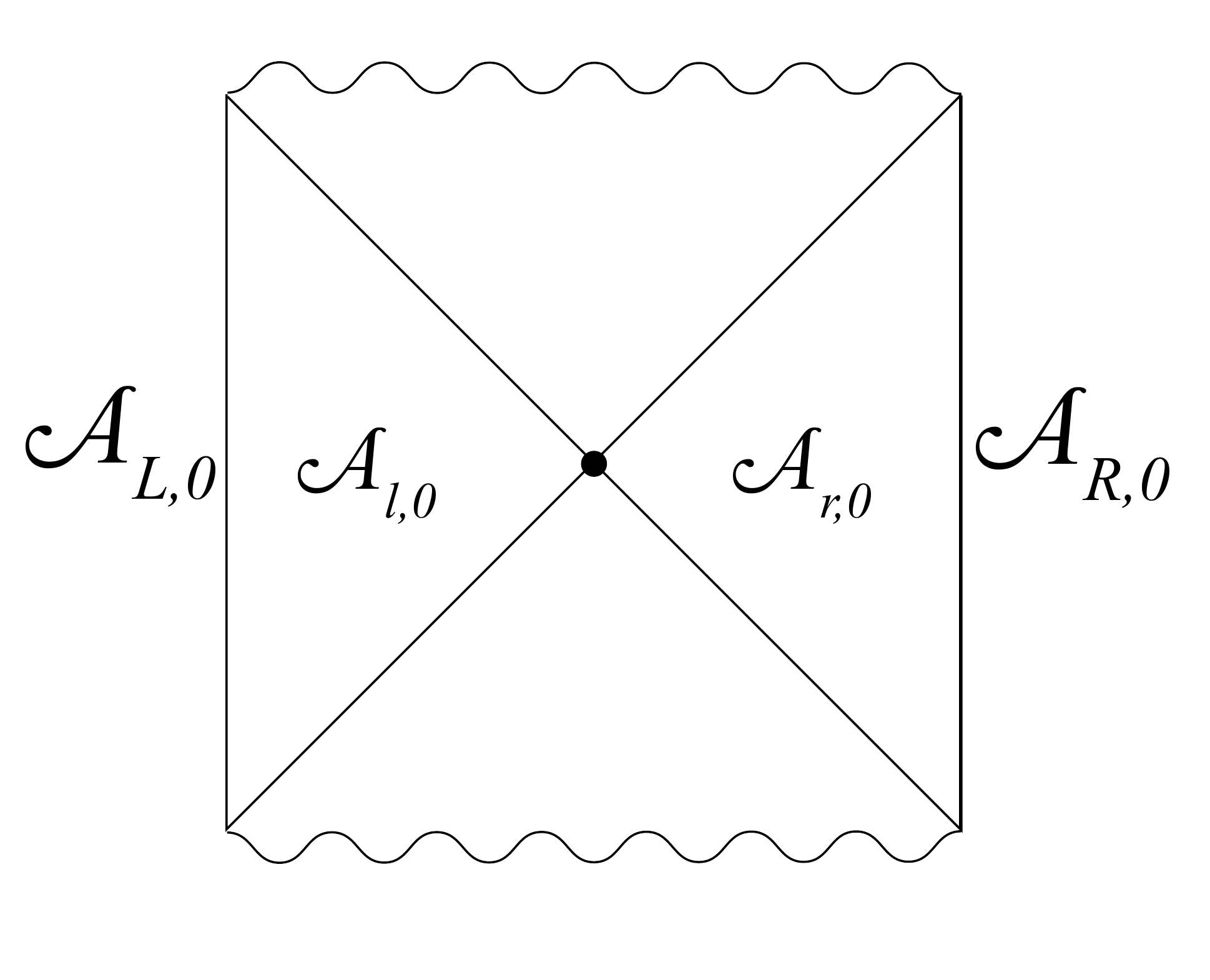} 
\end{center}
\caption{The large $N$ boundary algebras $\mathcal{A}_{R,0}$ and $\mathcal{A}_{L,0}$ are holographically dual to bulk quantum field theory algebras $\mathcal{A}_{r,0}$ and $\mathcal{A}_{\ell,0}$ associated to the right and left exterior respectively.}
\label{fig:ARrLl}
\end{figure}

The holographic dual of the thermofield double state above the Hawking-Page transition is a two-sided eternal black hole \cite{Maldacena:2001kr}. In the large $N$ limit, the black hole can be treated semiclassically using bulk quantum field theory on a classical curved spacetime background.\footnote{The bulk QFT includes quantized graviton excitations, but these can be treated like any other quantum field theory since we are in the zero coupling limit $N\to \infty$.} The extrapolate dictionary of AdS/CFT says that local single-trace boundary operators are dual to bulk quantum fields near asymptotic infinity. But the HKLL reconstruction procedure \cite{HKLL} allows one to rewrite \emph{any} bulk QFT observable in the black hole exterior in terms of bulk fields near asymptotic infinity.\footnote{This can be viewed as an example of the timelike tube theorem \cite{borcherstt, arakitt}, which says that the von Neumann algebra generated by operators in any small timelike tube describes the entire causal diamond of that tube.} We therefore conclude that the algebras $\mathcal{A}_{L,0}$ and $\mathcal{A}_{R,0}$ are dual to the algebras $\mathcal{A}_{\ell,0}$ and $\mathcal{A}_{r,0}$ describing bulk quantum fields in the left and right exterior respectively. This is depicted in Figure \ref{fig:ARrLl}. The algebra of operators in quantum field theory that are localised to a causal diamond is always Type III; the holographic dictionary therefore justifies our earlier claim that $\mathcal{A}_{L,0}$ and $\mathcal{A}_{R,0}$ are Type III algebras.

In fact, we made a somewhat stronger claim that $\mathcal{A}_{L,0}$ and $\mathcal{A}_{R,0}$ are Type III$_1$ \emph{factors}. A von Neumann algebra is a factor if its center (the intersection of the algebra with its commutant) consists solely of $c$-numbers. This is true for the algebra $\mathcal{A}_{R,0}$ by definition: we deliberately only included noncentral single-trace operators in its construction. What happens if we do add central single-trace operators to the large $N$ algebra? In general the set of single-trace central operators is finite-dimensional and related
to conserved charges of the theory.
For our purposes, the most important conserved charges are the left and right Hamiltonians $H_L$ and $H_R$. (We briefly discuss other conserved charges in Appendix \ref{app:otherscharges}.) The left and right Hamiltonians are single-trace operators, but they do not have a sensible large $N$ limit, because the thermal expectation value $\langle H_{R}\rangle_\beta$ diverges as $O(N^2)$, and similarly for $H_L$. Even if we subtract this expectation value, the fluctuations 
\begin{align}
\langle (H_{R} - \langle H_{R}\rangle_\beta)^2 \rangle \sim N^2
\end{align}
diverge for any state in the Hilbert space $\mathcal{H}_0$. To obtain a large $N$ limit with finite fluctuations, we need to use the rescaled operator $U = (H_R - \langle H_R\rangle_\beta)/N$. Only the operator $U$, and not $H_R$, is an appropriately normalized single-trace operator according to the convention used at the start of this section. However the operator $U$ is central at $N=\infty$ because
\begin{align}
[U, a] = -\frac{1}{N}i\partial_t a \to 0, \,\,\,\,\,\,\,\,\,\,\,\,\,\,\,\,\,\,\,~ \forall a \in \mathcal{A}_{R,0}. 
\end{align}
As a result, $U$ was not included in the algebra $\mathcal{A}_{R,0}$. However we are free to include it in an extended algebra 
    $\mathcal{A}_{R,0} \otimes \mathcal{A}_{U},$
where $\mathcal{A}_U$ is the algebra of bounded functions of $U$. 
There is a similar story for $H_L$ and more generally for any conserved charge; the  conserved charges have the general form $Q=N\,\Tr\,T$ for some $T$, so $\Tr\,T=Q/N$ is $1/N$ times a symmetry generator and is central in the large $N$ limit.

In the large $N$ limit, the operator $U$ has a continuous spectrum.
So the  algebra   $ \mathcal{A}_{R,0} \otimes \mathcal{A}_{U}$
acts on an extended large $N$ Hilbert space $\mathcal{H}_0 \otimes L^2(\mathbb{R})$, where $U$ acts on $L^2(\mathbb{R})$ as the position operator. The algebra $\mathcal{A}_{R,0} \otimes \mathcal{A}_U$ is still Type III$_1$, but it is no longer a factor, because it now contains the infinite-dimensional center $\mathcal{A}_U$. In the large $N$ limit, $\ket{\mathrm{TFD}}$ has Gaussian correlation functions for $U$, with a variance $\langle U^2 \rangle_\beta = -(1/N^2) \partial_\beta \langle H\rangle_\beta$ controlled by the heat capacity of the black hole. It follows that we should identify the thermofield double state with
\begin{align}
   \ket{\widehat\Psi} =  \int \,dU \,\left(\frac{N^2}{2 \pi|\partial_\beta \langle H\rangle_\beta|}\right)^{1/2} \exp\left( - \frac{N^2 U^2}{2|\partial_\beta \langle H \rangle_\beta|}\right)\,\ket{\Psi}\ket{U}.
\end{align}

Since $U$ is central, it is also contained in the commutant algebra 
\begin{align}
    (\mathcal{A}_{R,0} \otimes \mathcal{A}_{U})' = \mathcal{A}_{L,0} \otimes \mathcal{A}_U.
\end{align}
Because we defined $U$ using the right boundary Hamiltonian $H_R$, one might worry that the commutant algebra can no longer be identified with the algebra of operators on the left boundary. However this is not the case. The difference $\hat h =H_R - H_L$ between the right and left Hamiltonians  annihilates the thermofield double state, and its commutators with single-trace operators are themselves $O(1)$ single-trace operators. As a result, $\hat h = O(1)$ in the large $N$ limit for all states, and hence 
\begin{align}
    U = \frac{H_R - \langle H\rangle_\beta}{N} = \frac{H_L - \langle H\rangle_\beta + \hat h}{N} = \frac{H_L - \langle H\rangle_\beta}{N}+O(1/N).
\end{align} 
In the strict infinite $N$ limit the same operator $U$ therefore describes both the rescaled and subtracted left boundary energy and the rescaled and subtracted right boundary energy.

The commutation relations $[\hat h, a_R]= -i\partial_t a_R$ for $a_R \in \mathcal{A}_{R,0}$ and $[\hat h, a_L]= i\partial_t a_R$ for $a_L \in \mathcal{A}_{L,0}$ (along with $\hat h \ket{\mathrm{TFD}}=0$) mean that in the large $N$ limit $\hat h$ defines an operator acting on $\mathcal{H}_0$. In fact, the action of $\hat h$ on $\mathcal{H}_0$ is related to that of the modular operator $\Delta_\Psi$ for the state $\ket{\Psi}$ on the algebra $\mathcal{A}_{R,0}$ by $\beta \hat h = -\log \Delta_\Psi$ \cite{Leutheusser:2021ab,Leutheusser:2021aa, wittengcp}.\footnote{Here, the modular operator $\Delta_\Psi = S_\Psi^\dagger S_\Psi$ is defined via the antilinear Tomita operator $S_\Psi$ that satisfies  $S_\Psi a \ket{\Psi} = a^\dagger \ket{\Psi}$ for all $a \in \mathcal{A}_{R,0}$ (see for example \cite{Witten:2018wc}). For a Type I or II von Neumann algebra $\mathcal{A}$, the modular operator for a state $\ket{\Psi}$ is related to the density matrices $\rho \in \mathcal{A}$ and $\rho' \in \mathcal{A}'$ for $\ket{\Psi}$ by $\Delta_\Psi = \rho \rho'^{-1}$. } This relationship can be most easily seen by going to finite $N$ where the algebras are Type I and hence the modular operator is $\Delta_{\mathrm{TFD}} = \rho_L^{-1} \otimes \rho_R = \exp(-
\beta(H_R - H_L))$. The operator $\log \Delta_\Psi$ generates a group of outer automorphisms of the right boundary algebra $\mathcal{A}_{R,0}$ known as the modular flow; in this case, the modular flow is simply the group of time translations. This group will play an important role in the following.

So far we have described the strict $N \to \infty$ limit. In \cite{wittengcp}, perturbative corrections to this limit were considered by working in a formal power series in $1/N$. Once these corrections are included, the operator $U$ stops being central and the algebra becomes Type II$_\infty$. (Of course, since we are now working with algebras over the ring $\mathbb{C}[[1/N]]$ of formal power series, rather than over the complex numbers, it is unclear to what extent the usual classification of von Neumann algebras applies.) We will discuss this approach in Appendix \ref{app:corrections}. In the main body of this paper, we will instead take an alternative large $N$ limit, starting from a microcanonical rather than a canonical ensemble, and thereby obtain a Type II$_\infty$ algebra even in the strict $N \to \infty$ limit.

\subsection{Microcanonical ensemble \label{sec:micro}}   
In the microcanonical ensemble approach, we consider a narrow band of energy eigenstates centered around some energy $E_0$. The typical energy $E_0$ will be $O(N^2)$ as in the canonical ensemble above the Hawking-Page transition, but the fluctuations in energy about $E_0$ will only be $O(1)$, as opposed to $O(N)$ for the canonical 
ensemble. Our starting point is a microcanonical version of the thermofield double state \eqref{eq:tfd}. Specifically, let
\begin{align}
|\widetilde{\text{TFD}}\rangle = e^{-S(E_0)/2} \sum_{i}e^{-\beta (E_i -E_0)/2}f(E_i - E_0)|E_i\rangle_L |E_i\rangle_R,  \label{eq:microtfd}
\end{align} 
where $S(E_0)$ is the Bekenstein-Hawking entropy of a black hole with energy $E_0$, and $f(E-E_0)$ is any smooth invertible function that is independent of $N$ and satisfies
\begin{align}
    \int_{-\infty}^\infty dx |f(x)|^2 = 1
\end{align}
 It is easy to check that the state $\ket{\widetilde{\mathrm{TFD}}}$ will then be normalized at large $N$. For example, we could take $f(E- E_0)$ to be a Gaussian of any fixed $O(1)$ width $\sigma$:
\begin{align}
f(E - E_0) = (2 \pi \sigma^2)^{-1/4} e^{-(E - E_0)^2/4\sigma^2}. \label{microdist}
\end{align}
The choice of $f$ may seem somewhat arbitrary or artificial. Fortunately, as we shall see, the final algebra and Hilbert space that we construct will be independent of this choice.

At leading order in $1/N$, correlation functions of right-boundary noncentral single-trace operators for the state $|\widetilde{\text{TFD}}\rangle$ are thermal, just like for the thermofield double state $\ket{\text{TFD}}$. To see this, note that for any product of right-boundary noncentral single-trace operators $t_1 \dots t_n$ we have
\begin{align}
\braket{\widetilde{\mathrm{TFD}}|t_1 \dots t_n|\widetilde{\mathrm{TFD}}} & = e^{-S(E_0)} \sum_i |f(E_i - E_0)|^2 e^{-\beta(E_i - E_0)} \braket{E_i| t_i \dots t_n|E_i}
\\& = e^{-S(E_0)} \sum_i \int_{-\infty}^\infty dt \, F (t) e^{-(\beta + i t)(E_i - E_0)} \braket{E_i| t_i \dots t_n|E_i}
\\&= e^{-S(E_0)} \int_{-\infty}^\infty dt \, F (t) e^{(\beta + i t) E_0} Z(\beta + it) \braket{t_i \dots t_n}_{\beta+it},\label{eq:micro=can}
\end{align}
where $F$ is the Fourier transform of $|f|^2$. At large $N$, we can approximate this integral using the saddle point for $e^{(\beta + i t) E_0} Z(\beta + it)$ at $t = 0$, leading to $\braket{\widetilde{\mathrm{TFD}}|t_1 \dots t_n|\widetilde{\mathrm{TFD}}} \approx \braket{t_i \dots t_n)}_{\beta}$. It is important here that the operators $\{t_i\}$ are noncentral and hence that $\braket{U^2 t_1 \dots t_n}_\beta \to \braket{U^2}_\beta \braket{t_1 \dots t_n}_\beta$ as $N \to \infty$. Otherwise we would have $\partial_t^2 \braket{t_i \dots t_n)}_{\beta+it} = O(N^2)$, and hence $O(1)$ corrections to $\braket{t_1 \dots t_n}$ from the perturbative expansion about the saddle point. 

It follows that, in the large $N$ limit, both $|\widetilde{\text{TFD}}\rangle$ and $\ket{\text{TFD}}$ lead to the same Type III$_1$ algebra $\mathcal{A}_{R,0}$ for right boundary noncentral single-trace operators.\footnote{The reason that we previously identified the state $\ket{\Psi} \in \mathcal{H}_0$ with the large $N$ limit of $\ket{\text{TFD}}$, rather than $|\widetilde{\text{TFD}}\rangle$, is that this led to a natural identification  $\mathcal{A}_{L,0} = \mathcal{A}_{R,0}'$  between the commutant 
$\mathcal{A}_{R,0}'$  of     $\mathcal{A}_{R,0}$
    and  the algebra     $\mathcal{A}_{L,0}$ of single-trace left boundary operators. Unlike purely right-boundary correlators, left-right correlators of noncentral single-trace operators differ between $\ket{\mathrm{TFD}}$ and $\ket{\widetilde{\mathrm{TFD}}}$ even at large $N$. As a result, we cannot simultaneously identify the state $\ket{\Psi}$ with $\ket{\widetilde{\mathrm{TFD}}}$, the algebra $\mathcal{A}_{R,0}$ with the large $N$ limit of right-boundary single-trace operators, and $\mathcal{A}_{R,0}'$ with the large $N$ limit of left boundary operators (with modular conjugation using $J_\Psi$ time reversing and exchanging the left and right boundaries). We will understand how to interpret left boundary operators acting on the large $N$ limit of $|\widetilde{\text{TFD}}\rangle$ below.}
Unlike the canonical ensemble however, the state  $|\widetilde{\text{TFD}}\rangle$ has finite fluctuations of the ``renormalized'' Hamiltonian $h_R = H_R - E_0$ even in the large $N$ limit. Specifically, for $u,v\in \mathbb R$, we have
\begin{align}\label{eq:stdevh_R}
    \lim_{N\to\infty} \braket{\widetilde{\text{TFD}}|\,\Pi_{h_R}([u,v])\,|\widetilde{\text{TFD}}} = \int_{u}^v dh_R |f(h_R)|^2,
\end{align}
where $\Pi_{h_R}([u,v])$ projects onto $h_R \in [u,v]$.

The operator $h_R$ can therefore be included in a large $N$ algebra along with the noncentral single-trace operators. In contrast to the operator $U$ that we previous added to the canonical ensemble large $N$ algebra, $h_R$ is not central, since
\begin{align}
[h_R, a] = -i\partial_t a = O(1), ~ a \in \mathcal{A}_{R,0}. 
\end{align}
On the other hand, unlike the operator $\hat h = h_R - h_L$, $h_R$ does not preserve the Hilbert space $\mathcal{H}_0$ in the large $N$ limit. For example, the state $h_R \ket{\widetilde{\mathrm{TFD}}}$ cannot be prepared from $\ket{\widetilde{\mathrm{TFD}}}$ using finite products of noncentral single-trace operators, and so its large $N$ limit is not described by $\mathcal{H}_0$. Of course, the large $N$ limit of $h_R \ket{\widetilde{\mathrm{TFD}}}$ will be contained in a larger Hilbert space constructed using an algebra that includes $h_R$ -- just like adding $U$ to the canonical ensemble algebra led to an action of $\mathcal{A}_{R,0} \otimes \mathcal{A}_U$ on an extended large $N$ Hilbert space $\mathcal{H}_0 \otimes L^2(\mathbb{R})$.

To understand better the extended large $N$ Hilbert space and algebra that include $h_R$, it is helpful to decompose $h_R = h_L + \hat h$. We already know the action of $\hat h = -(1/\beta) \log \Delta_\Psi$ on $\mathcal{H}_0$ from Section \ref{sec:canon}. What about $h_L$? Just like the operator $U$ in Section \ref{sec:canon}, $h_L$ commutes with $\mathcal{A}_{R,0}$. Moreover operators $a \in \mathcal{A}_{R,0}$ are uncorrelated with functions $g(h_L)$:
\begin{align}
    \braket{\widetilde{\mathrm{TFD}}|\,a g(h_L)\,|\widetilde{\mathrm{TFD}}} &= e^{-S(E_0)} \sum_i g(E_i - E_0) |f(E_i - E_0)|^2 e^{-\beta(E_i - E_0)} \braket{E_i| a|E_i}
    \\&\to \braket{a}_\beta \int_{-\infty}^\infty dx\, g(x) \,|f(x)|^2 \,\,\,\,\text{   as }N\to \infty.
\end{align}
In the second step we used the large $N$ saddle point approximation to \eqref{eq:micro=can} with $|f|^2$ replaced by $g|f|^2$. The extended large $N$ Hilbert space is therefore $\mathcal{H} \cong \mathcal{H}_0 \otimes L^2(\mathbb{R})$, with the operator $h_L$ acting as the position operator $x$ on $L^2(\mathbb{R})$, while  the large $N$ limit of the state $\ket{\widetilde{\mathrm{TFD}}}$ identified with
\begin{align} \label{eq:hatpsidef}
   \ket{\widehat\Psi} = \int\,dx\,f(x)\, \ket{\Psi}\ket{x}. 
\end{align}
 Note that by acting with the operator $g(h_R) = f'(h_R) f^{-1}(h_R)$ we can map the state $|\widetilde{\text{TFD}}\rangle$ to the state
\begin{align}
    |\widetilde{\text{TFD}}'\rangle = g(h_R) |\widetilde{\text{TFD}}\rangle = e^{-S(E_0)/2} \sum_{i}e^{-\beta (E_i -E_0)/2}f'(E_i - E_0)|E_i\rangle_L |E_i\rangle_R, 
\end{align}
for any square-integrable function $f'$. The large $N$ limit of $|\widetilde{\text{TFD}}'\rangle$ is therefore identified with
\begin{align}
g(h_R) \ket{\widehat\Psi} =  g(h_L) \ket{\widehat\Psi} = \int\,dx\,f'(x)\, \ket{\Psi}\ket{x}.   
\end{align}
The extended large $N$ Hilbert space and algebra is therefore independent of the choice of $f$: once we include bounded functions $g(h_R)$ in our algebra, all possible choices of $f$ end up being included in the same large $N$ Hilbert space.

The large $N$ algebra $\mathcal{A}_R$ acting on $\mathcal{H}$ that describes right boundary operators is generated by $\mathcal{A}_{R,0}$ along with
\begin{align}
    \beta h_R = \beta (h_L + \hat h) = \beta x + h_{\Psi}~,
\end{align}
where $h_{\Psi} = - \log \Delta_\Psi$. This 
 algebra is known in the mathematics literature as the crossed product $\mathcal{A}_{R,0} \rtimes \mathbb{R}_h$ of the algebra $\mathcal{A}_{R,0}$ by the modular group for the state $\ket{\Psi}$. It is a standard fact that for any Type III$_1$ von Neumann factor $\mathcal{A}_{R,0}$ and cyclic separating state $\ket{\Psi}$, the algebra $\mathcal{A}_R$ is a Type II$_\infty$ factor \cite{Takesakicp} (see \cite{wittengcp} for an introduction). The primary difference between a Type II von Neumann algebra and a Type III von Neumann algebra is that Type II von Neumann algebras admit a trace $\tr$ -- a positive linear functional on operators in the algebra satisfying
\begin{align} \label{eq:tracecond}
    \tr[a b] = \tr[b a]~,
\end{align}
for any pair of operators $a,b$ in the algebra. In the case of the algebra $\mathcal{A}_{R}$, the trace of an operator $\hat a \in \mathcal{A}_R$ can be defined as\footnote{The fact that this functional satisfies the condition \eqref{eq:tracecond} is nontrivial. See \cite{wittengcp} for details (note that $x$ in \cite{wittengcp} corresponds to $\beta x$ in our notation).}
\begin{align}\label{eq:crossedtrace}
    \tr[\hat a] = \int_{-\infty}^\infty dx\, e^{\beta x} \braket{\Psi|\hat a |\Psi}.
\end{align}
Note that the expectation value $\braket{\Psi|\hat a |\Psi}$ is in general a nontrivial function of $x$; this is necessary for the integral to converge. A generic operator in $\hat a \in \mathcal{A}_R$ can be expanded as 
\begin{align}
\hat a = \int_{-\infty}^\infty ds \, a_0(s) e^{i s (x + \hat h)}    
\end{align}
for $a_0(s) \in \mathcal{A}_{R,0}$. The integral in \eqref{eq:crossedtrace} will converge if the Fourier transform of the function $\braket{\Psi|a_0(s) e^{i s \hat h}|\Psi}$ decays sufficiently quickly as $x \to +\infty$. 

We emphasize that this trace is not the same as the standard trace on $\mathcal{H}$, which will be infinite for any $a \in \mathcal{A}_R$. However the condition \eqref{eq:tracecond} means that we should think of \eqref{eq:crossedtrace} as a renormalised (i.e. infinitely-rescaled) version of this trace. The difference between a Type II$_1$ algebra and a Type II$_\infty$ algebra is simply whether this trace is finite for all operators -- including in particular the identity operator (Type II$_1$ algebras) -- or is only finite for a dense set of operators in the weak/strong operator topologies (Type II$_\infty$ algebras). In our case we have
\begin{align}
    \tr[1] = \int_{-\infty}^\infty dx\, e^{\beta x} = + \infty~,
\end{align}
and so the algebra is Type II$_\infty$. Because the algebra $\mathcal{A}_R$ is a factor, the trace $\tr$ is unique up to rescaling. However unlike for the traditional trace on a Hilbert space, which has a natural normalisation where the trace of a projector onto a pure state $\ket{\psi}\bra{\psi}$ is one, there is no canonical normalisation of the trace on a Type II$_\infty$ algebra. In fact, the algebra $\mathcal{A}_R$ has a one-parameter family of outer automorphisms\footnote{Since $x = h_L = H_L - E_0$, this has a physical interpretation as a shift in the renormalization constant $E_0$ to $E_0 - c$.}
$x\to x+c$; this automorphism rescales $\tr$ by $e^c$.

The existence of a trace allows us to define density matrices for the algebra $\mathcal{A}_R$: given any state $\ket{\Phi} \in \mathcal{H}$, the density matrix $\rho_\Phi \in \mathcal{A}_R$ is defined by
\begin{align} \label{eq:densitymatrixdef}
    \tr[\rho_\Phi a] = \braket{\Phi|a |\Phi} \,\,\,\forall a \in \mathcal{A}_R.
\end{align}
The existence and uniqueness of $\rho_\Phi$ follows from the fact that the inner product on $\mathcal{A}_R$ defined by $\braket{a|b} = \tr[a^\dagger b]$ is nondegenerate. Importantly, if we can define density matrices, we can also define entropies, by the usual formula
\begin{align}
    S(\Phi)_{\mathcal{A}_R} = - \tr[\rho_\Phi \log \rho_\Phi] = -\braket{\Phi| \log \rho_\Phi |\Phi}.
\end{align}
Again, this entropy is not the ``real'' entanglement entropy of the algebra $\mathcal{A}_R$, which would be infinite for all states $\ket{\Phi}$. Instead, it should be thought of as a renormalized entropy where we have subtracted a divergent constant piece. If we rescale the trace $\tr \to e^c \tr$ as discussed above, then \eqref{eq:densitymatrixdef} implies that the density matrix  is rescaled by $\rho_\Phi \to e^{-c} \rho_\Phi$. It follows that
\begin{align}
S(\Phi)_{\mathcal{A}_R} \to S(\Phi)_{\mathcal{A}_R} + c.
\end{align}
Because the choice of normalization for $\tr$ was essentially arbitrary, this means that the entropy of any individual state $\Phi$ is also dependent on an arbitrary normalization choice; essentially it depends on our choice of renormalization scheme. However the difference between the entropies of two states $\ket{\Phi_1}$ and $\ket{\Phi_2}$ is a physical quantity that is independent of this choice.

So far we have focussed on the right boundary algebra $\mathcal{A}_R$. What about the left boundary? Because $\mathcal{A}_{L,0} = \mathcal{A}_{R,0}'$ does not commute with $\hat h$, $\mathcal{A}_{L,0}$ is not contained in the left boundary algebra $\mathcal{A}_L = \mathcal{A}_R'$. Instead, $\mathcal{A}_L$ is generated by $x = h_L $ together with $e^{i p \hat h} \mathcal{A}_{L,0} e^{-i p \hat h}$, where $p$ is canonically conjugate to $x$ and acts only on $L^2(\mathbb{R})$.\footnote{One can rewrite the algebras $\mathcal{A}_L$ and $\mathcal{A}_R$ in a more symmetric form by conjugating with $e^{i p \hat h/2}$, but the effect is only to make all formulas strictly less convenient.} The conjugation by $e^{i p \hat h}$ ensures that $h_L$ generates time evolution of the left boundary algebra (i.e. $[h_L, a_L] = -i \partial_t a_L$ for all $a_L \in \mathcal{A}_L$), while $[h_R,\mathcal{A}_L] = 0$, as at finite $N$.

To summarize: the same Type III$_1$ von Neumann factor $\mathcal{A}_{R,0}$ describes the large $N$ limit of noncentral single-trace operators in both the canonical and microcanonical ensembles. However, if we try to add the boundary Hamiltonian (renormalized and rescaled as necessary) to this algebra then the two ensembles behave very differently. In the canonical ensemble, the operator with a sensible large $N$ limit is $U = (H_R-E_0)/N$. This leads to a Type III$_1$ von Neumann algebra with an infinite-dimensional center consisting of bounded functions of $U$. In the microcanonical ensemble, energy fluctuations are finite and so the operator with a sensible large $N$ limit is instead $h_R = H_R - E_0$ with no factor of $1/N$. The resulting algebra is a Type II$_\infty$ von Neumann factor that is the crossed product of $\mathcal{A}_{R,0}$ by the modular group of the state $\ket{\Psi}$.

\subsection{Bulk interpretation \label{sec:interp}}

The discussion above becomes clearer once we understand the bulk description of the Hilbert space $\mathcal{H}$ and the algebra $\mathcal{A}_R$. We already argued that the Hilbert space $\mathcal{H}_0$ is dual to the Hilbert space of low energy quantum field theory on the two-sided black hole background, with the algebra $\mathcal{A}_{R,0}$ dual to the algebra of QFT operators $\mathcal{A}_{r,0}$ in the right exterior. The extended Hilbert space $\mathcal{H}$ contains an additional mode $L^2(\mathbb{R})$. What is its bulk interpretation?

Since the renormalized left boundary Hamiltonian $h_L$ acts on $\mathcal{H}$ as $x$, it generates translations of the conjugate variable $p$ on $L^2(\mathbb{R})$. The bulk dual of $h_L$ is the renormalized left ADM mass, whose bulk action generates time translations of the left boundary. It follows that we should identify the variable $p$ with the timeshift $\Delta$ between the left and right boundaries, which is a physical observable in the zero-coupling limit of quantum gravity that (along with the ADM masses at each boundary) is not present in quantum field theory in a fixed black hole background. The timeshift $\Delta$ is defined as follows.\footnote{For an accessible discussion of this mode in the context of pure JT gravity, see \cite{Harlow:2018tqv}.} Schwarzschild coordinates give a preferred timeslice (defined by fixed Schwarzschild time) extending from any boundary time $t_R$ to the bifurcation surface. This slice can be extended to a unique zero-extrinsic-curvature slice that continues to the left boundary. If the Schwarzschild slice reaches the left boundary at time $t_L$ then the timeshift is $\Delta = t_R + t_L$; the symmetry of the Schwarzschild solution under equal and opposite shifts of the time
on the two sides ensures that $\Delta$ is independent of the initial choice of $t_R$. This timeshift can be eliminated by a diffeomorphism, but only by a diffeomorphism that acts nontrivially at the asymptotic boundary. It is therefore a physical mode and not simply a gauge choice. The timeshift itself is not measurable locally on either the left or the right; however it is conjugate to both the left and right ADM masses, which can be locally measured. 

The full large $N$ Hilbert space $\mathcal{H}$ is therefore holographically dual to the Hilbert space that describes QFT modes in the black hole background, together with the timeshift mode $L^2(\mathbb{R})$. The algebra $\mathcal{A}_R$ is dual to the algebra $\mathcal{A}_r$ of bulk observables that can be measured at the right asymptotic boundary. By  HKLL reconstruction, or equivalently the timelike tube theorem, this algebra includes the full algebra $\mathcal{A}_{r,0}$ of QFT modes in the right exterior, along with the (renormalized) right ADM mass $h_R$.

This bulk interpretation helps explain why the left boundary algebra $\mathcal{A}_L$ is generated by $h_L$ and $e^{i p\hat h} \mathcal{A}_{L,0} e^{-i p \hat h}$ rather than $\mathcal{A}_{L,0}$. The algebra $\mathcal{A}_L$ is holographically dual to the algebra $\mathcal{A}_{\ell}$ of bulk operators accessible at the left asymptotic boundary. Naively this should include the algebra $\mathcal{A}_{\ell,0}$ of bulk QFT operators in the left exterior, which we previously argued was holographically dual to $\mathcal{A}_{L,0}$. However, local quantum field theory operators in quantum gravity are not gauge invariant, and hence are not physical observables, unless they are gravitationally dressed, e.g. to an asymptotic boundary. In Section \ref{sec:canon}, the bulk geometry was completely fixed, and so gravitational dressing did not change the algebra. However we are now working in a Hilbert space where the timeshift between the two boundaries is allowed to fluctuate. As a result, an operator dressed using the right boundary time is different from the same bulk QFT operator dressed using the left boundary time. Since we are working in a gauge where $h_L$ commutes with operators acting on $\mathcal{H}_0$, operators in $\mathcal{A}_{L,0}$ are, by definition, gravitationally dressed to the right boundary. As a result, they cannot be measured from the left asymptotic boundary, even though they act on bulk quantum fields in the left exterior. Instead the operators in $\mathcal{A}_L$ need to be gravitationally dressed to the left boundary. To switch the gravitational dressing of operators in $\mathcal{A}_{L,0}$ from the right boundary to the left, we need to evolve them in Schwarzschild time by an amount equal to the timeshift $\Delta = p$ between the two boundaries. This is exactly what the conjugation by $e^{i p \hat h}$ achieves.

The timeshift $p$ is canonically conjugate to both the left and right renormalized Hamiltonians $h_L$ and $h_R$. It follows that they obey uncertainty relations
\begin{align}
\Delta p \Delta h_{L/R} \geq \frac{1}{2}~.
\end{align}
If we want to construct states where the bulk geometry is approximately fixed, rather than being a superposition of different time evolutions of the black hole geometry, then the fluctuations in the timeshift $p$ need to be small. As a result the fluctuations in both $h_L$ and $h_R$ must be parametrically large. If a state $\ket{\widehat\Phi} \in \mathcal{H}$ has $\braket{\Delta p} = O(\varepsilon)$ for some small $\varepsilon$, we say that $\ket{\Psi}$ is semiclassical. Such states will be the focus of Section \ref{sec:sgen}, where we show that the entropy of a semiclassical state is approximately equal to the generalized entropy of its bulk dual. 

The extreme limit of a semiclassical state is the thermofield double state, which has $\Delta h_L = \Delta h_R = O(N)$ and $\Delta p = O(1/N)$. One could in principle make $\Delta p$ even smaller by making $\Delta h_{L/R}$ even larger, but then fluctuations in the horizon area $\Delta A = O(\Delta E/N^2)$ become larger than the fluctuations in $\Delta p$ and the overall geometry becomes less semiclassical rather than more. The thermofield double state itself does not have a large $N$ limit in $\mathcal{H}$, since in the limit $N \to \infty$ its wavefunction becomes a delta function $\delta(p)$. A delta function is not a normalizable state in $L^2(\mathbb{R})$. This is why the canonical ensemble has a different algebra in the large $N$ limit; the fact that the large $N$ algebra is Type III rather than Type II is a direct consequence of the divergent fluctuations in the energy, and hence in the entanglement of $\ket{\mathrm{TFD}}$, that follow from the fluctuations in the timeshift being infinitely small.  We will therefore instead work with states where $\Delta p = O(\varepsilon)$ is small but finite in the large $N$ limit, and only afterwards take the limit $\varepsilon \to 0$.

\subsection{The area operator and the curious case of the vanishing center}\label{sec:curious}

It has previously been argued \cite{Harlow:2015lma,Harlow:2016vwg, Dong:2018seb,Akers:2018fow} that the bulk algebra associated to the right exterior of a two-sided black hole should contain the area $A$ of the bifurcation surface as a nontrivial central element. From a boundary perspective, however, the algebra of operators associated to the right boundary is the full algebra of operators on the right CFT and so it is clear that no nontrivial central element can exist. The tension between the apparent existence of a nontrivial center in the bulk algebra, but not in the boundary algebra, is known as the factorisation problem \cite{Harlow:2015lma}.\footnote{Henry Maxfield has suggested that the problem described here -- the existence of a center in the bulk algebra, but not the boundary algebra, associated to each asymptotic boundary of an Einstein-Rosen bridge -- should be known as the factorisation problem (with an s) in order to distinguish it from the related factorization problem -- where quantum gravity partition functions appear not to factorize on a product of disconnected spacetime boundaries thanks to contributions from spacetime wormholes \cite{Maldacena:2004rf, Saad:2019lba}. We are supportive of this suggestion.}

In pure JT gravity \cite{Harlow:2018tqv}, the algebra of observables at either the left or right boundaries is commutative and consists solely of bounded function of the ADM mass. As a result, all asymptotic boundary operators are central. Because of the absence of matter fields, the left and right ADM masses are equal, and are an invertible function of the horizon area. This is all in complete accordance with the story from \cite{Harlow:2015lma,Harlow:2016vwg, Dong:2018seb,Akers:2018fow}. That story is also consistent with our discussion of the canonical ensemble, where the operator $U = h_L/N = h_R/ N$ is central and is a linear function of the horizon area at leading order in $1/N$. However, in the microcanonical ensemble, the center of the algebra $\mathcal{A}_R$ -- and, more importantly, of the isomorphic dual bulk algebra $\mathcal{A}_r$ -- consists solely of c-numbers. The factorisation problem has vanished.\footnote{Of course, one still needs to understand how nonperturbative gravitational interactions change the bulk algebra from the Type II factor $\mathcal{A}_r$ into the Type I factor $\mathcal{B}(\mathcal{H}_\mathrm{CFT})$ of bounded operators on the finite $N$ boundary CFT Hilbert space. However, since both $\mathcal{A}_r$ and $\mathcal{B}(\mathcal{H}_\mathrm{CFT})$ are factors, this problem presumably needs a different name.} What happened?

The area $A$ of an extremal surface is canonically conjugate to the boost angle $s$ across that surface \cite{Banados:1993qp, Carlip:1994gc}, with the commutation relation $[A/4G,s] = i$. In the canonical ensemble, fluctuations $\Delta A$ in the area have magnitude $O(1/N)$ and hence the natural large $N$ operator is $U/T = (A-A_0)/N$, where $A_0$ is the saddle point horizon area of a black hole at temperature $T$. Since $G= O(1/N^2)$, the boost angles generated by this operator are perturbatively small in the large $N$ limit. In contrast, in the microcanonical ensemble $\Delta A = O(G)$, and hence the natural large $N$ operator proportional to $A$ appears to be $(A - A_0)/4G$ itself. Such an operator would generate $O(1)$ boost angles across the horizon, even at large $N$.

In the absence of the bulk quantum fields in $\mathcal{H}_0$, this would be consistent with our results, with the boost angle $s$ directly related to the timeshift $p$ by $p = \beta s$. However, if we try to change the boost angle $s$ while keeping the state of the matter fields in each exterior fixed (i.e. dressing each exterior to its respective boundary), the state of the QFT modes will become singular, and no longer be described by the continuum Hilbert space $\mathcal{H}_0$. The best that one can do is to construct a nonsingular operator that boosts the left exterior relative to the right exterior everywhere except very close to the horizon. The operator that does this can be though of a slightly smeared out version of $A/4G$, just like local operators need to be smeared to produce finite continuum operators in ordinary QFT. However this smeared out area operator is no longer central because it won't commute with operators (in at least one of the left or right exterior) that are sufficiently close to the horizon.

A similar story shows up in gauge theories. Naively, the electric or magnetic flux across the boundary of a region lies in the center of the algebra associated to that region. However, to obtain a well-defined operator in a continuum we need to smear that flux slightly in spacetime. In pure $U(1)$ gauge theory (analogous to pure JT gravity) the Gauss law constraint means that the smeared operator remains central. However this is not true in theories where all (electric and magnetic) matter charges are present (analogous to the existence of matter fields and/or gravitons); the algebra associated to the region should then presumably be a crossed product algebra that is a von Neumann factor.\footnote{We are unclear on whether this claim has previously been stated in this form anywhere in the literature; however see \cite{Harlow:2015lma, Harlow:2018tng} for closely related discussions with the same basic conclusions.} 

A more precise version of this argument is the following. Formally, we could try to write the boost generator $\hat h$ on the QFT Hilbert space $\mathcal{H}_0$ as $\hat h = \hat h_r - \hat h_l$ where $\hat h_r \in \mathcal{A}_{r,0}$ generates boosts of the right exterior and $-\hat h_\ell \in \mathcal{A}_{l,0}$ generates boosts in the left exterior.\footnote{The sign difference here comes from the fact that boosts which evolve the right exterior forwards in time evolve the left exterior backwards in time.} If the operators $\hat h_r$ and $\hat h_\ell$ existed, the operator $h_R - \hat h_r = h_L - \hat h_\ell$ would be central and would be holographically dual to $(A-A_0)/4G \beta$. However it is a standard fact about quantum field theory that the operators $\hat h_\ell$ and $\hat h_r$ do not exist, precisely because they act as a one-sided boost and hence create singular states. It is only \emph{after} we include the $L^2(\mathbb{R})$ mode that we can split $\hat h = h_R - h_L$. 

The existence of such a splitting is at the heart of the difference between a Type II algebra and a Type III algebra. Since $\beta \hat h = - \log \Delta_\Psi$, the fact that $h_\ell$ and $h_r$ don't exist means that $\log \Delta_\Psi$ cannot be decomposed into the sum of an operator in $\mathcal{A}_{R,0}$ and one in $\mathcal{A}_{R,0}'$.
In contrast, the modular operator $\Delta_{\widehat\Psi}$ of the state $\ket{\widehat\Psi}$ defined in \eqref{eq:hatpsidef} for the algebra $\mathcal{A}_R$ satisfies\footnote{See \cite{wittengcp} for a derivation.}
\begin{align} \label{eq:cqmodular}
\log \Delta_{\widehat\Psi} = \log[\Delta_{\Psi} |f(x + \hat h)|^2 |f(x)|^{-2}] = \left[-\beta h_R + \log |f(h_R)|^2\right] - \left[-\beta h_L  + \log |f(h_L)|^2\right]~.
\end{align}
We can therefore split $\log \Delta_{\widehat\Psi}$ into an element of $\mathcal{A}_R$ and an element of $\mathcal{A}_L$. A modular operator can be factored in this way if and only if the algebra is Type I or Type II. For a Type I or II factor $\mathcal{A}$, we have 
\begin{align} \label{eq:modopsplit}
    \log \Delta_\Phi = \log \rho_\Phi  - \log \rho_\Phi'\,,
\end{align}
where $\rho_\Phi$ is the density matrix of $\ket{\Phi}$ on $\mathcal{A}$ and $\rho_\Phi'$ is the density matrix of $\ket{\Phi}$ on $\mathcal{A}'$.\footnote{For Type II factors, this statement is true so long as the relative normalization of the traces on $\mathcal{A}$ and $\mathcal{A}'$ is chosen correctly (otherwise there will be an additional constant term).} We therefore conclude that\footnote{Recall that the overall scaling of $\rho_{\widehat\Psi}$ is determined by the condition $\tr[\rho_{\widehat\Psi}] = 1$.}
\begin{align}
    \rho_{\widehat\Psi} = e^{-\beta h_R} |f(h_R)|^2~. \label{eq:cqdensity}
\end{align}
It can be easily verified that this indeed satisfies \eqref{eq:densitymatrixdef}. This equation will prove useful in Section \ref{sec:sgen}.

Before moving on, we make a couple of final observations about the relationship between the bulk algebras $\mathcal{A}_{r,0}$ and $\mathcal{A}_r$ and holography. Even though the bulk QFT algebra $\mathcal{A}_{r,0}$ is holographically dual to a large $N$ boundary algebra $\mathcal{A}_{R,0}$, it does not itself know about anything holographic. The same algebra $\mathcal{A}_{r,0}$ would appear in quantum field theory in curved spacetime, with no gravity present at all. This is not true for the algebra $\mathcal{A}_r$. This larger algebra knows that the ADM energy $h_R$ can be measured as an operator at infinity, which is a purely gravitational phenomenon. Naively, the fact that in gravitational theories energy can be measured at asymptotic infinity might not seem any more profound than the fact that in gauge theories charge can be measured at asymptotic infinity. However the asymptotic boundary operators in a gauge theory with compact gauge group do not change the algebra from Type III$_1$, whereas in gravity one obtains a Type II$_\infty$ algebra and hence a notion of entropy. And as we are about to see, that is enough for the Bekenstein-Hawking entropy of a black hole to begin to appear. One might therefore say that the algebra $\mathcal{A}_r$ is already ``proto-holographic.'' Of course, to truly obtain a holographic theory we need the asymptotic boundary algebra to be not Type II but Type I. And the step from Type II to Type I is presumably much harder than getting from Type III to Type II, since it requires knowledge of the nonperturbative finite-coupling algebra of observables in quantum gravity.

\section{Entropy equals generalized entropy \label{sec:sgen}}
We have shown that the right boundary algebra $\mathcal{A}_R$ that describes the large $N$ limit of the microcanonical ensemble is a Type II$_\infty$ von Neumann factor, and hence that one can define density matrices and entropies for it. Moreover, the algebra $\mathcal{A}_R$ and the large $N$ Hilbert space $\mathcal{H}$ on which it acts have simple semiclassical bulk duals. In this section we relate the entropy of semiclassical states for the algebra $\mathcal{A}_R$ to the generalized entropy of their bulk duals. We find that the two agree up to parametrically small corrections.

Recall from Section \ref{sec:interp} that we say a state $\ket{\widehat\Phi} \in \mathcal{H}$ is semiclassical if the fluctuations $\Delta p$ in the timeshift $p$ between the two boundaries satisfy 
\begin{align}
\Delta p = O(\varepsilon)
\end{align}
for some parametrically small $\varepsilon \ll 1$. The uncertainty relation
\begin{align}
    \Delta x \Delta p \geq \frac{1}{2}
\end{align}
then requires $\Delta x \geq O(1/\varepsilon)$. A priori, a semiclassical state $\ket{\widehat\Phi}$ can have a large typical value $p_0$ for the timeshift while still having small fluctuations around this value. However, if we apply the unitary $e^{-ip_0 x}$ to the state then we can produce a new semiclassical state $\ket{\widehat\Phi'} = e^{-ip_0 x} \ket{\widehat\Phi}$. Since $e^{-i p_0 x} \in \mathcal{A}_L$ commutes with everything in $\mathcal{A}_R$, $\ket{\widehat\Phi'}$ has the same density matrix and entropy for $\mathcal{A}_R$ as $\ket{\widehat\Phi}$. A simple calculation shows that \begin{align}
    e^{ip_0 x}p e^{-ip_0 x} = p - p_0,
\end{align} 
which means that the typical value of $p$ for the state $\ket{\widehat\Phi'}$ is close to zero. 

As a result, without loss of generality we can assume $p = O(\varepsilon)$ (i.e. $p_0 = 0$). We will also assume, as is conventional in generalized entropy calculations, that the state $\ket{\Phi} \in \mathcal{H}_0$ of the bulk quantum fields is fixed (i.e. independent of the ADM mass). The general form of the semiclassical states that we consider is therefore
\begin{align}
    |\widehat{\Phi}\rangle = \int_{-\infty}^{\infty}dx ~ \varepsilon^{1/2}g(\varepsilon x) |\Phi\rangle |x\rangle, \label{eq:semclass} 
\end{align}
for arbitrary $\ket{\Phi} \in \mathcal{H}$ and normalized $g(x) \in L^2(\mathbb{R})$.

To state our formula for the von Neumann entropy $S(\widehat{\Phi})_{\mathcal{A}_R}$ we first need to define the relative modular operator $\Delta_{\Psi|\Phi}$. The relative Tomita operator $S_{\Psi|\Phi}$ is an antilinear operator defined via
\begin{align} \label{eq:relmoddef}
 S_{\Psi|\Phi} a\ket{\Phi} =  a^\dagger \ket{\Psi},
\end{align} 
for all operators $a \in \mathcal{A}_{R,0}$.\footnote{More precisely it is the closure of this operator. This means that we also define $S_{\Psi|\Phi} \lim_{n \to \infty} a_n\ket{\Phi} := \lim_{n \to \infty} S_{\Psi|\Phi} a_n\ket{\Phi}$ whenever both limits exist.} The relative modular operator is $\Delta_{\Psi|\Phi} = S_{\Psi|\Phi}^\dagger S_{\Psi|\Phi}$; in the special case where $\ket{\Psi} = \ket{\Phi}$ then the relative modular operator $\Delta_{\Psi|\Psi}$ reduces to the ordinary modular operator $\Delta_\Psi$. For a Type I and II von Neumann factor $\mathcal{A}$ we have 
\begin{align}\label{eq:deltafromdens}
\Delta_{\Psi|\Phi} = \rho_\Psi \rho_\Phi'^{-1}
\end{align}
where $\rho_\Psi$ is the density matrix of $\ket{\Psi}$ on $\mathcal{A}$ and $\rho_\Phi'$ is the density matrix of $\ket{\Phi}$ on $\mathcal{A}'$. For a Type III algebra such as $\mathcal{A}_{R,0}$, the right-hand side of \eqref{eq:deltafromdens} is not well defined, but we can still define $\Delta_{\Psi|\Phi}$ via \eqref{eq:relmoddef}.

We are now ready to state the main result of this section. We claim that
\begin{align}
    S(\widehat{\Phi})_{\mathcal{A}_R} = \langle \widehat{\Phi}|\beta h_R|\widehat{\Phi}\rangle - \langle \widehat{\Phi}|h_{\Psi|\Phi}|\widehat{\Phi}\rangle - \langle \widehat{\Phi}|\log \varepsilon |g(\varepsilon h_R)|^2|\widehat{\Phi}\rangle + O(\varepsilon). \label{eq:entrform} 
\end{align}
Here $h_{\Psi|\Phi} = - \log \Delta_{\Psi|\Phi}$. Later in this section, we will provide two independent derivations of this result. However, we first explain the relationship between the right-hand side of \eqref{eq:entrform} and generalized entropy. Recall that $\ket{\Psi}$ is dual to the Hartle-Hawking state of the bulk quantum fields. The second term in \eqref{eq:entrform} is simply the relative entropy\footnote{It is easy to check that for Type I or II von Neumann algebras, \eqref{eq:relentdef} reduces to the usual definition of relative entropy via \eqref{eq:deltafromdens}. For Type III algebras, \eqref{eq:relentdef} is the standard definition of relative entropy.} \begin{align}\label{eq:relentdef}
S_{\text{rel}}(\Phi||\Psi) = - \langle \Phi|\log \Delta_{\Psi|\Phi}|\Phi\rangle.
\end{align}
It was previously shown by Wall \cite{Wall:2011hj} that this relative entropy is precisely the difference
\begin{align}
  S_{\text{rel}}(\Phi||\Psi) =   \Delta S_\mathrm{gen} = S_\mathrm{gen}(\infty) - S_\mathrm{gen}(b)
\end{align}
between the generalized entropy $S_\mathrm{gen}(b)$ of the entire right exterior and the generalized entropy $S_\mathrm{gen}(\infty)$ of the exterior of a cut in the black hole horizon, in the limit where that cut is taken to future infinity.\footnote{In fact, Wall's result was much more general than this, and can be used to compute the generalized entropy of any horizon cut. We will only need the special case of the generalized entropy $S_\mathrm{gen}(b)$ of the bifurcation surface $\partial b$.}

Since this argument is crucial to our derivation, we will briefly review it here. We first note that, in asymptotically anti-de Sitter space, any bulk quantum state on a black hole background becomes indistinguishable from $\ket{\Psi}$ at sufficiently late times, because perturbations
fall in across the horizon. Hence
\begin{align}\label{eq:entatinf}
    S_\mathrm{bulk}(\infty)_{\Phi} = S_\mathrm{bulk}(\infty)_{\Psi} = S_\mathrm{bulk}(b)_{\Psi},
\end{align}
where in the second step we use the fact that $\Psi$ is independent of time.
We are temporarily pretending that the entropies $S_\mathrm{bulk}$ are well defined; our final answer will of course only involve well-defined finite quantities. Now let $v$ be an affine parameterization for the black hole horizon, such that $v=0$ is the bifurcation surface. In the semiclassical limit of small $O(G)$ perturbations around the Schwarzschild black hole solution, Raychaudhuri's equation becomes
\begin{align}
    \partial_v \theta_{v} = - 8 \pi G T_{vv} + O(G^2),
\end{align}
where $\theta_v$ is the classical area expansion along the affine generator. Integrating by parts, and using the fact that for an event horizon $\theta_v \to 0$ as $v \to \infty$, leads to
\begin{align}\label{dolgo}
    \int_{0}^{\infty} dv\, \theta_v = \big[v \,\theta_v\big]^\infty_0 - \int_0^{\infty} dv \,v\,\partial_v \theta_v= 8\pi G \int_0^{\infty} dv \,v\, \langle \Phi| T_{vv}|\Phi\rangle . 
\end{align}
The derivative of the area is $\partial_v A=\int d\Omega \,\partial_v\theta$, where the integral is taken on the horizon cut defined by $v$, and $d\Omega$ is the area element on this cut.   So integrating the formula (\ref{dolgo}) over the horizon and over $v$, and using the fact that the one-sided boost generator discussed in section \ref{sec:curious} satisfies
$\beta \hat h_r=2\pi \int_0^\infty dv \int d\Omega\, T_{vv}$,
 we obtain
\begin{align}\label{eq:areadif}
    \frac{A(\infty)}{4G} - \frac{A(\partial b)}{4G} = \beta \braket{\Phi| \hat h_r |\Phi},
\end{align}
where $\hat h_r$ is the divergent one-sided boost generator discussed in Section \ref{sec:curious}. If density matrices existed for the algebra $\mathcal{A}_{R,0}$ (i.e. the algebra was not Type III) then we would conclude from comparison of \eqref{eq:modopsplit} and $h_\Psi = \beta \hat h = \beta (\hat h_r - \hat h_\ell)$ that
\begin{align} \label{eq:rhoonesidedboost}
    \log \rho_{\Psi} = -\beta \hat h_r + C
 \end{align}
 for some constant $C$. Comparing \begin{align}
 \braket{\Psi|\log \rho_\Psi |\Psi} = - S_\mathrm{bulk}(b)_\Psi
 \end{align} 
 with $\braket{\Psi|\hat h_r|\Psi} = 0$ leads to $C = -S_\mathrm{bulk}(b)_\Psi$. It follows that
\begin{align}\notag
   S_\mathrm{gen}(\infty) - S_\mathrm{gen}(b) &= \frac{A(\infty)}{4G} - \frac{A(\partial b)}{4G} +  S_\mathrm{bulk}(b)_\Psi - S_\mathrm{bulk}(b)_\Phi
   \\ \notag &=-\braket{\Phi|\log \rho_\Psi|\Phi} - S_\mathrm{bulk}(b)_\Phi
   \\&= S_{\text{rel}}(\Phi||\Psi). \label{monoto}
\end{align}
In the first equality we used \eqref{eq:entatinf}, while in the second we used \eqref{eq:areadif}. In the last equality we used the standard formula for relative entropy in terms of density matrices to rewrite the right hand side as a quantity that is well defined even in a Type III algebra.

To complete our proof that the right-hand side of \eqref{eq:entrform} is equal to the generalized entropy $S_\mathrm{gen}(b)$, it remains to show that
\begin{align}
    S_\mathrm{gen}(\infty) \stackrel{?}{=} \braket{\widehat\Phi| \beta h_R |\widehat\Phi} - \braket{\widehat\Phi|\log \varepsilon |g(\varepsilon h_R)|^2|\widehat \Phi} + \mathrm{const}.\label{eq:sgeninf}
\end{align}
Note that both the terms on the right hand side depend only on the energy distribution $p(h_R) = \varepsilon |g(\varepsilon h_R)|^2$ of the state $\ket{\widehat\Phi}$ and not on the state $\ket{\Phi}$ of the bulk quantum fields. This is expected since $\ket{\Phi}$ becomes indistinguishable from $\ket{\Psi}$ at sufficiently late times. At future infinity, because there is no matter left outside the horizon, the ADM mass $h_R$ directly controls the area of the black hole horizon. Over an $O(1)$ range of energies $h_R$, we have
\begin{align} \label{eq:areaenergy}
    \frac{A}{4G} = \frac{A_0}{4G} + \beta h_R ,
\end{align}
for some constant $A_0$ equal to the horizon area of a black hole with energy $E_0$. Hence the first term in \eqref{eq:sgeninf} describes the variation in $S_\mathrm{gen}(\infty)$ from variation in the horizon area. The second term meanwhile describes the entropy of \emph{fluctuations} in the area of the horizon (see e.g. \cite{Das:2001ic, Harlow:2016vwg, wittengcp}). Since the properties of the state $\ket{\widehat\Phi}$ at late times are completely determined by its energy distribution $p(h_R)$, these are the only possible sources of variation in $S_\mathrm{gen}(\infty)$; so indeed \eqref{eq:sgeninf} holds as desired.\footnote{If we had included additional charges, as in Appendix \ref{app:otherscharges}, when constructing the algebra $\mathcal{A}_R$, those would give additional contributions here.}

Putting everything together, we have found that \eqref{eq:entrform} can be written as 
\begin{align}
    S(\widehat{\Phi})_{\mathcal{A}_R} = S_{\text{gen}}(b) + \text{const,}
\end{align}
where the constant on the right hand side diverges in the $G \to 0$ limit. Since the left hand side is the von Neumann entropy of a Type II$_{\infty}$ factor, which is only defined up to a state-independent constant and is implicitly renormalized by an infinite counterterm, this is exactly the result that we were hoping for: \eqref{eq:entrform} really is the statement that boundary entropy in the large $N$ limit is equal bulk generalized entropy in the limit $G \to 0$.

What we have yet to do is provide any evidence that the formula \eqref{eq:entrform} for the boundary entropy is actually correct. We therefore now provide two different, and somewhat complementary, proofs of \eqref{eq:entrform}. 

\subsection{First derivation} \label{sec:firstderiv}
Our first approach to deriving \eqref{eq:entrform} is to show directly that the density matrix $\rho_{\widehat{\Phi}}$ for the state $\ket{\widehat{\Phi}}$ given in \eqref{eq:semclass} can be written approximately as
\begin{align} 
    \rho_{\widehat{\Phi}} \stackrel{?}{\approx} \varepsilon \overline{g}(\varepsilon h_R) e^{-\beta x}\Delta_{\Phi|\Psi}g(\varepsilon h_R) \label{eq:densop}
\end{align}
The approximation in \eqref{eq:densop}, and in all other approximate equalities in this section, is valid up to corrections suppressed by $O(\varepsilon)$. It will then follow immediately that the entropy of $\rho_{\widehat{\Phi}}$ is given by \eqref{eq:entrform}. 
To show \eqref{eq:densop}, we need to show  a) that $\rho_{\widehat{\Phi}}$ is positive, b) that $\rho_{\widehat{\Phi}}\in \mathcal{A}_r$ and c) that we have
\begin{align} \label{eq:densdef}
    \tr[\rho_{\widehat{\Phi}} \hat a] \approx \braket{\widehat{\Phi} |\hat a |\widehat{\Phi}},
\end{align} 
for all operators $\hat a \in \mathcal{A}_R$. Because density matrices on a Type II algebra may be unbounded, a more precise statement of the second condition is that $\rho_{\widehat{\Phi}}$ should be \emph{affiliated} to $\mathcal{A}_R$, meaning that bounded functions of $\rho_{\widehat{\Phi}}$ should be contained in $\mathcal{A}_R$. 

The positivity of $\rho_{\widehat\Phi}$ follows immediately from inspection of \eqref{eq:densop}. To see that $\rho_{\widehat{\Phi}}\in \mathcal{A}_r$, we need to use the Connes cocycle flow
\begin{align}\label{eq:ccflow}
    u_{\Phi|\Psi}(s) = \Delta_{\Phi|\Psi}^{is}\Delta_{\Psi}^{-is} = \Delta_{\Phi}^{is}\Delta_{\Psi|\Phi}^{-is}.
\end{align}
Some important properties of $u_{\Phi|\Psi}(s)$ are that for any real $s$, $u_{\Phi|\Psi}(s)$ is unitary and contained in the algebra $\mathcal{A}_{R,0}$ and that the two definitions given in \eqref{eq:ccflow} are equivalent. We have
\begin{align}
    [e^{-\beta x}\Delta_{\Phi|\Psi}]^{is} = u_{\Phi|\Psi}(s) \text{exp}\left(- is \beta h_R\right) \in \mathcal{A}_{R}.
\end{align}
Hence $e^{-\beta x}\Delta_{\Phi|\Psi}$, and thus also $\rho_{\widehat{\Phi}}$ are affiliated to $\mathcal{A}_R$.  

It remains to show \eqref{eq:densdef}. We can write a generic operator in $\mathcal{A}_R$ as 
\begin{align}
    \widehat{a} = \int_{-\infty}^{\infty}ds ~ a(s) e^{is (x+ \hat h)}, ~ a(s) \in \mathcal{A}_{R,0}.
\end{align}
Then, 
\begin{align}
    \langle \widehat{\Phi}|\widehat{a}|\widehat{\Phi}\rangle &= \int_{-\infty}^{\infty}dx \int_{-\infty}^{\infty}ds ~|\varepsilon g(\varepsilon x)|^2 e^{isx}\langle \Phi|a(s) e^{is \hat h}|\Phi\rangle
\end{align}
The integral over $x$ is exponentially small unless $s = O(\varepsilon)$. As a result, we can drop the factor $e^{is \hat h}$ while incurring only an $O(\varepsilon)$ error.\footnote{We cannot drop the $e^{isx}$ term because $\Delta x = O(1/\varepsilon)$.} We can also use \eqref{eq:relmoddef} to write\footnote{Recall that the adjoint $S^\dagger$ of an antilinear operator $S$ is defined by $\braket{\phi|S^\dagger|\psi} = \braket{\psi|S|\phi}$.}
\begin{align} \label{eq:Deltacond}
 \braket{\Psi|\Delta_{\Phi|\Psi} a(s) |\Psi} = \braket{\Psi|S_{\Phi|\Psi}^\dagger S_{\Phi|\Psi} a(s) |\Psi}= \braket{\Psi|S_{\Phi|\Psi}^\dagger a(s)^\dagger |\Phi} =     \braket{\Phi|a(s)|\Phi},
\end{align}
to relate $\braket{\Phi|a(s)|\Phi}$ to an expectation value in the state $\ket{\Psi}$ as appears  in the definition
of the trace $\tr$. Hence
\begin{align}\notag
    \langle \widehat{\Phi}|\widehat{a}|\widehat{\Phi}\rangle &\approx \int_{-\infty}^{\infty}dx \int_{-\infty}^{\infty}ds ~|\varepsilon g(\varepsilon x)|^2 e^{isx}\langle \Psi|\Delta_{\Phi|\Psi} a(s) |\Psi\rangle
    \\ \notag&\approx \int_{-\infty}^{\infty}dx \int_{-\infty}^{\infty}ds ~ \langle \Psi||\varepsilon g(\varepsilon (x+\hat h))|^2\Delta_{\Phi|\Psi} a(s) e^{is(x+\hat h)}|\Psi\rangle
    \\ \notag &\approx \int_{-\infty}^{\infty}dx ~ e^{\beta x} \,\langle \Psi|\varepsilon \overline{g}(\varepsilon (x+\hat h)) e^{-\beta x}\Delta_{\Phi|\Psi} g(\varepsilon(x + \hat h)) \hat a|\Psi\rangle
    \\&\approx \tr[\rho_{\widehat{\Phi}} \hat a]~.
\end{align}
In the second equality we used the fact that $\hat h \ket{\Psi} = 0$, while in the third equality we used the fact that $g(\varepsilon(x + \hat h))$ approximately commutes with $\Delta_{\Phi|\Psi}$ because it is a slowly varying function of $\hat h$.

With the explicit form of $\rho_{\widehat{\Psi}}$ in hand, it is straightforward to show that the von Neumann entropy is indeed given by \eqref{eq:entrform}. Because $g(\varepsilon(x + \hat h))$ approximately commutes with $\Delta_{\Phi|\Psi}$, we have
\begin{align} \notag
    \log \rho_{\widehat{\Phi}} &\approx - \beta x - h_{\Phi|\Psi} + \log \left[\varepsilon |g(\varepsilon h_R)|^2\right] + O(\varepsilon)
    \\&= -\beta h_R + h_{\Psi|\Phi} - h_\Phi + \log \left[\varepsilon |g(\varepsilon h_R)|^2\right]~,
\end{align}
where in the second equality we used the identity
\begin{align}\label{eq:hdiff}
    -i \partial_s u_{\Phi|\Psi}(s)\big \lvert_{s=0} = h_{\Psi} - h_{\Phi|\Psi} = h_{\Psi|\Phi} - h_\Phi~,
\end{align}
which follows directly from \eqref{eq:ccflow}. Since $\braket{\Phi|h_\Phi|\Phi} = 0$, \eqref{eq:entrform} follows immediately from $S(\widehat{\Phi})_{\mathcal{A}_R} = - \braket{\widehat{\Phi}|\log \rho_{\widehat{\Phi}} |\widehat{\Phi}}$.

\subsection{Second derivation}
An alternative approach to deriving \eqref{eq:entrform} uses the existence of a canonical isomorphism between the crossed product algebra $\mathcal{A}_R$ obtained by adjoining $\beta h_R = \beta x + h_\Psi$ to the algebra $\mathcal{A}_{R,0}$, and the algebra $\mathcal{A}_R^{(\Phi)}$ obtained by adjoining $\beta x + h_\Phi$ to $\mathcal{A}_{R,0}$ \cite{wittengcp}. Explicitly, \begin{align} 
    \mathcal{A}_R^{(\Phi)} = u'_{\Phi|\Psi}(p/\beta)\,\mathcal{A}_R\, u'_{\Phi|\Psi}(p/\beta)^{\dagger}, \label{eq:isomorph}
\end{align}
where $u'_{\Phi|\Psi}(s)$ is the Connes cocycle flow for the algebra $\mathcal{A}_{R,0}'$. Note that $\Delta_{\Phi|\Psi} = \Delta_{\Psi|\Phi}'^{-1}$ and hence $u'_{\Phi|\Psi}(s)= \Delta_{\Psi|\Phi}^{-is} \Delta_{\Psi}^{is}$.  To check that \eqref{eq:isomorph} is an isomorphism of algebras,  it suffices, since $u'_{\Phi|\Psi}(s)$ is invertible, to show that conjugation by $u'_{\Phi|\Psi}(p/\beta)$ maps a set of generators for $\mathcal{A}_R$ into $\mathcal{A}_R^{(\Phi)}$, while conjugation by $u'_{\Phi|\Psi}(p/\beta)^\dagger = u'_{\Psi|\Phi}(p/\beta)$ maps a set of generators for $\mathcal{A}_R^{(\Phi)}$ into $\mathcal{A}_R$. In fact, the symmetry between $\ket{\Psi}$ and $\ket{\Phi}$ means that we only have to check one direction. The algebra $\mathcal{A}_R$ is generated by $\mathcal{A}_{R,0}$ together with $e^{i s (\beta x + h_\Psi)}$ for arbitrary real $s$. Since $u'_{\Phi|\Psi} \in \mathcal{A}_{R,0}'$, we have 
\begin{align}
u'_{\Phi|\Psi}(p/\beta) \mathcal{A}_{R,0}u'_{\Phi|\Psi}(p/\beta)^\dagger = \mathcal{A}_{R,0} \subseteq \mathcal{A}_R^{(\Phi)}.
\end{align}
Meanwhile
\begin{align} \label{eq:isoproof}
\begin{split}
    u'_{\Phi|\Psi}(p/\beta)  e^{is (\beta x + h_\Psi)}   u'_{\Phi|\Psi}(p/\beta)^{\dagger} &= u'_{\Phi|\Psi}(p/\beta)  \Delta_{\Psi}^{-is}   u'_{\Phi|\Psi}(p/\beta -s)^{\dagger} e^{is \beta x} \\&= \Delta_{\Psi|\Phi}^{-ip/\beta} \Delta_{\Psi}^{ip/\beta}  \Delta_{\Psi}^{-is}   \Delta_{\Psi}^{-i(p/\beta - s)} \Delta_{\Psi|\Phi}^{i(p/\beta - s)}  e^{is \beta x}
    \\&= \Delta_{\Psi|\Phi}^{-is} e^{is \beta x} = u_{\Psi|\Phi}(-s) e^{is(\beta x + h_{\Phi})} \in \mathcal{A}_R^{(\Phi)}.
    \end{split}
\end{align}
So this is indeed the case. An important additional observation is that the isomorphism \eqref{eq:isomorph} is trace-preserving if we define the trace $\tr_\Phi$ on $\mathcal{A}_R^{(\Phi)}$ in the obvious way by replacing $\ket{\Psi}$ by $\ket{\Phi}$ in \eqref{eq:crossedtrace}. In other words  
\begin{align}
    \tr_{\Phi}\left(u'_{\Phi|\Psi}(p)\widehat{a} u'_{\Phi|\Psi}(p)^{\dagger} \right) = \tr[\widehat{a}], ~~~~~ \forall\widehat{a} \in \mathcal{A}_R \label{eq:isotrace}
\end{align}
We provide a proof of this in Appendix \ref{app:isotrace}. Normalized density matrices for the algebra $\mathcal{A}_R$ are therefore also normalized density matrices for the algebra $u'_{\Phi|\Psi}(p/\beta)^\dagger\,\mathcal{A}_R^{(\Phi)}\, u'_{\Phi|\Psi}(p/\beta)$.

We can therefore compute the entropy of the state $\ket{\widehat{\Phi}}$ for the algebra $\mathcal{A}_R$ by computing the entropy of the state $u'_{\Phi|\Psi}(p/\beta) \ket{\widehat\Phi}$ for the algebra $\mathcal{A}_R^{(\Phi)}$. However, since $p = O(\varepsilon)$ for the state $\ket{\widehat\Phi}$, we have
\begin{align} \label{eq:approxstate}
    u'_{\Phi|\Psi}(p/\beta) \ket{\widehat{\Phi}} \approx \ket{\widehat{\Phi}} = \int_{-\infty}^{\infty}dx ~ \varepsilon^{1/2} g(\varepsilon x) |\Phi\rangle |x\rangle.
\end{align}
The density matrix $\rho^{(\Phi)}_{\widehat\Phi}$ for the algebra $\mathcal{A}_R^{(\Phi)}$ and a state of the form \eqref{eq:approxstate} is known exactly \cite{wittengcp}, and was given (with $\ket{\Phi}$ replaced by $\ket{\Psi}$) in \eqref{eq:cqdensity}. We therefore have
\begin{align} \label{eq:cantbebothered}
    \log \rho^{(\Phi)}_{u'_{\Phi|\Psi}(p/\beta) \widehat\Phi} \approx \log \rho^{(\Phi)}_{\widehat\Phi} =  - (\beta x + h_\Phi) + \log\left[ \varepsilon |g(\varepsilon(x + h_\Phi/\beta))|^2\right].
\end{align}
Naively, one might think that we then have
\begin{align}
    S(\widehat{\Phi})_{\mathcal{A}_R} \stackrel{?}{\approx} - \braket{\widehat\Phi|\log \rho_{u'_{\Phi|\Psi}(p/\beta) \widehat\Phi}| \widehat\Phi} \approx \braket{\widehat\Phi|\beta x + h_\Phi - \log\left[ \varepsilon |g(\varepsilon(x + h_\Phi/\beta))|^2\right]|\widehat\Phi},
\end{align}
since $u'_{\Phi|\Psi}(p/\beta) \ket{\widehat{\Phi}} \approx \ket{\widehat{\Phi}}$. However this is not the case. The fluctuations in $x$ for the state $\ket{\widehat\Phi}$ are $O(1/\varepsilon)$. As a result, the $O(\varepsilon)$ difference between the state $\ket{\widehat\Phi}$ and the state $u'_{\Phi|\Psi}(p/\beta) \ket{\widehat\Phi}$ can lead to an $O(1)$ difference in the expectation value of $\beta x$. It is therefore crucial that we compute the expectation value in the state $u'_{\Phi|\Psi}(p/\beta) \ket{\widehat\Phi}$, rather than $\ket{\widehat\Phi}$.

Exchanging $\ket{\Phi}$ and $\ket{\Psi}$ in \eqref{eq:isoproof}, and taking a derivative with respect to $s$ at $s = 0$, we find that
\begin{align} \label{eq:xtrans}
 u'_{\Phi|\Psi}(p/\beta)^{\dagger}  (\beta x + h_\Phi)   u'_{\Phi|\Psi}(p/\beta) =  \beta x + h_{\Phi|\Psi}.
\end{align}
Hence
\begin{align}
    S(\widehat{\Phi})_{\mathcal{A}_R} &\approx \braket{\widehat\Phi|u'_{\Phi|\Psi}(p/\beta)^\dagger \log \rho_{u'_{\Phi|\Psi}(p/\beta) \widehat\Phi} u'_{\Phi|\Psi}(p/\beta)| \widehat\Phi}  
    \\&\approx \braket{\widehat\Phi|u'_{\Phi|\Psi}(p/\beta)^\dagger (\beta x + h_\Phi) u'_{\Phi|\Psi}(p/\beta)| \widehat\Phi}  - \braket{\widehat\Phi|\log\left[ \varepsilon |g(\varepsilon(x + h_\Phi/\beta))|^2\right]|\widehat\Phi}
    \\& \approx \braket{\widehat\Phi| \beta x + h_{\Phi|\Psi} - \log\left[ \varepsilon |g(\varepsilon h_R)|^2\right]|\widehat\Phi}
    \\& \approx\langle \widehat{\Phi}|\beta h_R|\widehat{\Phi}\rangle - \langle \widehat{\Phi}|h_{\Psi|\Phi}|\widehat{\Phi}\rangle - \langle \widehat{\Phi}|\log \varepsilon |g(\varepsilon h_R)|^2|\widehat{\Phi}\rangle~.
\end{align}
In the second equality we used the fact that the fluctuations in $\log\left[ \varepsilon |g(\varepsilon(x + h_\Phi/\beta))|^2\right]$ are $O(1)$ and hence that (unlike for the $\beta x$ term) we can compute its expectation value using the state $\ket{\widehat\Phi}$ while incurring only $O(\varepsilon)$ errors. In the third equality we used \eqref{eq:xtrans} along with the approximate equality
\begin{align}
\braket{\widehat\Phi|\log\left[ \varepsilon |g(\varepsilon(x + h_\Phi/\beta))|^2\right] |\widehat\Phi} \approx \braket{\widehat\Phi|\log\left[ \varepsilon |g(\varepsilon h_R)|^2\right]|\widehat\Phi}
\end{align}
which is valid since $g(\varepsilon y)$ is a slowly varying function of $y$. Finally in the last equality we used \eqref{eq:hdiff} and $\braket{\Phi|h_\Phi|\Phi}$.

\subsection{Comparison with previous derivations of the Bekenstein-Hawking entropy} \label{sec:BHentropy}
In the preceding arguments, we interpreted $\mathcal{A}_R$ as a large $N$ boundary CFT algebra. However, as explained in Section \ref{sec:interp}, the algebra $\mathcal{A}_R$ is holographically dual to the semiclassical bulk algebra $\mathcal{A}_r$ generated by the algebra $\mathcal{A}_{r,0}$ of bulk QFT operators in the right exterior, along with the renormalized right ADM mass $h_R$. Viewed in this light, the arguments in this section constitute a derivation of the Bekenstein-Hawking entropy $S_\mathrm{BH}$ from the effective field theory of quantum gravity. We started with an algebra of effective field theory modes describing everything outside the black hole in the $G \to 0$ limit, and we found that its entropy was uniquely determined up to a state-independent constant and included not only a contribution from the usual QFT entropy of modes outside the horizon, but also one proportional to the horizon area of the black hole -- the Bekenstein-Hawking entropy $S_\mathrm{BH}$.

From this perspective, there was nothing in our derivation that limits us to anti-de Sitter space. The same derivation applies equally well for black holes in asymptotically flat spacetimes.\footnote{There is a minor subtlety here related to the fact that the Hartle-Hawking state in asymptotically flat spacetimes has an IR divergence and hence does not exist in the bulk QFT Hilbert space. However one can still define a sequence of states that look like the Hartle-Hawking state up to arbitrarily large distances, and these can be used to create a sequence of approximations to \eqref{eq:crossedtrace}, which will converge for trace-class operators.} Indeed, as we explained in the companion paper \cite{Chandrasekaran:2022cip}, a variation of it can also be used to derive the entropy of the cosmological horizon of the static patch in de Sitter space, so long as one introduces an observer in the static patch.

It may therefore be helpful to make some brief comparisons to previous derivations of the Bekenstein-Hawking entropy, in order to clarify the advantages and disadvantages of this new explanation.

The original arguments for the form of the Bekenstein-Hawking entropy given by Hawking went as follows.\footnote{Significant inspiration for Hawking's work came from earlier arguments by Bekenstein \cite{Bekenstein:1973ur}, who advocated an entropic interpretation of the horizon area in order to produce a ``generalized second law'' that generalized both the ordinary second law of thermodynamics and Hawking's area theorem. However these arguments were not precise enough to fix the famous coefficient of $1/4$ in $S_\mathrm{BH}$.} The Hawking temperature $T_\mathrm{BH}$ can be derived in Lorentzian effective field theory by now-standard arguments that we will not review. One can then use the Clausius relation $dE = T_{BH}\, dS$ to obtain
\begin{align}\label{eq:dS}
    dS = \frac{dA}{4 G}.
\end{align}
Finally, so long as the entropy of small black holes (in Planck units) is finite, we can integrate \eqref{eq:dS} to derive the Bekenstein-Hawking entropy.\footnote{It is not obvious how to justify this last step in a computation of the de Sitter entropy, since you can only decrease the horizon area by adding matter to the static patch. But for black holes it seems relatively unproblematic.} We emphasize that this should really be thought of as a derivation of the Bekenstein-Hawking entropy as a classical thermodynamic entropy, rather than as a statistical entropy. At most it is suggestive of a statistical mechanical underpinning, in light of our understanding of the analogous situation in classical thermodynamics.

A more direct derivation follows from Gibbons and Hawking's Euclidean gravity argument \cite{PhysRevD.15.2752}. However these rely on an interpretation of the Euclidean gravity partition function as computing a trace over a canonical ensemble; such an interpretation is mostly justified post-hoc by the fact that it works. Moreover, except in the case of anti-de Sitter boundary conditions, we do not know the degrees of freedom that make up this canonical ensemble, and that have entropy $S_\mathrm{BH}$.\footnote{Again, this is particularly true in de Sitter space.} The recent derivations of the Page curve \cite{Penington:2019npb, Almheiri:2019vm}, which is equal to the Bekenstein-Hawking entropy at late times \cite{Page:1993wv, Page:1993df}, are computing a clearly formulated entropy, albeit the entropy of the radiation rather than the black hole itself, even in asymptotically flat spacetimes. However they still rely on the black magic of Euclidean gravity \cite{Penington:2019kki, Almheiri:2019qdq}.

Finally, we have the microscopic string theory derivations, starting with \cite{Strominger:1996sh}. These give much stronger conclusions -- approximate or  sometimes even exact \cite{Dabholkar:2014ema} microstate counts -- but also involve commensurately stronger assumptions. These assumptions include not only a specific microscopic theory but also specific, and normally highly supersymmetric, black holes. And even then, whatever geometrical descriptions of the microstates one obtains is not valid for parameters at which the microstates can be understood as black holes.

The derivation we have presented above can be thought of as a more precise upgrade of the original Hawking argument (albeit one that is somewhat abstract mathematically). As with the Hawking approach, we only require the very minimal assumption of Lorentzian effective quantum gravity. Again in common with Hawking, we only directly derive the variation \eqref{eq:dS} in the Bekenstein-Hawking entropy over a small range of energies, and only work in the $G \to 0$ limit where the number of microstates diverges. The primary improvement that we offer is a direct connection to the (statistical) von Neumann entropy of an operator algebra, rather than just a classical thermodynamic relationship between energy and temperature.

\section{The generalized second law \label{sec:gsl}}

We now turn our attention to the generalized second law of black hole thermodynamics. In a limit where distinct excitations of a black hole are separated by a time much larger than the thermal time, we will find that the generalized second law is precisely explained by the monotonicity of entropy for trace-preserving inclusions of algebras \cite{Longo:2022lod}.

\subsection{Trace-preserving inclusions and entropy} \label{sec:tracepreserve}

Let $\mathcal{B} \subseteq \mathcal{A}$ be a von Neumann subalgebra. Moreover let $\mathcal{B}$ and $\mathcal{A}$ be equipped with traces $\tr_{\mathcal{A}}$ and $\tr_{\mathcal{B}}$ respectively. If
\begin{align}
    \tr_{\mathcal{A}} \,b = \tr_{\mathcal{B}} \,b \,\,\,\,\, \forall\, b \in \mathcal{B}\,,
\end{align}
we say that the inclusion $\mathcal{B} \subseteq \mathcal{A}$ is trace preserving. It can then be shown \cite{Longo:2022lod} that for any state $\ket{\Psi} \in \mathcal{H}$, we have
\begin{align} \label{eq:monotonicity}
    S(\Psi)_{\mathcal{B}} \geq S(\Psi)_{\mathcal{A}}\,,
\end{align}
with equality if and only if the density matrix $\rho_{\mathcal{A}}$ of $\ket{\Psi}$ on $\mathcal{A}$ is contained in the subalgebra $\mathcal{B}$.

Before trying to understand this statement in full generality, let us first consider the simplest case where the algebras are finite type I von Neumann factors. In others words, let $\mathcal{H}_A \cong \mathcal{H}_B \otimes \mathcal{H}_R$ with $\mathcal{B} = \mathcal{B}(\mathcal{H}_B)$ and $\mathcal{A} = \mathcal{B}(\mathcal{H}_A)$. In this case, the inequality $S(B) \geq S(A) = S(BR)$ seems (and indeed is) obviously untrue. The source of this possible confusion is that, using the conventional normalisation of traces on finite-dimensional Hilbert spaces, we have
\begin{align}
    \Tr_A(b) = \Tr_B(b) \Tr_R(\mathds{1}) = d_R \Tr_B(b)\,,
\end{align}
and so $\mathcal{B} \subseteq \mathcal{A}$ is not trace preserving. To fix this, we need to define new traces $\tr_{\mathcal{A}}$ and $\tr_{\mathcal{B}}$ on the algebras $\mathcal{A}$ and $\mathcal{B}$ that are rescaled by a relative factor of $d_R$. A natural choice is to define $\tr_{\mathcal{A}}$ and $\tr_{\mathcal{B}}$ such that
\begin{align}
    \tr_{\mathcal{A}}\, \mathds{1} = \tr_{\mathcal{B}}\, \mathds{1} = 1 \,.
\end{align}
In other words, $\tr_{\mathcal{A}} = \frac{1}{d_A}\Tr_A$ and $\tr_{\mathcal{B}} = \frac{1}{d_B} \Tr_B$. The density matrices $\rho_{\mathcal{A}}$ and $\rho_{\mathcal{B}}$ normalized with respect to these new traces satisfy   $\rho_{\mathcal{A}} = d_A \rho_A$ and $\rho_{\mathcal{B}} = d_B \rho_B$. We therefore have
\begin{align}
\begin{split}
    S(\mathcal{A}) &= -\tr_{\mathcal{A}}(\rho_{\mathcal{A}} \log \rho_{\mathcal{A}}) = S(A) - \log d_A \\
    S(\mathcal{B}) &= -\tr_{\mathcal{B}}(\rho_{\mathcal{B}} \log \rho_{\mathcal{B}}) = S(B) - \log d_B\,.
\end{split}
\end{align}
With this normalisation of the trace, the entropies $S(\mathcal{A})$ and $S(\mathcal{B})$ are really \emph{entropy deficits} relative to the maximally mixed states on $\mathcal{H}_A$ and $\mathcal{H}_B$ respectively. The inequality $S(\mathcal{B}) \geq S(\mathcal{A})$ then says that the state on $\mathcal{B}$ cannot have a larger entropy deficit than the state on $\mathcal{A}$, or equivalently that
\begin{align}
    S(B) \geq S(BR) - \log d_R\,.
\end{align}
This statement should seem unremarkable to most readers, and indeed it has a simple proof
\begin{align} \label{eq:monotonicityproof}
    S(\mathcal{B}) - S(\mathcal{A}) = \tr_\mathcal{A} \left[\rho_\mathcal{A} (\log \rho_\mathcal{A} - \log \rho_{\mathcal{B}})\right] = S(\rho_\mathcal{A}||\rho_\mathcal{B}) \geq 0~.
\end{align}
In the formula $S(\rho_\mathcal{A}||\rho_\mathcal{B})$, we are interpreting the density matrix $\rho_{\mathcal{B}}$ as a density matrix for the larger algebra $\mathcal{A}$. This is only possible without rescaling because the inclusion is trace preserving. Relative entropy is positive for any von Neumann algebra, so \eqref{eq:monotonicityproof} is in fact valid for any trace-preserving inclusion of either Type I or II algebras; it is therefore a one line derivation of our original claim \eqref{eq:monotonicity}.

\subsection{Monotonicity and the generalized second law}

The generalized second law says that the generalised entropy of the black hole horizon monotonically increases with time. As information falls into the black hole it becomes inaccessible, at least from the perspective of finite complexity observables at large $N$. If the late-time observables form a strict subalgebra of the large $N$ boundary algebra, we could potentially explain the generalized second law using the monotonicity of entropy for trace-preserving inclusions of algebras.

\begin{figure}[t]
\begin{center}
  \includegraphics[width = 0.6\linewidth]{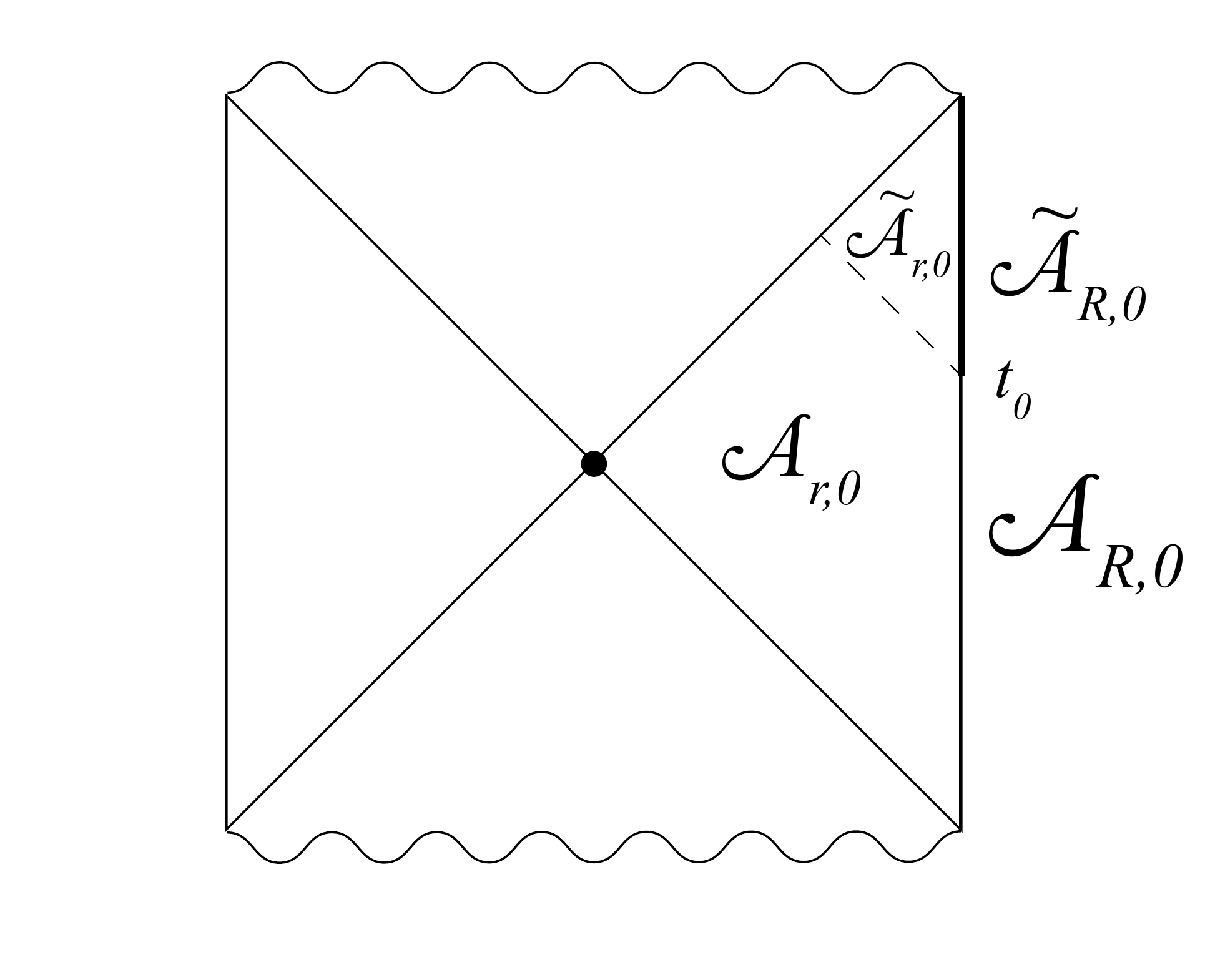} 
\end{center}
\caption{The boundary Type III subalgebra $\widetilde{\mathcal{A}}_{R,0} \subseteq \mathcal{A}_{R,0}$ describing noncentral single-trace operators at times $t>t_0$ is holographically dual to the bulk QFT subalgebra $\widetilde{\mathcal{A}}_{r,0} \subseteq \mathcal{A}_{r,0}$ describing the causal wedge of the timeband.}
\label{fig:nestedalgebras}
\end{figure}

Given a boundary time $t = t_0$, one can define a semi-infinite timeband subregion of the boundary spacetime consisting of times $t \in [t_0, +\infty]$. Any such boundary timeband has an associated Type III$_1$ algebra of operators $\widetilde{\mathcal{A}}_{R,0} \subseteq \mathcal{A}_{R,0}$ \cite{Leutheusser:2021aa}. One expects that this algebra is holographically dual to the bulk quantum field theory algebra $\widetilde{\mathcal{A}}_{r,0}$ associated to the causal wedge of the timeband, as shown in Figure \ref{fig:nestedalgebras}. Can we relate the entropy of the  algebra $\widetilde{\mathcal{A}}_{R,0}$, or more precisely the entropy of a related Type II subalgebra of $\mathcal{A}_R$, to the generalized entropy of the horizon cut bounding this causal wedge? If so, then we could explain the generalized second law using the monotonicity of trace-preserving inclusions.

Unfortunately, the obvious construction of a Type II algebra from $\widetilde{A}_R$, analogous to the construction of $\mathcal{A}_R$ from $\mathcal{A}_{R,0}$ in Section \ref{sec:micro}, does not lead to a subalgebra of $\mathcal{A}_R$ but to the algebra $\mathcal{A}_R$ itself. Since the renormalized Hamiltonian $h_R = H_R - E_0$ can be localised to any Cauchy slice, it should presumably be affiliated to a subalgebra of $\mathcal{A}_R$ associated to any timeband. However conjugation by $\exp[ih_R t]$ can be used to evolve operators in $\tilde{\mathcal{A}}_{R,0}$ backwards in time to single-trace operators outside the timeband, and thereby construct \emph{any} operator in the full algebra $\mathcal{A}_{R,0}$. And of course $\mathcal{A}_{R,0}$ and $h_R$ together generate the entire Type II right boundary algebra $\mathcal{A}_R$.

The best one can hope for is to construct a \emph{subspace} of operators $\widetilde{\mathcal{S}}_R$ that are localised to the timeband. This subspace will not be closed under operator multiplication and hence will not form an algebra. For example, one possible definition is\footnote{There are also a number of other sensible possibilities; we do not strongly advocate for any one particular option here.}
\begin{align}
    \widetilde{\mathcal{S}}_R = \,\mathrm{span}\,\{\,\tilde a\, f(h_R) \,\tilde a': \,\,f \in L^\infty(\mathbb{R})\,\,\,\,\mathrm{and}\,\,\,\,\tilde a, \tilde a' \in \widetilde{\mathcal{A}}_{R,0}\}.
\end{align} 
This subspace is large enough to include all operators that we would typically associate to the timeband, without being large enough to allow time evolution out of the timeband.  One possible physical justification for restricting to the subspace $\widetilde{\mathcal{S}}_R$ is the following. Recall that the canonically normalized large $N$ single-trace operator was not $h_R$ but $U = h_R/N$. Measuring $h_R$ to $O(1)$ precision is equivalent to measuring $U$ to $O(1/N)$ precision. It seems plausible that it might require a parametrically long time to implement such a measurement with sufficient precision. If so, it would not be possible to first measure $h_R$ and then any other boundary operator, all within the semi-infinite timeband $[t_0,+\infty]$. On the other hand, one could easily first measure any other single-trace operator and then measure $h_R$. Consequently, only operators of the form $\tilde a f(h_R) \tilde a'$ would have observable matrix elements, with the operators $\tilde a$ and $\tilde a'$ acting on the bra and ket respectively.

How can we define the entropy of a subspace of operators that does not form an algebra? One approach is the following. Given a trace-preserving inclusion of algebras $\mathcal{B} \subseteq \mathcal{A}$, we have
\begin{align} \label{eq:subalgmax}
S(\rho_\mathcal{B})_{\mathcal{B}} = \max_{\sigma_\mathcal{A}|_\mathcal{B} = \rho_\mathcal{B}} S(\sigma_\mathcal{A})_{\mathcal{A}}~,
\end{align}
where the maximisation is over density matrices $\sigma_\mathcal{A}$ whose restriction to $\mathcal{B}$ is $\rho_\mathcal{B}$.  To see this, note that the left-hand side upper bounds the right-hand side by \eqref{eq:monotonicity}, and this bound is saturated by $\sigma_\mathcal{A} = \rho_\mathcal{B}$. Given a subspace $\mathcal{S} \subseteq \mathcal{A}$ (with a trace on $\mathcal{S}$ inherited from the trace on $\mathcal{A}$), it is therefore natural to define the entropy of $\mathcal{S}$ by analogy with the right-hand side of \eqref{eq:subalgmax}. Specifically, we define
\begin{align} \label{eq:subspaceentropy}
S(T)_{\mathcal{\rho}_\mathcal{A}} = \sup_{\sigma_\mathcal{A}|_{\mathcal{S}} = \rho_\mathcal{A}|_\mathcal{S}} S(\sigma_\mathcal{A})_{\mathcal{A}}~.
\end{align}
Here we are supremising over density matrices $\sigma_\mathcal{A}$ such that
\begin{align} \label{eq:subspacedensitycondition}
 \forall s \in \mathcal{S} \,\,\,\,\,\,\tr(s \sigma_\mathcal{A}) = \tr(s \rho_\mathcal{A})~.
\end{align}
This is sometimes called the ``coarse-grained entropy'' of the subspace $\mathcal{S}$, although this distinction is somewhat misleading.\footnote{In practice it may be helpful to allow some small error in \eqref{eq:subspacedensitycondition}.} If $\mathcal{S}$ is a subalgebra, and the inclusion is trace-preserving, then \eqref{eq:subspaceentropy} agrees with the usual definition of von Neumann entropy. On the other hand, if $\mathcal{S}$ is not a subalgebra then there is no other definition of entropy for \eqref{eq:subspaceentropy} to be distinguished from.

In \cite{Bousso:2019dxk}, building off previous work on the classical limit in \cite{Engelhardt:2018aa}, it was argued that the maximum generalised entropy of the bifurcation surface, over all semiclassical states that are indistinguishable from a semiclassical state $\ket{\psi}$ on a timeband $[t_0,+\infty]$, is equal to the generalised entropy in the state $\ket{\psi}$ of the \emph{apparent horizon} on a lightsheet fired in from $t_0$. The distinction between the apparent and event horizons here is important: even though in our setup the separation between the two horizons is $O(G)$,\footnote{In \cite{Engelhardt:2018aa, Bousso:2019dxk}, more general situations were considered where the separation between the two horizons could be $O(1)$.} and hence they coincide in the limit $N \to \infty$, the generalized entropy gradient (along an infalling lightsheet) diverges as $O(1/G)$ in the same limit. As a result the difference between the generalized entropies of the two horizons is $O(1)$. (In contrast, the difference in generalised entropy between the causal wedge of the entire right boundary and the entanglement wedge of the right boundary goes to zero in the $N \to \infty$ limit because generalized entropy gradients perturbatively close to an extremal surface do not diverge.)

Given the results in Section \ref{sec:sgen}, we expect that one can reinterpret \cite{Bousso:2019dxk} as an argument that the generalised entropy of the apparent horizon is the entropy of an appropriate subspace $\widetilde{\mathcal{S}}_R$ of boundary observables -- at least so long as the supremum is restricted to the semiclassical states for which the formula \eqref{eq:entrform} applies. It would be nice to show that this somewhat artificial restriction of the supremum is not required. However we do not know any argument to that effect.

Instead, we will focus on a simpler setting, where the black hole is allowed to temporarily equilibrate over a time period that is much longer than its inverse temperature $\beta$. In this case, there are indeed boundary subalgebras (and not just subspaces of operators) that are localised at late times. And the entropies of these subalgebras are holographically dual to the generalised entropies of appropriate cuts of the black hole horizon. As hoped, the generalized second law is a direct consequence of the monotonicity of entropy under trace-preserving inclusions.

\subsection{The algebra at infinity}

As a first example, we can consider the subalgebra $\mathcal{A}_\infty \subseteq \mathcal{A}_R$ of operators that can be measured on the boundary at future infinity. This algebra consists solely of bounded functions of the conserved charges of the theory, which for simplicity we are currently taking to be only the (renormalized) Hamiltonian $h_R$. 

States on $\mathcal{A}_\infty$ are positive linear functionals
\begin{align} \label{eq:commutestate}
    \langle f \rangle_p = \int_{-\infty}^\infty\, d h_R p(h_R) f(h_R)~,
\end{align}
with $p(h_R)$ any nonnegative function with integral one. If the integral of $p$ is not equal to one, then \eqref{eq:commutestate} still defines a positive linear functional or ``weight''. This weight is said to be finite if the integral of $p$ is finite, or semifinite (as for the trace $\tr$ on a Type II$_\infty$ factor) if it is not.

Since the algebra $\mathcal{A}_\infty$ is commuting, any weight $p$ is tracial, satisfying $\braket{ab}_p = \braket{ba}_p$ for all $a,b \in \mathcal{A}_\infty$. However, if we want the inclusion $\mathcal{A}_\infty \subseteq \mathcal{A}_R$ to be trace-preserving, we are forced to define the trace $\tr_\infty$ to agree with the trace $\tr_R$ on $\mathcal{A}_R$. In other words,
\begin{align}
    \tr_\infty[f(h_R)] = tr_R[f(h_R)] = \int dx e^{\beta x} \braket{\Psi | f(x+h) |\Psi} = \int dh_R e^{\beta h_R} f(h_R)~.
\end{align}
The trace $\tr_\infty$ is therefore a weight $p$ with $p(h_R) = \exp(\beta h_R)$. With respect to this trace, the density matrix $\rho(h_R)$ of a state $p(h_R)$ is the operator 
\begin{align}
\rho(h_R) = e^{-\beta h_R} p(h_R).
\end{align}
Finally, we can compute the entropy of the state $p$ by
\begin{align}
    S(p)_{\mathcal{A}_\infty} = - \tr[\rho \log \rho] = \int_{-\infty}^\infty dh_R \,p(h_R) [\beta h_R - \log p(h_R)]~.
\end{align}
As we argued below \eqref{eq:sgeninf}, this is the generalized entropy of the black hole horizon at future infinity. Monotonicity of trace-preserving inclusions say that the generalized entropy at future infinity has to be larger than the generalized entropy at the bifurcation surface. That is true becasue the relative entropy that appears
on the right hand side of eqn. (\ref{monoto}) is positive.  This is our first taste of the generalized second law.

\subsection{Multiple timebands and parametrically large time gaps} \label{sec:plarge}

So far we have derived an ``infinitely coarse-grained'' version of the generalised second law, which says that the generalised entropy at infinity should be larger than the generalised entropy at the bifurcation surface, from the monotonicity of trace-preserving inclusions. We would like to go further and talk about intermediate times as well. Our strategy for doing so will be to introduce a parametrically large time gap, much longer than the thermal timescale $\beta$, between two locations for possible boundary operator insertions. At intermediate times, the black hole is always allowed to equilibrate, without any excitations. We will then be able to derive a generalized second law relating the generalized entropy of the black hole horizon at a) the bifurcation surface, b) the intermediate period between the two sets of operator insertions, and c) future infinity.

\begin{figure}[t]
\begin{center}
  \includegraphics[width = 0.7\linewidth]{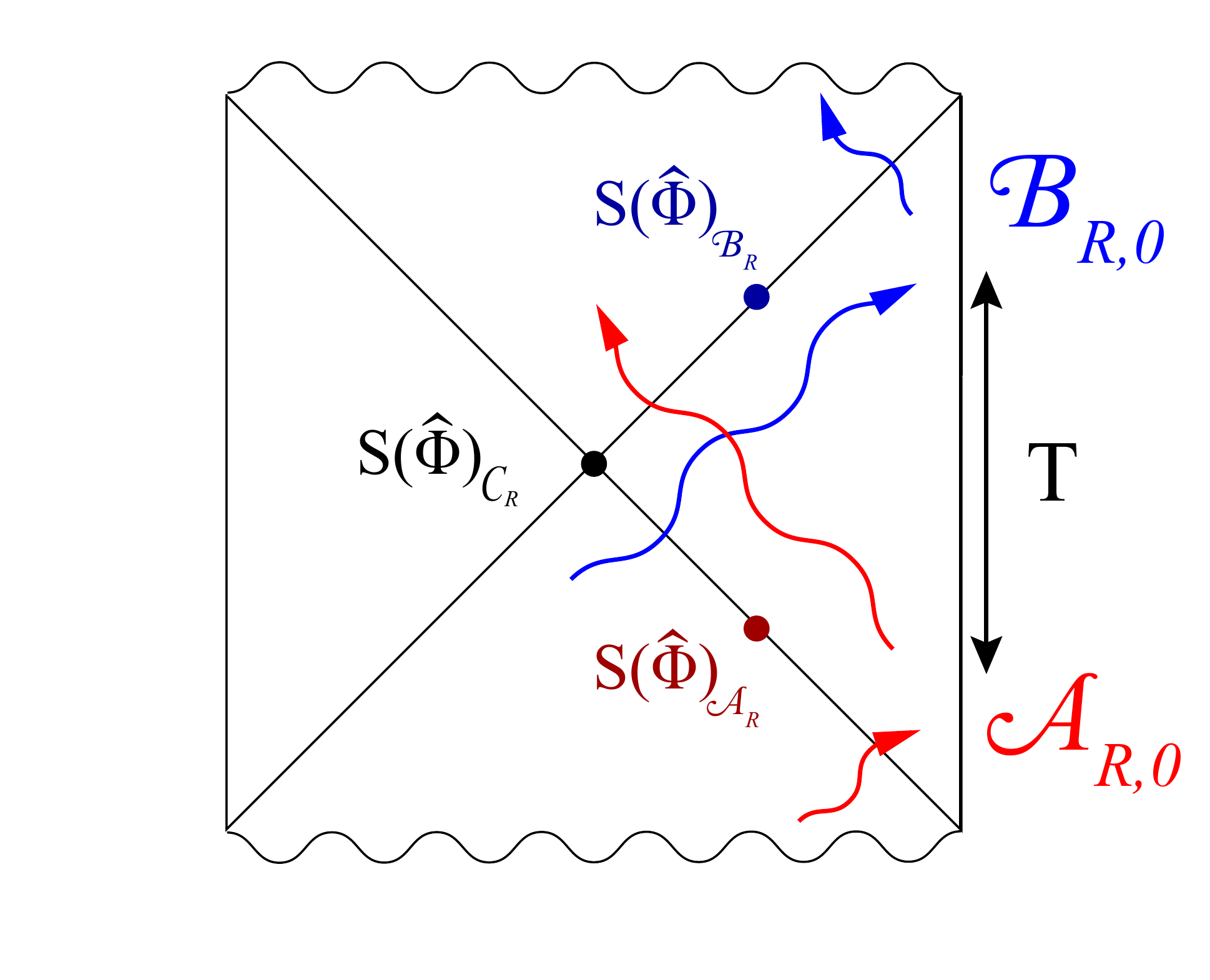} 
\end{center}
\caption{The large $N$ algebra $\mathcal{C}_{R,0}$ is generated by the isomorphic and commuting algebras $\mathcal{A}_{R,0}$ and $\mathcal{B}_{R,0}$ which describe early and late time excitations respectively. The two algebras are separated by a time $T$, which diverges at large $N$ such that $\beta \ll T \ll t_\mathrm{scr}$. The generalized entropy of the bifurcation surface computes the entropy $S(\widehat\Phi)_{\mathcal{C}_R}$ of the Type II algebra $\mathcal{C}_R$ generated by $\mathcal{C}_{R,0}$ and $h_R$, while the generalized entropies of the white and black hole horizons, at intermediate times between the two sets of modes (marked by the indicated cuts of the future and
past horizons), compute respectively the entropies of algebras $\mathcal{A}_R$ and $\mathcal{B}_R$ (generated by $h_R$ and either $\mathcal{A}_{R,0}$ or $\mathcal{B}_{R,0}$).}
\label{fig:GSL}
\end{figure}

Concretely, let $T = T(N)$ be a smooth function of $N$ such that $T(N) \to \infty$ in the $N \to \infty$ limit. For the moment, we shall assume that $T$ diverges less quickly than the scrambling time $t_\mathrm{scr} = \beta/2\pi \log S_\mathrm{BH}$ so that at large $N$ we have $t_\mathrm{scr} \gg T \gg \beta$. We will discuss times $T$ longer than the scrambling time in Section   \ref{sec:scrambling}. As in Section \ref{sec:canon}, we want to construct an algebra out of the large $N$ limit of correlation functions for the thermofield double state $\ket{\mathrm{TFD}}$. Previously the algebra $\mathcal{A}_{R,0}$ was generated by local operators $a(t)$ at any finite time $t$, in a limit where the inverse temperature $\beta$ remained finite at large $N$. Instead, as shown in Figure \ref{fig:GSL} our algebra will now be generated by two sets of operators: the first set are local single-trace operators $a(x,t)$ that act at an arbitrary finite time $t$ -- these generate the algebra $\mathcal{A}_{R,0}$ -- whereas the second set are operators $b(x', T + t')$ that act at a time $T + t'$, with $T$ fixed and divergent as defined above, while $t'$ is arbitrary but finite. The large $N$ algebra $\mathcal{B}_{R,0}$ generated by this second set of operators is isomorphic to $\mathcal{A}_{R,0}$, because $\ket{\mathrm{TFD}}$ is invariant under time translations, but it is distinct from $\mathcal{A}_{R,0}$ as a subalgebra of the larger algebra of operators $\mathcal{C}_{R,0}$ generated by the combination of early- and late-time operators.

In the large $N$ limit, correlation functions of operators $a_i(t_i)$ and operators $b_j(T+t_j)$ factorise into a product of early time and late time operators.\footnote{It is important here that $\beta \ll T \ll t_\mathrm{scr}$ in the large $N$ limit.} So for example
\begin{align}
    \lim_{N\to \infty} \braket{a_1(t_1) b_2(T+ t_2) a_3(t_3) b_4(T+t_4)}_\beta = \lim_{N\to\infty} \braket{a_1(t_1) a_3(t_3) }_\beta \braket{ b_2(t_2)  b_4(t_4)}_\beta~.
\end{align}
This result is independent of the ordering of early-time operators $a_i(t_i)$ relative to late-time operators $b_j(T+t_j)$. Consequently, the algebra $\mathcal{C}_{R,0}$ generated by both early and late time operators is simply the tensor product $\mathcal{C}_{R,0} \cong \mathcal{A}_{R,0} \otimes \mathcal{B}_{R,0}$. The large $N$ Hilbert space $\mathcal{H}_C$ on which $\mathcal{C}_{R,0}$ acts is also the tensor product $\mathcal{H}_C \cong \mathcal{H}_A \otimes \mathcal{H}_B$ of the large $N$ Hilbert spaces $\mathcal{H}_A$ and $\mathcal{H}_B$ on which $\mathcal{A}_{R,0}$ and $\mathcal{B}_{R,0}$ act. The state $\ket{\mathrm{TFD}}$ is identified with the product state $\ket{\Psi}_A \ket{\Psi}_B \in \mathcal{H}_C$. In close analogy to Section \ref{sec:canon}, the algebras $\mathcal{A}_{R,0}$ and $\mathcal{B}_{R,0}$ are holographically dual to algebras $\mathcal{A}_{r,0}$ and $\mathcal{B}_{r,0}$ of early and late time bulk QFT operators in the right exterior, while the commutant algebra $\mathcal{C}_{R,0}' \cong \mathcal{A}_{R,0}' \otimes \mathcal{B}_{R,0}'$ describes the large $N$ left boundary algebra and is dual to bulk QFT operators in the left exterior.

As in Section \ref{sec:micro}, we can switch to the microcanonical ensemble without affecting the large $N$ algebra $\mathcal{C}_{R,0}$ and then add the renormalised Hamiltonian $h_R = H_R - E_0$ to our algebra. Because the Hamiltonian is conserved, it generates time evolution for operators both in $\mathcal{A}_{R,0}$ and in $\mathcal{B}_{R,0}$; we do not have separate Hamiltonians $h_R$ for each algebra. The operator $\beta \hat h = \beta (h_R - h_L)$ generates modular flow for the state $\ket{\Psi}_A$ on the algebra $\mathcal{A}_{R,0}$, \emph{and} for the state $\ket{\Psi}_B$ on the algebra $\mathcal{B}_{R,0}$. It therefore generates modular flow for the algebra $\mathcal{C}_{R,0}$ and the state $\ket{\Psi}_A\ket{\Psi}_B$. 

The large $N$ algebra $\mathcal{C}_R$ generated by $\mathcal{C}_{R,0}$ and $h_R$ is therefore the crossed product of the Type III$_1$ factor $\mathcal{C}_{R,0} \cong \mathcal{A}_{R,0} \otimes \mathcal{B}_{R,0}$ by a modular flow. Consequently, it is a Type II$_\infty$ von Neumann factor. The algebra $\mathcal{C}_R$ contains a Type II subalgebra $\mathcal{A}_R$ that is generated by $\mathcal{A}_{R,0}$ and $h_R$, and a subalgebra $\mathcal{B}_R$ generated by $\mathcal{B}_{R,0}$ and $h_R$. These subalgebras are each isomorphic to the crossed product algebra $\mathcal{A}_R$ defined in Section \ref{sec:micro} using the obvious identifications of $h_R$ and $\mathcal{A}_{R,0} \cong \mathcal{B}_{R,0}$. Note however that with this identification $h_L = x = h_R - \hat h_\mathcal{A}$ (where $\beta \hat h_\mathcal{A}$ generates modular flow for the state $\ket{\Psi}$ on the algebra $\mathcal{A}_{R,0}$) as previously defined in Section \ref{sec:micro} is not the same as the operator $h_L = x = h_R - \hat h_\mathcal{A} - \hat h_\mathcal{B} \in \mathcal{C}_R'$ in our current construction.

We define the trace $\tr_\mathcal{C}$ on the algebra $\mathcal{C}_R$ as before by
\begin{align}
    \tr_\mathcal{C}[\hat c] = \int_{-\infty}^\infty \, dx\,e^{\beta x} \bra{\Psi}_A \braket{\Psi|_B \,\hat c\,|\Psi}_A\ket{\Psi}_B.
\end{align}
For any operator $\hat b \in \mathcal{B}_R \subseteq \mathcal{C}_R$, we can write
\begin{align}
    \hat b = \int_{-\infty}^\infty \, ds \, b_0(s) e^{i s (x + \hat h_\mathcal{A} + \hat h_\mathcal{B} )}.
\end{align}
Then
\begin{align}
    \tr_\mathcal{C}[\hat b] &= \int \, dx ds\,e^{\beta x + i s x} \bra{\Psi}_A \braket{\Psi|_B \,b_0(s) e^{i s (\hat h_\mathcal{A} + \hat h_\mathcal{B} )}\,|\Psi}_A\ket{\Psi}_B
    \\&= \int \, dx ds\,e^{\beta x + i s x} \braket{\Psi|_B \,b_0(s) e^{i s \hat h_\mathcal{B} }\,|\Psi}_B
    \\&= \tr_{\mathcal{B}} [\hat b],
\end{align}
where $\tr_\mathcal{B}$ is defined as for $\mathcal{A}_R$ in Section \ref{sec:micro}. It follows that, with the standard traces $\tr_\mathcal{B}$ and $\tr_\mathcal{C}$, the inclusion $\mathcal{B}_R \subseteq \mathcal{C}_R$ is trace-preserving; the same is true by symmetry for $\mathcal{A}_R \subseteq \mathcal{C}_R$.

The arguments in Section \ref{sec:tracepreserve} show that any state $\ket{\widehat\Phi} \in \mathcal{H}_A \otimes \mathcal{H}_B \otimes L^2(R)$ satisfies
\begin{align} \label{eq:BR>CR}
    S(\widehat\Phi)_{\mathcal{B}_R} \geq S(\widehat\Phi)_{\mathcal{C}_R}~.
\end{align}
As in Section \ref{sec:sgen}, let $\ket{\widehat\Phi}$ be a semiclassical state of the form
\begin{align}
    \ket{\widehat\Phi} = \int dx \varepsilon^{1/2} g(\varepsilon x) \ket{\Phi} \ket{x}~,
\end{align}
for some $\ket{\Phi} \in \mathcal{H}_C$. We want to argue that \eqref{eq:BR>CR} explains the generalized second law for the black hole horizon. (By symmetry, the corresponding inequality with $\mathcal{B}_R$ replaced by $\mathcal{A}_R$ should explain the generalized second law for the white hole horizon.) We know from the results of Section \ref{sec:sgen} that
\begin{align} \label{eq:sgenbif}
    S(\widehat\Phi)_{\mathcal{C}_R} \approx \braket{h_R} - \braket{\log\left[\varepsilon |g(\varepsilon h_R)|^2\right]} - S_\mathrm{rel}(\Phi||\Psi \otimes \Psi)_{\mathcal{A}_{R,0}\otimes \mathcal{B}_{R,0}}~,
\end{align}
which is equal to the generalized entropy of the bifurcation surface of the black hole. We can also apply the same arguments to show that
\begin{align} \label{eq:sgenint}
    S(\widehat\Phi)_{\mathcal{B}_R} \approx \braket{h_R} - \braket{\log\left[\varepsilon |g(\varepsilon h_R)|^2\right]} - S_\mathrm{rel}(\Phi||\Psi)_{\mathcal{B}_{R,0}}~.
\end{align}
From the point of view of the modes in $\mathcal{H}_B$, the modes in $\mathcal{H}_A$ are exponentially close to the white hole horizon as a function of the time gap $T$. The generalized entropy of a horizon cut that is far in the past of the modes in $\mathcal{B}_{R,0}$, but far in the future of the modes $\mathcal{A}_{R,0}$, is therefore exponentially close to the \emph{hypothetical} generalized entropy of the bifurcation surface for a state with the same energy distribution $p(h_R)$ as $\widehat{\Phi}$, but with the bulk quantum fields in a state  $\ket{\Psi}_A \ket{\widetilde\Phi}_B$ that is indistinguishable from $\ket{\Phi}$ on $\mathcal{B}_{R,0}$ and indistinguishable from the Hartle-Hawking state $\ket{\Psi}$ on $\mathcal{A}_{R,0}$. Note that it does not matter here whether we are talking about a cut of the apparent horizon or the event horizon here, for the same reason that it doesn't matter for the generalized entropy of the bifurcation surface: for a (temporarily) equilibrated horizon, the two generalized entropies agree up to corrections that vanish in the limit of large $T$. The relative entropy satisfies
\begin{align}
    S_\mathrm{rel}(\Psi_A \otimes \Phi_B||\Psi_A \otimes \Psi_B)_{\mathcal{A}_{R,0} \otimes \mathcal{B}_{R,0}} = S_\mathrm{rel}(\Phi||\Psi)_{\mathcal{B}_{R,0}}.
\end{align}
So the formula \eqref{eq:sgenbif} for the generalized entropy of the bifurcation surface reduces to \eqref{eq:sgenint} -- if the modes in $\mathcal{A}_{R,0}$ are in the vacuum state. It follows that \eqref{eq:sgenint} is equal to the generalized entropy of the horizon after the modes in $\mathcal{A}_{R,0}$ have fallen in, but before the modes in $\mathcal{B}_{R,0}$ fall in. Monotonicity of entropy for the trace-preserving inclusions $\mathcal{A}_\infty \subseteq\mathcal{B}_R \subseteq \mathcal{C}_R$ therefore directly accounts for the monotonicity of generalized entropy along the black hole horizon.

\subsection{One-sided black holes}

So far in this paper we have focused on constructing algebras for two-sided black holes dual to microcanonical or canonical versions of the thermofield double state. However the same trick of introducing a parametrically large timescale $T$ that we just used in Section \ref{sec:plarge} can also be applied to a one-sided black hole formed from collapse.

\begin{figure}[t]
\begin{center}
  \includegraphics[width = 0.9\linewidth]{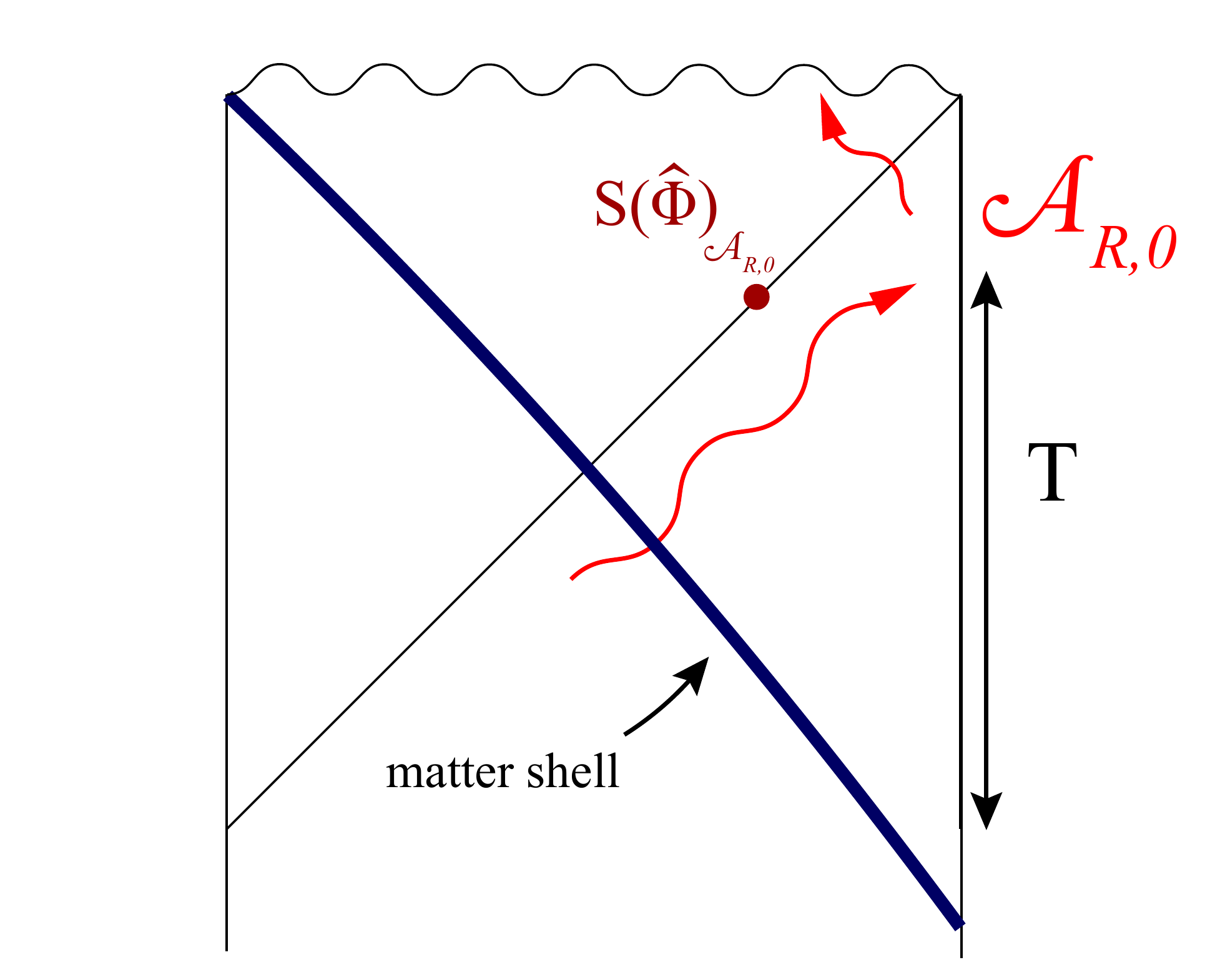} 
\end{center}
\caption{If a one-sided black hole formed from collapse is allowed to equilibrate for a parametrically long time $T$, the late time large $N$ algebra $\mathcal{A}_{R,0}$ is unable to detect that the state is not thermal. It is therefore a Type III algebra, and we can construct a Type II algebra $\mathcal{A}_R$ as usual by adding the renormalized Hamiltonian $h_R$. The entropy of this algebra is equal to the generalized entropy of the equilibrated black hole horizon before the matter in $\mathcal{A}_{R,0}$ falls in.}
\label{fig:onesidedBH}
\end{figure}

The strategy is the following. Our starting point is any fixed, semiclassical method of forming a black hole, such as a collapsing shell of matter with mass $E_0 = O(N^2)$. The details of the black hole formation process do not matter; all that we need is for it to have a sensible large $N$ limit.\footnote{It is important here that we only trying to describe a single microstate (and excitations around it) semiclassically, rather than attempting to give a semiclassical description of \emph{all} black hole microstates. As emphasized recently in \cite{Schlenker:2022dyo}, the latter cannot be possible, since the number of microstates grows exponentially each time we increase $N$ to $N+1$. } We then allow the black hole to equilibrate for a parametrically large time $T \gg \beta$, as shown in Figure \ref{fig:onesidedBH}.

We can now consider the large $N$ algebra built out of single-trace operators acting on this one-sided black hole after we allow it to equilibrate. Because the late-time Hawking radiation physics does not depend on the details of the black hole formation process, correlation functions of single-trace operators will look exactly thermal in the strict large $N$ limit, and hence the large $N$ algebra of noncentral single-trace operators will be $\mathcal{A}_{R,0}$. If the semiclassical construction of the black hole produces $O(1)$ energy fluctuations in the large $N$ limit, we can add the operator $h_R = H_R - E_0$ and thereby construct the same Type II$_\infty$ algebra $\mathcal{A}_R$ that first appeared in Section \ref{sec:micro}. To be clear: at finite $N$, we can always conjugate a late-time operator $a$ by $e^{-i H T}$ and thereby produce an operator that can see the matter which formed the black hole. However, because we are adjusting $T$ so that $T\to\infty$ in the large $N$ limit, finite time evolution at large $N$ preserves the post-equilibration algebra $\mathcal{A}_{R,0}$, just like it preserved the algebras $\mathcal{A}_{R,0}$ and $\mathcal{B}_{R,0}$ in Section \ref{sec:plarge}, and cannot be used to see the pre-collapse state. As usual, the entropy of the algebra $\mathcal{A}_R$ for semiclassical states of the late time fields will be approximately equal to the generalized entropy of the black hole after the initial equilibration but before any late time excitations.

It is worth pausing for a minute to examine this result. At finite $N$, single-trace operators (including the Hamiltonian $H_R$) generate the entire boundary CFT algebra. States built by applying single-trace operators to the microcanonical thermofield double $|\widetilde{\text{TFD}}\rangle$ have a large but finite entropy for this right boundary algebra; this entropy diverges as $N \to \infty$. Up to the state-independent arbitrary constant involved in the renormalization of this divergence, the large $N$ limit of the finite $N$ entropy agrees with the entropy of the large $N$ Type II$_\infty$ algebra $\mathcal{A}_R$. In contrast, something superficially much weirder happens for the one-sided black hole. At any finite $N$, the one-sided black hole is a pure state of the boundary algebra, and hence has zero entropy. And yet, even though the large $N$ limit of zero is zero, the large $N$ algebra generated by single-trace operators is still Type III (if we don't include $h_R$) or Type II (if we do) and thus has divergent entanglement.  

What is this algebra entangled with? A Type II or III von Neumann algebra always has a commutant algebra of the same type. The late time bulk algebra $\mathcal{A}_{r,0}$ that is holographically dual to $\mathcal{A}_{R,0}$ has a commutant algebra $\mathcal{A}_{r,0}'$ that describes the bulk quantum field theory modes in the interior of the black hole. The mysterious emergence of a Type II or III algebra at large $N$ is therefore intimately related to the emergence of the black hole interior.

 The resolution of this mystery is that the large $N$ algebra $\mathcal{A}_R$ only knows about the large $N$ limit of finite products of single-trace operators. To see that the black hole state is pure rather than thermal, we need to consider operators that have parametrically large complexity. For example, to write a pre-collapse operator in terms of a late-time operator conjugated by $\exp(-i h_R T)$, we need to write $\exp(-i h_R T)$ as a product of operators such as $\exp(-i \beta h_R)$ that are contained in $\mathcal{A}_{R}$ at large $N$. But the number of operators in this product is $O(T)$ and so diverges at large $N$.  Similarly, the interior operators that purify Hawking modes are described by highly complex interior operators \cite{Papadodimas:2012aq, Papadodimas:2013jku, Engelhardt:2021qjs}. At large $N$ these operators are not contained in $\mathcal{A}_{R,0}$ and instead are described by the commutant algebra $\mathcal{A}_{R,0}'$.
 
To be clear: modulo details of the mathematical formalism being used, the discussion in the last paragraph is mostly not new. 
Mirror operators associated to the action of simple boundary modes on an equilibrium state were introduced in
 \cite{Papadodimas:2012aq}. The primary new feature that we have found here is the presence of a Type II von Neumann algebra $\mathcal{A}_R$, and hence of density matrices and entropies for the simple boundary degrees of freedom.

How then should we interpret the entropy of the algebra $\mathcal{A}_R$? Suppose we consider the \emph{subspace} of single-trace operators (including $h_R$ but with no products allowed). If we define an entropy for this subspace as in \eqref{eq:subspaceentropy} (perhaps with some small errors allowed) it will generally be nonzero even for a one-sided black hole. Indeed it was argued in \cite{Engelhardt:2018aa, Bousso:2019dxk} that this subspace entropy should be equal to the generalized entropy of the black hole horizon. Of course, the large $N$ algebra $\mathcal{A}_R$ knows about more than just one-point functions of single-trace operators. However, the same result should be true for sufficiently large $N$ even if we allow products of up to $k$ single-trace operators for any fixed finite $k$. Of course, if we first take $k \to \infty$ while holding $N$ fixed we recover the entire right boundary algebra and hence the answer of zero entropy that we had previously. However if we first take $N \to \infty$ and then $k \to \infty$ we should obtain a nonzero entropy that is equal to the generalized entropy of the black hole. We expect that this latter order of limits is being computed by the entropy of the large $N$ Type II$_\infty$ algebra $\mathcal{A}_R$.\footnote{An interesting open question is whether the entropy of the Type II$_1$ algebra introduced in \cite{Chandrasekaran:2022cip} describing the static patch in de Sitter space comes from the entropy of a finite $G$ algebra, as for a two-sided black hole, from the entropy of a subspace of finite $G$ observables, as for a one-sided black hole, or from something else entirely.}

\section{Scrambling and long wormholes} \label{sec:scrambling}

When considering algebras separated by parametrically large time gaps in Section \ref{sec:plarge}, it was important that the timegap $T$ was much smaller than the scrambling time in the limit $N \to \infty$, otherwise operators in the two algebras wouldn't commute. This leads to an obvious question: what happens if $T \gg t_\mathrm{scr}$? Rather than a tensor product of algebras as we found for $T \ll t_\mathrm{scr}$, we will find that the resulting algebra is a \emph{free product}. Bulk states no longer resemble the fixed two-sided black hole geometry, but instead have long wormholes supported by high energy shocks.

\subsection{Free Products of von Neumann algebras}

Abstractly, the free product $\mathcal{A} * \mathcal{B}$ of two algebras $\mathcal{A}$ and $\mathcal{B}$ (as with the perhaps more familiar case of a free product of groups) is defined by the universal property that any pair of maps $f_A: \mathcal{A} \to \mathcal{X}$ and $f_B: \mathcal{B} \to \mathcal{X}$, with $\mathcal X$ an arbitrary third algebra, factor uniquely through the canonical embeddings $i_A: \mathcal{A} \hookrightarrow \mathcal{A} * \mathcal{B}$ and $i_B: \mathcal{B} \hookrightarrow \mathcal{A} * \mathcal{B}$. In other words, there always exists a unique map $f: \mathcal{A} * \mathcal{B} \to \mathcal{X}$ such that $f_A = f \circ i_A$ and $f_B = f \circ i_B$.\footnote{In the language of category theory, $\mathcal{A} * \mathcal{B}$ is the coproduct of $\mathcal{A}$ and $\mathcal{B}$.}

More concretely, $\mathcal{A} * \mathcal{B}$ is spanned by alternating strings of operators in $\mathcal{A}$ and $\mathcal{B}$, e.g. $a_1 b_2 a_3 b_4 \dots b_n$. Operators of this form are known as monomials. In general, the initial and final operators in a monomial $m$ may lie in either algebra $\mathcal{A}$ or algebra $\mathcal{B}$; the two will lie in the same algebra for $n$ odd and different algebras for $n$ even. However in equations we shall often write monomials that start with $a$ and end with $b$; these are meant to be illustrative of an arbitrary monomial unless stated otherwise. Monomials satisfy the equivalence relations
\begin{align} \label{eq:equiv}
\begin{split}
    \dots a_{i-1} \,(c b_i) \,a_{i+1} \dots &= c (\dots a_{i-1} \,b_i \,a_{i+1} \dots ) \\ 
    \dots b_{i-1} \,(c a_i) \,b_{i+1} \dots &= c (\dots b_{i-1} \,a_i \,b_{i+1} \dots ) \\ 
    \dots a_{i-1}\, (b_i + b_i') \,a_{i+1} \dots &= (\dots a_{i-1} \,b_i \,a_{i+1} \dots ) + (\dots a_{i-1} \,b_i'\, a_{i+1} \dots )\\
    \dots b_{i-1} \,(a_i + a_i')\, b_{i+1} \dots &= (\dots b_{i-1} \,a_i\, b_{i+1} \dots ) + (\dots b_{i-1}\, a_i' \,b_{i+1} \dots )\\
    \dots a_{i-1} \,\mathds{1}_{B} \,a_{i+1} \dots &= \dots (a_{i-1} a_{i+1}) \dots \\
    \dots b_{i-1} \,\mathds{1}_{A} \,b_{i+1} \dots &= \dots (b_{i-1} b_{i+1}) \dots ,
    \end{split}
\end{align}
where $c$ is any c-number, but are otherwise independent. Multiplication of monomials acts by string composition.

Now suppose that $\mathcal{A}$ and $\mathcal{B}$ are von Neumann algebras, acting on Hilbert spaces $\mathcal{H}_A$ and $\mathcal{H}_B$ respectively, and let $\ket{\Psi_A} \in \mathcal{H}_A$ and $\ket{\Psi_B} \in \mathcal{H}_B$ be cyclic and separating states. The states $\ket{\Psi_A}$ and $\ket{\Psi_B}$ allow us to define an  \emph{irreducible monomial} $m = a_1 b_2 a_3 \dots$ as a monomial that satisfies $\braket{\Psi_A | a_i |\Psi_A} = 0$ and $\braket{\Psi_B| b_i |\Psi_B} = 0$ for all $i$. Using the equivalence relations \eqref{eq:equiv}, it is possible to write any operator in $\mathcal{A} * \mathcal{B}$ as a sum of irreducible monomials together with a multiple of the identity.\footnote{By \eqref{eq:equiv}, the identity on $\mathcal{A}$ is equivalent to the identity on $\mathcal{B}$.} We can construct a distinguished state $\Psi$ for the algebra $\mathcal{A} * \mathcal{B}$ by defining
\begin{align} \label{eq:freeprodrep}
\braket{m}_{\Psi} = 0
\end{align}
for any irreducible monomial $m$. This state is faithful because $\ket{\Psi_A}$ and $\ket{\Psi_B}$ are separating. We can then use the GNS construction to create a Hilbert space $\mathcal{H}$ as the completion of a vector space spanned by states of the form $m \ket{\Psi}$, where we identify the Hilbert space state $\ket{\Psi}$ with the state $\Psi$ of the free product algebra. The inner product on $\mathcal{H}$ is defined by \begin{align}
\braket{\Psi| m_1^\dagger m_2 |\Psi} = \braket{m_1^\dagger m_2}_\Psi.     
\end{align}
Finally, we can construct a von Neumann algebra $(\mathcal{A} * \mathcal{B})''$ as the double commutant (or equivalently the weak closure) of the free product algebra with respect to its natural action on $\mathcal{H}$. Note that unlike the original free product algebra $\mathcal{A} * \mathcal{B}$ the von Neumann algebra $(\mathcal{A} * \mathcal{B})''$ depends explicitly on the choice of states $\ket{\Psi_A}$ and $\ket{\Psi_B}$ through the definition of an irreducible monomial. If $\mathcal{A}$ and $\mathcal{B}$ are type III$_1$, then $(\mathcal{A} * \mathcal{B})''$ is also type III$_1$ \cite{barnett1995free}. In a mild abuse of notation, we shall henceforth refer to the von Neumann algebra $(\mathcal{A} * \mathcal{B})''$ as simply $\mathcal{A} * \mathcal{B}$, since the original free product algebra $\mathcal{A} * \mathcal{B}$ will play no further role. 

The structure of the Hilbert space $\mathcal{H}$ can be understood as follows. Let $\mathcal{H}_A^* \cong \mathcal{H}_A \ominus \ket{\Psi_A}$ be the orthogonal complement of $\ket{\Psi_A}$ in $\mathcal{H}_A$, and similarly $\mathcal{H}_B^* \cong \mathcal{H}_B \ominus \ket{\Psi_B}$. We then have
\begin{align} \label{eq:hilbertdecomp}
    \mathcal{H} \cong \ket{\Psi} \oplus \mathcal{H}_A^* \oplus \mathcal{H}_B^* \oplus (\mathcal{H}_A^* \otimes \mathcal{H}_B^*) \oplus (\mathcal{H}_B^* \otimes \mathcal{H}_A^*) \oplus (\mathcal{H}_A^* \otimes \mathcal{H}_B^* \otimes \mathcal{H}_A^*) \oplus \dots,
\end{align}
where we have a direct sum over all possible alternating tensor products of $\mathcal{H}_A^*$ and $\mathcal{H}_B^*$. Here $\ket{\Psi}$ represents the one-dimensional Hilbert space spanned by $\ket{\Psi}$. To see this, recall that $\ket{\Psi_{A}}$ is cyclic and separating, and so states of the form $a\ket{\Psi_A}$ with $a \in \mathcal{A}$ and $\braket{\Psi_A | a |\Psi_A} = 0$ are dense in $\mathcal{H}_A^*$ (and similarly for $\mathcal{B}$). It follows from \eqref{eq:freeprodrep} that any irreducible monomial $m =a_1 b_2 \dots b_n$ satisfies
\begin{align}\notag
    \braket{\Psi | b_n^\dagger \dots b_2^\dagger a_1^\dagger a_1 b_2 \dots b_n |\Psi} &= \braket{\Psi_A | a_1^\dagger a_1 |\Psi_A} \braket{\Psi | b_n^\dagger \dots b_2^\dagger  b_2 \dots b_n |\Psi} \\
    &= \braket{\Psi_A | a_1^\dagger a_1 |\Psi_A} \braket{\Psi_B | b_2^\dagger b_2 |\Psi_B} \dots \braket{\Psi_B | b_n^\dagger b_n |\Psi_B}.\label{eq:innerprodPsi}
\end{align}
Moreover $\braket{\Psi|m^\dagger m'|\Psi} = 0$ unless $m$ and $m'$ satisfy two conditions: a) they have the same length and b) they start with an element of the same algebra. We can therefore naturally identify
\begin{align}
    a_1 b_2 \dots \ket{\Psi} \cong  a_1 \ket{\Psi_A} \otimes b_2 \ket{\Psi_B} \otimes \dots \in \mathcal{H}_A^* \otimes \mathcal{H}_B^* \dots,
\end{align}
and thereby obtain the isomorphism of Hilbert spaces given in \eqref{eq:hilbertdecomp}.

As hinted at in the brief abstract definition of a free product that we gave above, there exist canonical embeddings $i_A: \mathcal{A} \hookrightarrow \mathcal{A} * \mathcal{B}$ and $i_B: \mathcal{B} \hookrightarrow \mathcal{A} * \mathcal{B}$. These embeddings map $\mathcal{A}$ and $\mathcal{B}$ to the respective subalgebras of single operator monomials in $\mathcal{A} * \mathcal{B}$. The algebras $\mathcal{A}$ and $\mathcal{B}$ act on $\mathcal{H}$ as follows. By identifying $\ket{\Psi}$ with $\ket{\Psi_A}$, we have $\mathcal{H}_A \cong \ket{\Psi} \oplus \mathcal{H}_A^*$. Similarly, we can identify the tensor products in \eqref{eq:hilbertdecomp} of the form $\mathcal{H}_B^* \otimes \mathcal{H}_A^* \dots$ with $\ket{\Psi_A} \otimes \mathcal{H}_B^* \otimes \mathcal{H}_A^* \dots$ Then \eqref{eq:hilbertdecomp} becomes
\begin{align} \label{eq:subalgebraaction}
\mathcal{H} \cong \mathcal{H}_A \otimes \left[ \mathds{C} \oplus \mathcal{H}^*_B \oplus (\mathcal{H}^*_B \otimes \mathcal{H}^*_A) \oplus (\mathcal{H}^*_B \otimes \mathcal{H}^*_A \otimes \mathcal{H}^*_B) \oplus \dots\right].
\end{align}
The algebra $\mathcal{A}$ acts on this leftmost copy of $\mathcal{H}_A$. An analogous story applies for $\mathcal{B}$.

Finally, we should briefly discuss the commutant algebra $(\mathcal{A} * \mathcal{B})'$. For any monomial $m\in \mathcal{A} * \mathcal{B}$ we can define an operator $m' \in (\mathcal{A} * \mathcal{B})'$ by
\begin{align} \label{eq:freeprodcommut}
    m' \tilde m \ket{\Psi} = \tilde m m^\dagger \ket{\Psi},
\end{align}
for any $\tilde m \in \mathcal{A} * \mathcal{B}$. It follows directly from this definition that $m_1' m_2' = (m_1 m_2)'$. Finite sums of such operators are dense in $(\mathcal{A} * \mathcal{B})'$. Heuristically we can therefore think of $(\mathcal{A} * \mathcal{B})'$ as the algebra of monomials acting on $\mathcal{H}$ ``from the right''.

\subsection{Out-of-time-order correlators and long wormholes}

How does this construction relate to the algebra generated by two large $N$ algebras $\mathcal{A}_{R,0}$ and $\mathcal{B}_{R,0}$ separated by a time $T$ that is much greater than the scrambling time? The key feature of quantum scrambling is that for $T \gg t_\mathrm{scr}$ any ``out-of-time-ordered'' correlation function of the form
\begin{align} \label{eq:OTOC}
    \braket{\Psi_{TFD}| a_1(t_1) b_{2}(T + t_{2}) a_{3}(t_{3}) \dots b_n( T + t_n) |\Psi_{TFD}} \to 0
\end{align}
whenever $\braket{a_1} = \braket{b_{2}} = \dots \braket{b_n} = 0$. For $n=4$, with $a_i$ and $b_i$ local operator insertions, the detailed form of \eqref{eq:OTOC} has been derived in numerous systems \cite{Shenker:2013pqa, Shenker:2014tu,Maldacena:2015vu,Maldacena:2016vs}. However, the decay of \eqref{eq:OTOC} at large $N$and $T \gg t_\mathrm{scr}$ is much more general than this and applies for any value of $n$ (for $n$ odd, the first and last operator insertions will either both be at late times or both be at early times).\footnote{In holographic theories, this result follows from the bulk picture discussed below. For detailed calculations at $n=6$ see \cite{Haehl:2021tft}.} The times $t_i$ can be arbitrary but should be fixed in the $N\to\infty$ limit, while the operators $a_i, b_i$ may in general be a product of local operators inserted at different spacetime points centered around the time $t_i$.  

By the universal property of the free product, the large $N$ algebra generated by $\mathcal{A}_{R,0}$ and $\mathcal{B}_{R,0}$ must be a quotient of the free product algebra $\mathcal{A}_{R,0} *\mathcal{B}_{R,0}$. However the equation \eqref{eq:OTOC} is identical to \eqref{eq:freeprodrep} which defined the canonical representation of the free product algebra. Since this representation is faithful, the large $N$ algebra is in fact $\mathcal{A}_{R,0} * \mathcal{B}_{R,0}$ itself; this algebra acts on a large $N$ Hilbert space isomorphic to \eqref{eq:hilbertdecomp} where the large $N$ limit of $\ket{\mathrm{TFD}}$ is identified with the distinguished state $\ket{\Psi}$.

As with the previous large $N$ constructions that we have seen, the algebra $\mathcal{A}_{R,0} * \mathcal{B}_{R,0}$, and the Hilbert space $\mathcal{H}$ that it acts on, have a very natural bulk interpretation. Suppose we start with the thermofield double state, and then apply some operator $a \in \mathcal{A}_{R,0}$ that creates a low energy excitation. In coordinates where operators in the algebra $\mathcal{A}_{r,0}$ dual to $\mathcal{A}_{R,0}$ act at times $t = O(1)$, this creates only a perturbatively small backreaction on the spacetime, which vanishes in the limit $N \to \infty$. However in the natural coordinates for studying bulk operators in the algebra $\mathcal{B}_{r,0}$ dual to $\mathcal{B}_{R,0}$, where the operators in $\mathcal{B}_{r,0}$ act at $t = O(1)$ and operators in $\mathcal{A}_{r,0}$ act at $t \approx - T$, modes in $\mathcal{H}_A$ are highly boosted into a high energy shockwave that creates a large backreaction \cite{Shenker:2013pqa}. In fact, any bulk state in $\mathcal{H}_A$ that is orthogonal to the Hartle-Hawking state has a divergent backreaction (in the limit of large $N$) on the black hole horizon in this frame. This is because the Hartle-Hawking state is the unique state of zero null energy on the black hole horizon. Any nonzero null energy creates focusing by Raychaudhuri's equation and so pushes the event horizon outwards by an $O(G)$ amount in the $\mathcal{A}_{r,0}$ frame. In a boosted frame, where the shock falls into the black hole at a time $t_\mathrm{shock} \ll 0$ this backreaction grows exponentially. It becomes parametrically large when $t_\mathrm{shock} \ll -t_\mathrm{scr}$.

So far, this distinction between coordinates where $a$ acts at time $t= O(1)$ and creates perturbatively small backreaction (which we call the natural frame for the $\mathcal{H}_A$ modes) and coordinates where $a$ acts at $t\approx - T$ and the backreaction appears very large (the natural frame for the $\mathcal{H}_B$ modes) is purely a gauge choice. However suppose we now apply an operator in $b \in \mathcal{B}_{r,0}$ that takes the modes in $\mathcal{H}_B$ to some state that is orthogonal to the Hartle-Hawking state -- i.e. $\braket{\Psi|b|\Psi} = 0$. In the natural frame for the $\mathcal{H}_A$ modes, this operator creates a large backreaction on the white hole horizon that hides the original excited modes in $\mathcal{H}_A$. Instead, any measurement on $\mathcal{A}_{r,0}$ will measure a ``fresh set'' of modes, which will be in the Hartle-Hawking state. To see the original modes that were excited with $a \in \mathcal{A}_{R,0}$, we must first undo the action of $b \in \mathcal{B}_{r,0}$ and thereby remove the shock hiding them. Of course, we could have equivalently worked in the natural frame for the $\mathcal{H}_B$ modes, where the backreaction created by $b$ is small. But in this frame the $\mathcal{H}_A$ modes are exponentially close to the white hole horizon, and so this small backreaction is enough to hide them. In either frame, one of the two excitations created a large backreaction, and so the resulting geometry looks very different from the original two-sided black hole. For example, the length of any geodesic connecting the two boundaries (at any boundary times) is much longer than in an ordinary two-sided black hole \cite{Shenker:2013yza}; in this sense, the wormhole has become long.

\begin{figure}[t]
\begin{center}
  \includegraphics[width = 0.95\linewidth]{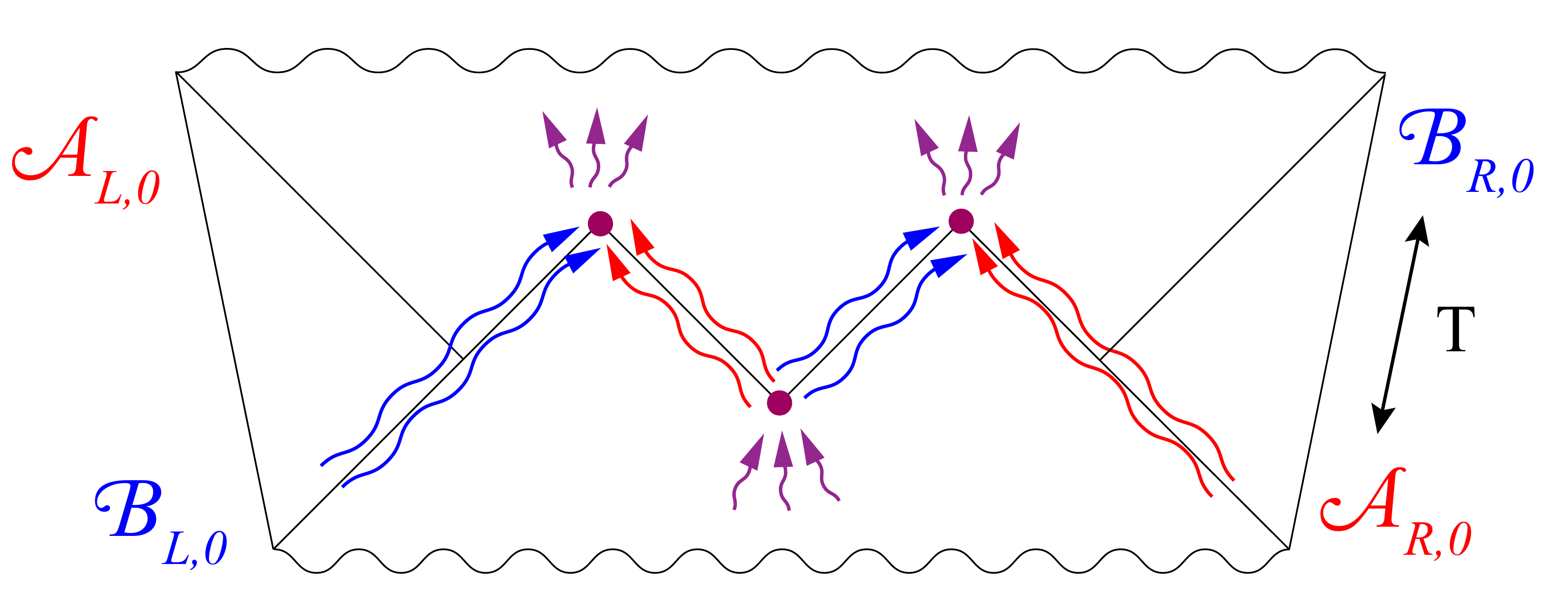} \end{center}
\caption{A long wormhole supported by four shocks. The free product algebra $\mathcal{A}_{R,0} * \mathcal{B}_{R,0}$ describes shocks created at the right boundary, while the commutant algebra $\mathcal{A}_{L,0} * \mathcal{B}_{L,0}$ describes shocks at the left boundary. The collisions between shocks (purple) are transplanckian and so cannot be described by semiclassical physics.}
\label{fig:multipleshocks}
\end{figure}

As shown in Figure \ref{fig:multipleshocks}, this process can continue indefinitely. Each time we excite modes in $\mathcal{H}_A$ or $\mathcal{H}_B$, we create a high energy shockwave from the perspective of the other frame. This frame hides the modes previously visible in $\mathcal{B}_{r,0}$ or $\mathcal{A}_{r,0}$ behind a horizon, and replaces them by new modes that are always in the Hartle-Hawking state. In this way, one can create arbitrarily long wormholes supported by shocks.\footnote{We can now see why the out-of-time-ordered correlators in \eqref{eq:OTOC} vanished in the large $N$ limit. Because the one-point functions were all zero, the operators $a_i$ and $b_i$ all mapped the Hartle-Hawking state of $\mathcal{H}_A$ or $\mathcal{H}_B$ into an orthogonal state. Hence by applying $n$ such operators in a row, we created a long wormhole supported by $n$ shocks, which was itself orthogonal to the original thermofield double state in the large $N$ limit.} The decomposition of the bulk Hilbert space into sectors supported by different numbers of shocks is the same as the decomposition of $\mathcal{H}$ given in \eqref{eq:hilbertdecomp}. The commutant algebra $(\mathcal{A}_{R,0} * \mathcal{B}_{R,0})' \cong \mathcal{A}_{L,0} * \mathcal{B}_{L,0}$ describes shocks created at the left boundary in accordance with \eqref{eq:freeprodcommut}.

\subsection{Type II algebras and entropies} \label{sec:scrambleII}

As in Section \ref{sec:micro}, we can construct Type II algebras by switching to the microcanonical ensemble and adding the renormalized Hamiltonian $h_R = H_R - E_0$ to the large $N$ algebra. At finite $N$, $h_R$ generates modular flow of the thermofield state $\ket{\mathrm{TFD}}$ for the right boundary algebra. At large $N$, we therefore expect that $h_R$ should generate modular flow of the state $\ket{\Psi} \in \mathcal{H}$ for the free product algebra $\mathcal{A}_{R,0} * \mathcal{B}_{R,0}$. This is indeed the case, but to show that it is true will require us to introduce a couple of auxiliary definitions. Let the operator $U$ be defined by
\begin{align}
    U a_1 b_2 \dots b_n \ket{\Psi} = b_n a_{n-1} \dots a_1 \ket{\Psi},
\end{align}
for any irreducible monomial $m = a_1 b_2 \dots b_n$. It follows immediately from \eqref{eq:innerprodPsi} that the operator $U$ is unitary. Next let $S_{\Psi_A}$ and $S_{\Psi_B}$ be the Tomita operators for $\ket{\Psi_A}$ and $\ket{\Psi_B}$ on $\mathcal{A}_{R,0}$ and $\mathcal{B}_{R,0}$ respectively, and let $S_{\Psi_A}^*$ and $S_{\Psi_B}^*$ be their restrictions to $\mathcal{H}_A^*$ and $\mathcal{H}_B^*$.\footnote{The operators $S_{\Psi_A}$ and $S_{\Psi_B}$ preserve the subspaces $\mathcal{H}_A^*$ and $\mathcal{H}_B^*$ respectively because $\braket{\Psi| a |\Psi} = 0$ implies $\braket{\Psi| a^\dagger |\Psi} = 0$.} It follows that for any irreducible monomial $m = a_1 \dots b_n$, the Tomita operator $S_\Psi$ for the state $\ket{\Psi}$ and the algebra $\mathcal{A}_{R,0} * \mathcal{B}_{R,0}$ satisfies
\begin{align} 
    U S_\Psi m \ket{\Psi} &= U m^\dagger \ket{\Psi}
    \\& = a_1^\dagger \dots b_n^\dagger \ket{\psi}
    \\& = [S_{\Psi_A}^* \otimes \dots S_{\Psi_B}^*] m \ket{\Psi}.\label{eq:deltapsicalc}
\end{align}
In other words, the operator $U S_\Psi$ acts diagonally with respect to the decomposition of $\mathcal{H}$ given in \eqref{eq:hilbertdecomp}, and within each sector it acts as a product of Tomita operators for $\ket{\Psi_A}$ and $\ket{\Psi_B}$. It immediately follows that the modular operator $\Delta_\Psi = S_\Psi^\dagger S_\Psi = S_\Psi^\dagger U^\dagger U S_\Psi$ decomposes into a direct sum of operators of the form $\Delta_{\Psi_A}^* \otimes \dots \Delta_{\Psi_B}^*$ where $\Delta_{\Psi_{A/B}}^* = S_{\Psi_{A/B}}^{*\dagger} S_{\Psi_{A/B}}^*$ is the restriction of $\Delta_{\Psi_{A/B}}$ to $\mathcal{H}_{A/B}^*$.\footnote{Since Tomita operators $S_\Psi$, $S_{\Psi_A}^*$ and $S_{\Psi_B}^*$ are unbounded, it is important here that the restriction of $S_\Psi$ to a sector from \eqref{eq:hilbertdecomp} has the same domain as $[S_{\Psi_A}^* \otimes \dots S_{\Psi_B}^*]$, in addition to agreeing on the intersection of their domains.}
The modular flow generated by $h_\Psi = - \log \Delta_\Psi$ acts on a monomial $m = a_1 \dots b_n$ as a product of the modular flows generated by $h_{\Psi_A}$ or $h_{\Psi_B}$ acting on the monomial's constituent elements $a_1$ to $b_n$. Since the same modular flows are generated by the renormalized Hamiltonian $h_R$, we conclude that the algebra generated by $h_R$ and $\mathcal{A}_{R,0}* \mathcal{B}_{R,0}$ is simply the crossed product
\begin{align}
    \mathcal{C}_R = (\mathcal{A}* \mathcal{B}) \rtimes \mathbb{R}_{h_\Psi}~,
\end{align}
of $\mathcal{A}_{R,0}* \mathcal{B}_{R,0}$ by the action of $h_\Psi$. As in the case $T \ll  t_\mathrm{scr}$ discussed in Section \ref{sec:gsl}, this algebra contains Type II von Neumann subalgebras $\mathcal{A}_R$ (and $\mathcal{B}_R$) that are generated by $\mathcal{A}_{R,0}$ (or $\mathcal{B}_{R,0}$) and $h_R$ and that are isomorphic to the crossed product algebra $\mathcal{A}_R$ defined in Section \ref{sec:micro}.

Applying the results of Sections \ref{sec:sgen} and \ref{sec:gsl}, we find that states of the form
\begin{align} \label{eq:widehatPhiscramble}
    \ket{\widehat \Phi} = \int dx\, \varepsilon^{1/2} g(\varepsilon x) \ket{\Phi} \ket{x}~,
\end{align}
where $\beta x = \beta h_L = \beta h_R - h_\Psi$ and $\ket{\Phi} \in \mathcal{H}$ is an arbitrary state, satisfy
\begin{align}
    S(\widehat \Phi)_{\mathcal{B}_R} \approx \beta \braket{h_R} - \braket{\log[\varepsilon |g(\varepsilon h_R)|^2]} - S_\mathrm{rel}(\Phi||\Psi)_{\mathcal{B}_{R,0}}.
\end{align}
As in Section \ref{sec:gsl}, this is equal to the generalised entropy of the black hole horizon after the rightmost $\mathcal{A}_{r,0}$ modes fall in, but before the rightmost $\mathcal{B}_{r,0}$ modes do.\footnote{In accordance with \eqref{eq:subalgebraaction}, if the rightmost excitation consists of $\mathcal{H}_A$ modes, then the ``rightmost $\mathcal{B}_{r,0}$ modes'' that $\mathcal{B}_{R,0}$ encodes are fresh modes in the state $\ket{\Psi}$ that were created by backreaction from the $\mathcal{H}_A$ excitation -- not the $\mathcal{B}_{r,0}$ modes that would have been visible in the absence of that excitation.} A similar result applies for $\mathcal{A}_{R}$ with the black hole horizon replaced by the white hole horizon. 

This is all fairly intuitive: with access to only $\mathcal{A}_R$ or $\mathcal{B}_R$ we have no way to see behind the horizon and into the long wormhole. A more interesting result is the entropy of the entire algebra $\mathcal{C}_R$ generated by both early and late time excitations. This is again a crossed product algebra, and we have the superficially similar-looking formula
\begin{align} \label{eq:vnentropyscrambling}
    S(\widehat\Phi)_{\mathcal{C}_R} \approx \braket{h_R} - \braket{\log[\varepsilon p(\varepsilon h_L)]} - S_\mathrm{rel}(\Phi||\Psi)_{\mathcal{A}_{R,0} * \mathcal{B}_{R,0}}~.
\end{align}
Unfortunately, the bulk interpretation $S_\mathrm{rel}(\Phi||\Psi)_{\mathcal{A}_{R,0} * \mathcal{B}_{R,0}}$ is far less clear than that of $S_\mathrm{rel}(\Phi||\Psi)_{\mathcal{A}_{R,0}}$ or $S_\mathrm{rel}(\Phi||\Psi)_{\mathcal{B}_{R,0}}$. Unlike the ordinary modular operator $\Delta_\Psi$, the relative modular operator $\Delta_{\Phi|\Psi}$ does not generically act diagonally with respect to the decomposition \eqref{eq:hilbertdecomp}. Rather than acting as a bulk quantum field theory operator within a fixed background bulk geometry, it can change the bulk geometry by adding or removing shocks. We shall see in Section \ref{sec:longwormentinterp} that this is related to the existence -- generically -- of multiple quantum extremal surfaces within a long wormhole, with $O(1)$ differences betweeen their generalized entropies.

However, for states of the $\ket{\Phi} = a \ket{\Psi}$ with $a \in \mathcal{A}_{R,0}$, we can provide a simple interpretation to \eqref{eq:vnentropyscrambling}.  (A similar analysis applies for states $\ket{\Phi}=b\ket{\Psi}$,
with $b\in \mathcal {B}_{R,0}$.) The relative Tomita operator $S_{\Psi|\Phi}$ acts on a state $a' \ket{\Phi} \in \ket{\Psi} \oplus \mathcal{H}_A^*$ by
\begin{align}
S_{\Psi|\Phi} a' \ket{\Phi} = a'^\dagger \ket{\Psi}.
\end{align}
It therefore preserves the subspace $\mathcal{H}_A \cong \ket{\Psi} \oplus \mathcal{H}_A^* \subseteq \mathcal{H}$ and acts on it as the relative Tomita operator $S_{\Psi_A|a \Psi_A}$. It is also easy to check that $S_{\Psi|\Phi}$ preserves the orthocomplement of $\mathcal{H}_A$ in $\mathcal{H}$. As a result, the relative entropy $S_\mathrm{rel}(\Phi||\Psi)_{\mathcal{A}_{R,0} * \mathcal{B}_{R,0}}$ in \eqref{eq:vnentropyscrambling} is simply the relative entropy $S_\mathrm{rel}(a \Psi_A||\Psi_A)_{\mathcal{A}_{R,0}}$ of the modes in $\mathcal{A}_{R,0}$. Since the modes in $\mathcal{B}_{R,0}$ are in the Hartle-Hawking state in the state $\ket{\Phi}$, the entropy $S(\widehat\Phi)_{\mathcal{C}_R}$ is simply the generalized entropy of the bifurcation surface.

Because the state $\ket{\Phi}$ does not have a long wormhole, this may not seem very interesting. After all, what else could the entropy $S(\widehat\Phi)_{\mathcal{C}_R}$ be? However, it is important to remember that the entropy of an algebra is invariant under unitaries in either the algebra or its commutant; analogously, the generalized entropy of a bulk region is unchanged by unitaries that either act within the region or that are spacelike-separated from it. By acting on the state $\ket{\Phi}$ with products of unitaries in $\mathcal{A}_{R,0}$, $\mathcal{B}_{R,0}$, $\mathcal{A}_{L,0}$ and $\mathcal{B}_{L,0}$, we can make states with very long wormholes, with a large number of unitary shocks on either side of the original nonunitary shock created by $a$. The entropy of the algebra $\mathcal{C}_R$ will still be equal to the generalized entropy of the quantum extremal surface that lies  within the original nonunitary shock created by $a$. Unlike the algebras $\mathcal{A}_R$ and $\mathcal{B}_R$, the entropy of the algebra $\mathcal{C}_R$ can know about the state of quantum fields deep in a long wormhole, outside of the causal wedge of the right boundary. This is of course consistent with standard ideas about entanglement wedge reconstruction \cite{Czech:2012bh,maximin,Headrick:2014cta,Dong:2016eik,Faulkner:2017vdd, Cotler:2017erl}: by acting with unitaries that are localized to one of the two boundaries, we cannot change the entanglement wedge, and so the original quantum extremal surface should still determine the boundary entanglement entropy, as we found.

\subsection{Computing R\'{e}nyi entropies} \label{sec:renyi}

Most states $\ket{\Phi} \in \mathcal{H}$ with a long wormhole supported by $n$ shocks can not be created by acting with unitary operators in $\mathcal{A}_{R,0}*\mathcal{B}_{R,0}$ and its commutant; consequently they do not have entanglement entropies that can be so easily computed. We can make more progress by computing instead a simpler quantity, the R\'{e}nyi 2-entropy
\begin{align}
    S_2(\widehat\Phi)_{\mathcal{C}_R} = - \log \tr[\rho_{\widehat\Phi}^2] = - \log \braket{\widehat\Phi| \rho_{\widehat\Phi}|\widehat\Phi}.
\end{align}
Like von Neumann entropies, R\'{e}nyi entropies are computed using the same formula in both Type I and II von Neumann factors, with the entropy in a Type II$_\infty$ factor defined only up to an arbitrary state-independent constant controlled by the scaling of the trace $\tr$. Again, this state-independent constant is a relic of the subtraction of a divergent constant in order to renormalize the entropy; in fact, because the divergent part of the entanglement in a Type II algebra necessarily has a flat spectrum, the same divergent constant is subtracted for both von Neumann entropies and R\'{e}nyi entropies.

Using the approximation \eqref{eq:densop} for the density matrix $\rho_{\widehat\Phi}$, along with the fact that $g(\varepsilon h_R) \approx g(\varepsilon x)$ in the limit $\varepsilon \to 0$, we find
\begin{align} \label{eq:renyiformula}
    e^{-S_2(\widehat\Phi)_{\mathcal{C}_R}} \approx \int_{-\infty}^\infty \,dx\, \varepsilon^2 |g(\varepsilon x)|^4\, e^{-\beta x}\, \braket{\Phi|\Delta_{\Phi|\Psi}|\Phi}~.
\end{align}
However there is an important subtlety that we need to be somewhat careful about here. The approximation \eqref{eq:densop}, along with the substitution $g(\varepsilon h_R) \approx g(\varepsilon x)$ above, relied crucially on the assumption that the derivative $dg(\varepsilon x)/dx = O(\varepsilon)$ in the limit $\varepsilon \to 0$. This is true for any smooth function $g$ so long as $\varepsilon x$ is held fixed as we take the small $\varepsilon$ limit. The wavefunction $\ket{\widehat\Phi}$ and the von Neumann entropy $S(\widehat\Phi)_{\mathcal{C}_R}$ are both dominated by $x \sim O(1/\varepsilon)$, and so in von Neumann entropies computations the approximation is valid. However if the integral \eqref{eq:renyiformula} is to converge then $|g(\varepsilon x)|^4$ must decay faster than $e^{\beta x}$ when $x \to - \infty$, for any $\varepsilon > 0$. Moreover the dominant contribution to the integral will always come from a region where $|g(\varepsilon x)|^4 e^{-\beta x}$ is approximately constant and hence where $dg(\varepsilon x)/dx = O(1)$. If $|g(x)|^2 \sim \exp(-x^2/2\sigma^2)$ is Gaussian, this occurs at $x \sim \beta \sigma^2/2 \varepsilon^2$. As a result, we cannot trust the approximation \eqref{eq:densop} when computing the dominant contribution to the R\'{e}nyi 2-entropy. 

We will not be overly concerned about this however. We are only computing the R\'{e}nyi 2-entropy as an easier-to-calculate proxy for the von Neumann entropy. Since the latter is dominated by the same range of $x \sim O(1/\varepsilon)$ that dominates the wavefunction itself, we are more interested in the contribution to the R\'{e}nyi entropy from $x \sim O(1/\varepsilon)$, where our approximations can be trusted, than the contribution from $x \sim \beta \sigma^2/2 \varepsilon^2$ that dominates the actual R\'{e}nyi entropy calculation but which involves only a tiny tail of the wavefunction.

It is worth pausing for a moment to explain the relationship between the discussion above and the standard lore in AdS/CFT \cite{Dong:2016fnf} that R\'{e}nyi entropies are computed using areas in \emph{backreacted geometries} that are very different from the semiclassical geometry of the state in question itself. As shown in \cite{Dong:2018seb, Akers:2018fow, Dong:2019piw}, the R\'{e}nyi 2-entropy version of the QES prescription, which can be derived using the replica trick, is simply
\begin{align} \label{eq:RenyiQES}
    e^{-S_2(B)} = \braket{e^{-A(\partial b)/4G} \rho_b}.
\end{align}
Here $b$ is the entanglement wedge of $B$ (the right exterior in the case of a two-sided black hole with $B$ the right boundary), and the expectation value is taken for the bulk dual of the boundary state (not for some backreacted version of that bulk dual). More explicitly, the density matrix $\rho_b$ in \eqref{eq:RenyiQES} should be defined as
\begin{align}
    \rho_b =\int d\mathrm{A}\,\ket{\mathrm A}\bra{\mathrm A}\, p(\mathrm A)\, \rho_b(\mathrm A), 
\end{align}
where $p(\mathrm A)$ is the probability that the surface $\partial b$ has area $A(\partial b) = \mathrm A$, $\rho_b(\mathrm A)$ is the normalized density matrix of the bulk quantum fields conditioned on $A(\partial b) = \mathrm A$, and $\ket{\mathrm A}\bra{\mathrm A}$ is a projector onto $A(\partial b) = \mathrm A$.
As usual, all UV issues involved in defining such a density matrix (e.g. the Type III nature of the bulk QFT algebra) are expected to be renormalized by the presence of the $e^{-A(\partial b)/4G}$ term. For example, as shown in Section \ref{sec:firstderiv}, when $b$ is the right exterior of a black hole, the full expression $e^{-A(\partial b)/4G} \rho_b$ is equal to the density matrix of the bulk algebra $\mathcal{A}_r$ dual to $\mathcal{A}_R$, which is well defined even though neither $e^{-A(\partial b)/4G}$ nor $\rho_b$ exists on its own.

An intuitive interpretation of \eqref{eq:RenyiQES} is the following: one can obtain the usual QES prescription by replacing $\log \rho_B$ by $\log \rho_b - A(\partial b)/4G$ in the formula $S(B) = - \braket{\log \rho_B}$; the same substitution in the formula $e^{-S_2(B)} = \braket{\rho_B}$ leads to \eqref{eq:RenyiQES}. Because $e^{-A/4G}$ is a highly nonlinear function of $A$, $\braket{e^{-A/4G}}$ is very different from $e^{-\braket{A}/4G}$. In fact the dominant contribution to the former comes from the tiny tail of the wavefunction where $A \ll \braket{A}$; the exponentially small amplitude of this tail is more than cancelled out by its exponentially larger value of $e^{A/4G}$. This is the source of the famous backreaction! In contrast, the contribution to $e^{-S_2}$ from the peak of the wavefunction does not involve any backreaction at all, but it gives a negligibly small contribution to the final answer. That said, that small contribution is in many ways more physical relevant, and is certainly more closely related to the von Neumann entropy than the dominant contribution from the backreacted geometry. It is the analogous contribution from the peak of the wavefunction that we will be able to compute using \eqref{eq:renyiformula}.

A priori, the formula \eqref{eq:renyiformula} does not seem any more hopeful than the formula \eqref{eq:vnentropyscrambling} for the von Neumann entropy. However it turns out that, unlike the difficult term $S_\mathrm{rel}(\Phi||\Psi) = - \braket{\Phi|\log \Delta_{\Psi|\Phi}|\Phi}$ in \eqref{eq:vnentropyscrambling}, the computation of $\braket{\Phi|\Delta_{\Phi|\Psi}|\Phi}$ is tedious but not too difficult. The key features of the calculation are nicely illustrated by a state
\begin{align} \label{eq:twoshockstate}
    \ket{\Phi} = \sum_i a_{i} b_i \ket{\Psi}
\end{align}
containing only two shocks. Here each term $a_i b_i$ is an irreducible monomial as usual, while the sum over $i$ is included to allow the two shocks to be entangled with one another. From now on we shall drop the explicit sum and instead use Einstein summation convention. We also assume that the state $\ket{\Phi}$ is normalized, if necessary by absorbing a normalization constant into the definition of $a_i b_i$. The generalization of \eqref{eq:twoshockstate} to a state containing $n$ shocks is conceptually straightforward, but is presentationally challenging because of the number of terms involved. We will therefore focus our arguments on the two-shock state $\ket{\Phi}$ and only briefly comment on its generalization. We have
\begin{align}\notag
    \braket{\Phi| \Delta_{\Phi|\Psi}|\Phi} &= \braket{\Psi|b_i^\dagger a_i^\dagger S_{\Phi|\Psi}^\dagger S_{\Phi|\Psi} a_j b_j |\Psi }
    = \braket{\Psi|b_i^\dagger a_i^\dagger S_{\Phi|\Psi}^\dagger b_j^\dagger a_j^\dagger |\Phi }
    \\ \notag&= \overline{\braket{\Phi|a_i b_i S_{\Phi|\Psi} a_j b_j |\Psi }}
    =\braket{\Phi| a_i b_i b_j^\dagger a_j^\dagger |\Phi }
    \\&=  \braket{\Psi| b_i^\dagger a_i^\dagger a_j b_j b_k^\dagger a_k^\dagger a_l b_l |\Psi } = \left\lVert b_i^\dagger a_i^\dagger a_j b_j \ket{\Psi }\right\rVert^2. \label{eq:quadsum}
\end{align}
The state $b_i^\dagger a_i^\dagger a_j b_j \ket{\Psi }$ can be expanded in the decomposition \eqref{eq:hilbertdecomp} by writing $a_i^\dagger a_j = \braket{a_i^\dagger a_j} + \tilde a_{i,j}$ where $\braket{a_i^\dagger a_j} = \braket{\Psi|a_i^\dagger a_j|\Psi}$. We then obtain
\begin{align}\notag
    \braket{\Phi| \Delta_{\Phi|\Psi}|\Phi} &=  \left\lVert \braket{a_i^\dagger a_j} b_i^\dagger b_j \ket{\Psi }\right\rVert^2 + \left\lVert b_i^\dagger \tilde a_{i,j} b_j \ket{\Psi }\right\rVert^2
    \\&= \braket{a_i^\dagger a_j}\braket{a_k^\dagger a_l} \braket{b_i^\dagger b_j b_k^\dagger b_l} + \left[\braket{a_i^\dagger a_j a_k^\dagger a_l} -  \braket{a_i^\dagger a_j} \braket{a_k^\dagger a_l}\right] \braket{b_i^\dagger b_l} \braket{b_j b_k^\dagger}, \label{eq:bigexpand}
\end{align}
where all expectation values are in the state $\ket{\Psi}$. The corresponding formula for a state containing $n$ shocks still involves a sum over four indices $i,j,k,l$ that range over the superposition of operators used to create the shocks as in \eqref{eq:twoshockstate}. However for each value of $i,j,k,l$ we now have $n$ positive terms and $(n-1)$ negative terms. Each positive term is associated to a specific shock $C$, and contains a factor of $\braket{c_i^\dagger c_j c_k^\dagger c_l}$ from the operators $\{c_i\}$ (in either $\mathcal{A}_{R,0}$ or $\mathcal{B}_{R,0}$) that created that shock. It also contains a factor of $\braket{\ell_i^\dagger \ell_l} \braket{\ell_j \ell_k^\dagger}$ for each shock to the left of $C$ (i.e created before $C$), created by the operators $\{\ell_i\}$, and a factor of $\braket{r_i^\dagger r_j}\braket{r_k^\dagger r_l}$ for each shock to the right of $c$, created by the operators $\{r_i\}$. For example, the first term in \eqref{eq:bigexpand} is associated to the $\mathcal{B}_{R,0}$ shock, while the positive component of the second term is associated to the $\mathcal{A}_{R,0}$ shock. The $(n-1)$ negative terms are associated to gaps between shocks and contain a factor of $\braket{\ell_i^\dagger \ell_l} \braket{\ell_j \ell_k^\dagger}$ for each shock to the left of the gap and a factor of $\braket{r_i^\dagger r_j}\braket{r_k^\dagger r_l}$ for each shock to the right of the gap.

The expansion in \eqref{eq:bigexpand} can be rewritten in a somewhat suggestive form that, as we will see in Section \ref{sec:longwormentinterp}, has a very natural bulk interpretation in terms of a sum over quantum extremal surfaces. Let $\ket{\widetilde{\Phi}} \in \mathcal{H}_A \otimes \mathcal{H}_B$ be defined by
\begin{align} \label{eq:twoshockstatephi}
   \ket{\widetilde\Phi} = \sum_i  a_i \ket{\Psi_A} b_i \ket{\Psi_B}.
\end{align}
In other words, $\ket{\widetilde\Phi}$ is the same state as $\ket{\Phi}$ except we are now treating $\mathcal{H}_A^* \otimes \mathcal{H}_B^*$ as a subspace of $\mathcal{H}_A \otimes \mathcal{H}_B$ rather than as a subspace of $\mathcal{H}$. One can think of $\ket{\widetilde\Phi}$ as the state of the bulk quantum fields on the two-shock background geometry. Our goal will be to rewrite \eqref{eq:bigexpand} entirely in terms of $\ket{\widetilde\Phi}$, without explicit reference to $\{a_i\}$ or $\{b_i\}$. The reduced density matrix $\tilde\rho_A$ of $\ket{\widetilde\Phi}$ on $\mathcal{H}_A$ can be written as
\begin{align}
    \tilde\rho_A = \left[\braket{\Psi_B| b_j^\dagger b_i |\Psi_B}\right] \,\,a_i \ket{\Psi_A}\bra{\Psi_A} a_j^\dagger.
\end{align}
We therefore have
\begin{align}\notag
    \braket{\widetilde \Phi| \tilde\rho_A \otimes \Delta_{\Psi_B} |\widetilde \Phi} &= \braket{ a_i^\dagger a_j}\braket{a_k^\dagger a_l} \braket{b_k^\dagger b_j} \braket{b_i^\dagger \Delta_{\Psi_B} b_l}
    \\& = \braket{ a_i^\dagger a_j}\braket{a_k^\dagger a_l} \braket{b_k^\dagger b_j} \braket{b_l b_i^\dagger}.
\end{align}
If we relabel $i \leftrightarrow k$ and $j \leftrightarrow l$ then this is exactly the negative term in \eqref{eq:bigexpand}. In a state with $n$ shocks, the negative terms can similarly be interpreted as the expectation value of a product of modular operators $\Delta_\Psi$ on all shocks to the left of the relevant gap, tensored with the reduced density matrix of the state on the shocks to the right of the gap.

What about the positive terms? Let $\Delta_{\widetilde\Phi|\Psi}^A$ be the relative modular operator of $\ket{\widetilde\Phi}$ relative to $\ket{\Psi_A}$ for the algebra $\mathcal{A}_{R,0}$. We have
\begin{align}\notag
    \braket{\widetilde \Phi| \Delta_{\widetilde\Phi|\Psi}^A \otimes \Delta_{\Psi_B} |\widetilde \Phi} &= \braket{\Psi_A|a_i^\dagger \Delta_{\widetilde\Phi|\Psi}^A a_j|\Psi_A} \braket{\Psi_B|b_i^\dagger \Delta_\Psi b_j|\Psi_B}
    \\ \notag&= \braket{\widetilde\Phi|a_j a_i^\dagger|\widetilde\Phi} \braket{ b_j b_i^\dagger}
    \\&= \braket{a_i^\dagger a_j a_k^\dagger a_l} \braket{ b_i^\dagger b_l}\braket{ b_j b_k^\dagger}. \label{eq:secondpositiveterm}
\end{align}
This is exactly the second positive term in \eqref{eq:bigexpand}. 
Concerning the first positive term, one might expect that it would be related to the second by a simple
exchange of $A$ and $B$, but an examination of the index structure in eqn. (\ref{eq:bigexpand}) reveals that this
is not the case.  In fact, we will have to work harder to find 
an analogous formula for the first positive term (and the logic of the resulting formula will be more
clear in section \ref{sec:longwormentinterp}.    It is helpful to define the relative modular operator $\Delta_{\widetilde\Phi|\Psi}^{B|A}$ for the state $\ket{\widetilde\Phi}$ relative to $\ket{\Psi_A}\ket{\Psi_B}$ -- on the algebra $\mathcal{B}_{R,0} \otimes \mathcal{B}(\mathcal{H}_A)$. Here $\mathcal{B}(\mathcal{H}_A)$ is the algebra of \emph{all} bounded operators (not just those in $\mathcal{A}_{R,0}$) on the Hilbert space $\mathcal{H}_A$. Because the state $\ket{\Psi_A}\ket{\Psi_B}$ is not separating for the algebra $\mathcal{B}_{R,0} \otimes \mathcal{B}(\mathcal{H}_A)$ (i.e. it can be annihilated by operators in that algebra), there is an additional subtlety in the definition of the relative Tomita operator $S_{\widetilde\Phi|\Psi}^{B|A}$. Specifically, the defining equation \eqref{eq:relmoddef} should be replaced by
\begin{align}
    S_{\widetilde\Phi|\Psi}^{B|A} \alpha \ket{\Psi_A}\ket{\Psi_B} = P \alpha^\dagger \ket{\widetilde\Phi} \,\,\,\,\,\, \forall \alpha \in \mathcal{B}_{R,0} \otimes \mathcal{B}(\mathcal{H}_A),
\end{align}
where $P \in \mathcal{B}_{R,0} \otimes \mathcal{B}(\mathcal{H}_A)$ is the smallest projector such that $P \ket{\Psi_A}\ket{\Psi_B} = \ket{\Psi_A}\ket{\Psi_B}$. In this case, $P = \ket{\Psi_A} \bra{\Psi_A}$. Setting
$\Delta_{\widetilde\Phi|\Psi}^{B|A}=S^{\dagger],B|A}_{\widetilde\Phi|\Psi}S^{B|A}_{\widetilde\Phi|\Psi}$, we
therefore have
\begin{align}\notag
    \braket{\widetilde \Phi| \Delta_{\widetilde\Phi|\Psi}^{B|A} |\widetilde \Phi} &= \bra{\Psi_A} a_i^\dagger \braket{\Psi_B| b_i^\dagger \,S_{\widetilde\Phi|\Psi}^{B|A  \dagger}S_{\widetilde\Phi|\Psi}^{B|A}\, a_j |\Psi_A} b_j\ket{\Psi_B} 
    \\ \notag&= \bra{\Psi_A} a_i^\dagger \braket{\Psi_B| b_i^\dagger \,S_{\widetilde\Phi|\Psi}^{B|A  \dagger} P a_j^\dagger b_j^\dagger |\widetilde \Phi}
    \\ \notag & = \braket{\widetilde\Phi|a_j b_j P a_i^\dagger b_i^\dagger |\widetilde\Phi}
    \\&= \braket{b_i^\dagger b_j b_k^\dagger b_l} \braket{a_i^\dagger a_j} \braket{a_k^\dagger a_l}, 
\end{align}
which is indeed the first term in \eqref{eq:bigexpand}. We conclude that
\begin{align} \label{eq:threeterms}
    \braket{\Phi|\Delta_{\Phi|\Psi}|\Phi} = \braket{\widetilde\Phi| \Delta_{\widetilde\Phi|\Psi}^{B|A} - \tilde\rho_A \otimes \Delta_{\Psi_B} + \Delta_{\widetilde\Phi|\Psi}^A \otimes \Delta_{\Psi_B}|\widetilde\Phi}.
\end{align}
For a state with $n$ shocks, the $n$ positive terms are equal to the expectation values of a product of modular operators $\Delta_\Psi$ for all shocks to the left of the shock $c$ associated to the term in question, together with a relative modular operator $\Delta_{\widetilde\Phi|\Psi}^{C|R}$ for the algebra consisting $\mathcal{A}_{R,0}$ (or $\mathcal{B}_{R,0}$) acting on the shock $C$ tensored with all bounded operators acting on shocks to the right of $c$. The form of the $(n-1)$ negative terms was already discussed above.

For our purposes, the rationale behind the definition of $\Delta^{B|A}_{\widetilde\Phi|\Psi}$ is as follows.  In ordinary quantum mechanics, consider a bipartite system $XY$ and a state $\Psi_{XY}$.   Write  $\rho_{\Psi,X}$, $\rho_{\Psi,Y}$ 
for the density matrices of $\Psi_{XY}$ when reduced to $X, Y$, and suppose  that $\rho_{\Psi,Y}$ is invertible but $\rho_{\Psi,X}$ is not.   Let $\widetilde\Phi_{XY}$ be another state.   Comparing to the situation considered in the last paragraph, the algebra of operators on system $X$ corresponds to ${\mathcal B}_{R,0}\otimes {\mathcal B}({\mathcal H}_A)$, the algebra of operators on system $Y$ corresponds to the commutant of this
algebra, which is ${\mathcal B}_{L,0}$, and $P$ corresponds to the projector on the support of $\rho_{X,\Psi}$.
  With $S_{\widetilde\Phi|\Psi}$ defined as in the text, including the projection operator $P$, and with $\Delta_{\widetilde\Phi|\Psi}=S^\dagger_{\widetilde\Phi|\Psi}S_{\widetilde\Phi|\Psi}$, one has, just as in
  \eqref{eq:deltafromdens}, where the density matrices were assumed to be invertible,
  \begin{equation}\Delta_{\widetilde\Phi|\Psi}=\rho_{\widetilde\Phi,X}\otimes \rho_{\Psi,Y}^{-1}.\label{goodrel}\end{equation} 
In section \ref{sec:longwormentinterp}, we will use 
this last relation, or rather its
analog in the gravity problem, to interpret a
certain product of bulk
density matrices in terms of the modular operator $\Delta^{B|A}_{\widetilde\Phi|\Psi}$.   This will be a step
in making contact with eqn. \eqref{eq:threeterms}.

\subsection{A bulk interpretation?} \label{sec:longwormentinterp}
Each of the three terms in \eqref{eq:threeterms} is the expectation value of a bulk QFT operator in the Hilbert space $\mathcal{H}_A \otimes \mathcal{H}_B$.  This suggests that the full formula should have a natural bulk interpretation, which we will now try to understand. 

\begin{figure}[t]
\begin{subfigure}{.48\textwidth}
  \centering
 \includegraphics[width = 0.97\linewidth]{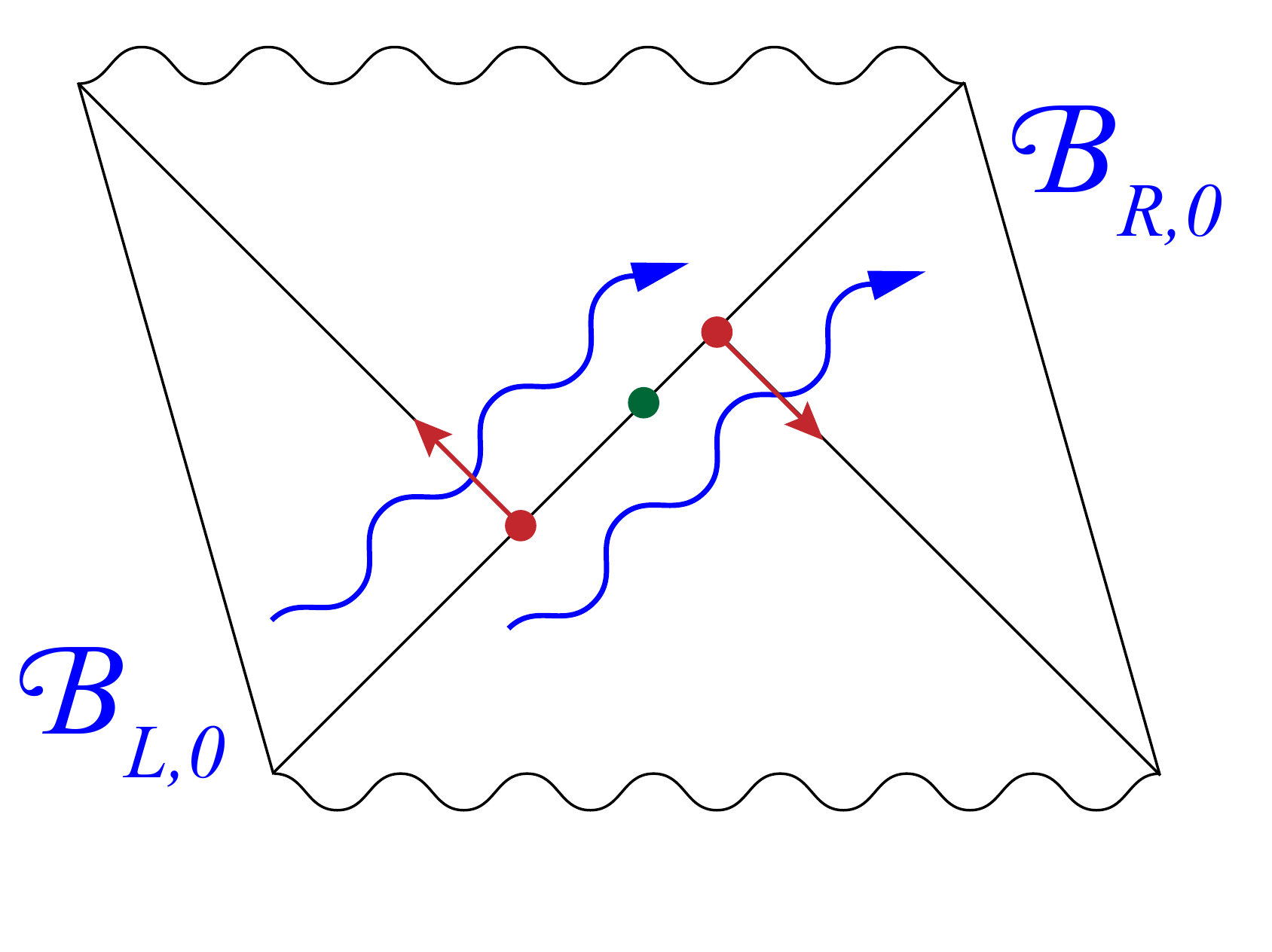}
\end{subfigure}
\begin{subfigure}{.48\textwidth}
  \centering
 \includegraphics[width = 0.95\linewidth]{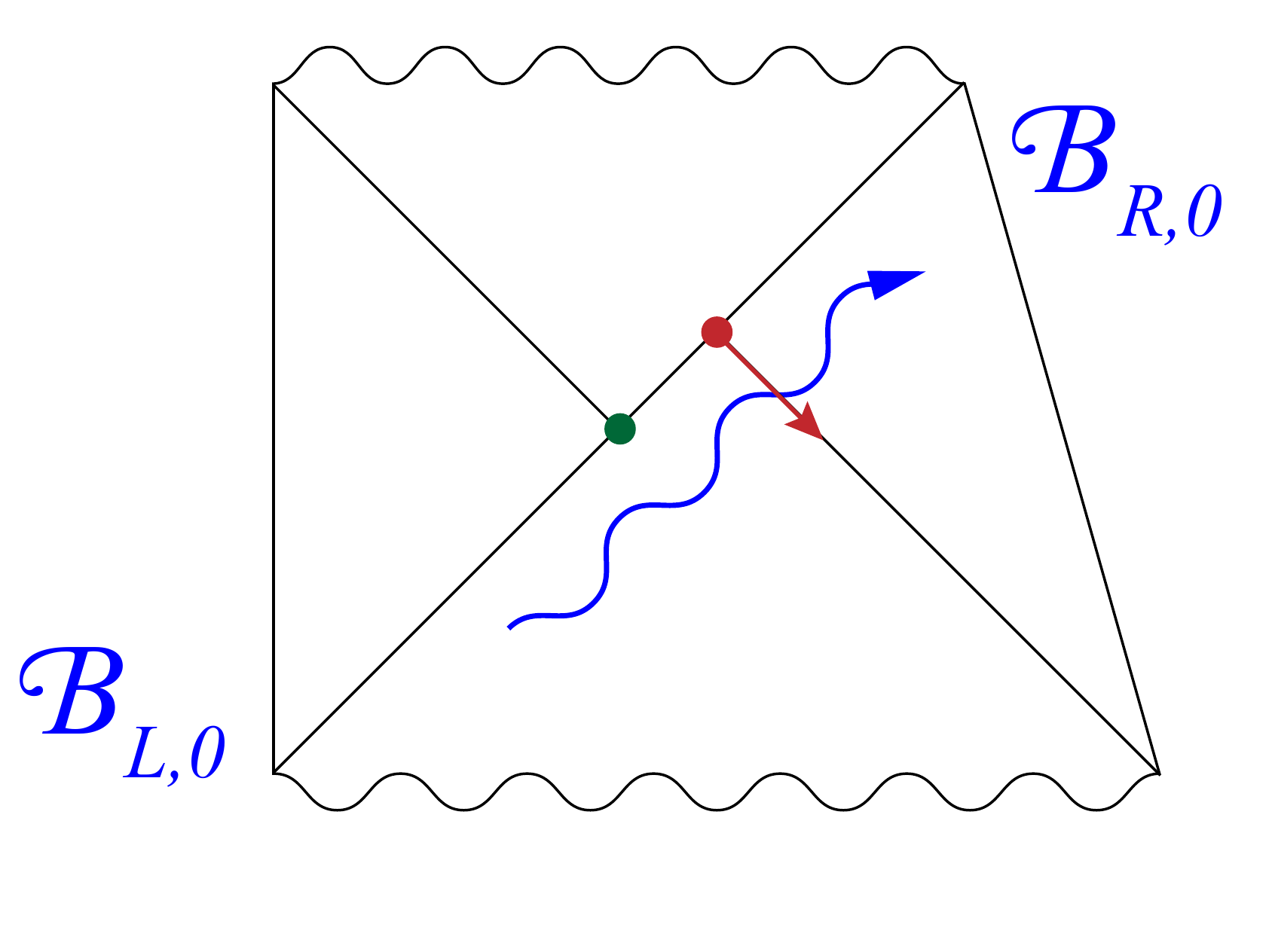}
\end{subfigure}
\caption{In the natural frame for $\mathcal{H}_A$, an operator $b_k \in \mathcal{B}_{R,0}$ creates a high energy shock and hence a large separation between the left black hole horizon and the right white hole horizon. The left white hole horizon and right black hole horizon coincide in this frame up to perturbative corrections. \emph{Left:} Generically, the quantum extremal surface (green) lies on this horizon in between the left and right causal surfaces (red); the red arrows show the direction of increasing generalized entropy on the causal surfaces. The QES and both causal surfaces have approximately the same generalized entropy. \emph{Right:} If the operator $b_k$ is unitary, then the modes in $\mathcal{B}_{\ell,0}$ remain in the Hartle-Hawking state, and the left causal surface is itself extremal.}
\label{fig:singleshock}
\end{figure}

To do so, we need to understand the  quantum extremal surfaces that generically exist for a black hole state containing two shocks. As a warm up however, we will consider the case where there is only a single shock, created by some operator $b_k \in \mathcal{B}_{R,0}$. In the natural frame of the $\mathcal{H}_B$ modes, this state is a slightly perturbed version of an empty two-sided black hole, and has a single quantum extremal surface at the bifurcation surface of the black hole. However, since we will later be interested in states where there are two shocks, it is helpful to instead study the same one-shock state in the natural frame for the $\mathcal{H}_A$ modes. In this frame, the operator $b_k$ creates a very high energy shock near the right black hole horizon. The backreaction of this shock creates a large separation between right white hole event horizon and left black hole event horizon, as shown in Figure \ref{fig:singleshock}. The right causal surface (i.e. the intersection of the right white and black hole horizons) and the left causal surface have a large, approximately lightlike separation, but have the same generalized entropy (up to $O(G)$ corrections). Generically, the right causal surface has a strictly positive quantum expansion in the rightwards, future-oriented null direction, while the left causal surface has negative expansion in the same direction.\footnote{The quantum expansion of a surface describes the functional derivative of generalized entropy at that surface \cite{Bousso:2015mna}. The quantum expansion on a causal horizon, in the direction of asymptotic infinity, is always positive by the generalized second law.} The single quantum extremal surface then lies on the (right black hole/left white hole) horizon on a cut strictly intermediate between the two causal surfaces. The sole exception to this generic statement occurs when either the modes in $\mathcal{B}_{\ell,0}$, or those in $\mathcal{B}_{r,0}$ are in exactly the Hartle-Hawking vacuum. In these cases, either the left or right causal surface is itself quantum extremal. For example, if the operator $b_k$ creating the shock is exactly unitary then the modes in $\mathcal{B}_{\ell,0}$ are not excited and hence the left causal surface is quantum extremal.

\begin{figure}[t]
\begin{center}
  \includegraphics[width = 0.8\linewidth]{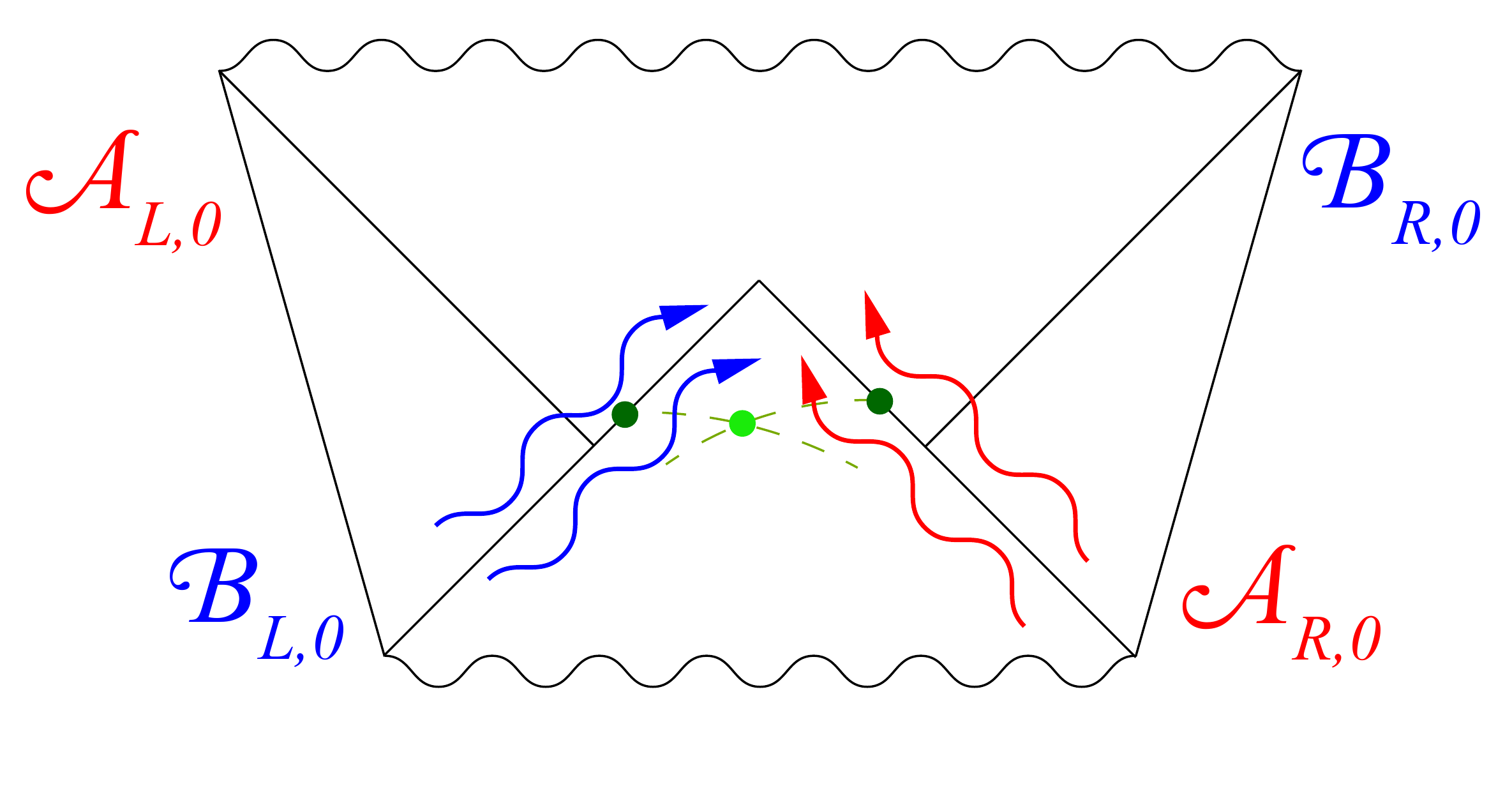} 
\end{center}
\caption{In a wormhole containing two shocks there are two ``throat'' quantum extremal surfaces (dark green). One lies inside the left shock and entirely to the left of the right shock; the other lies inside the right shock and entirely to the right of the left shock. A ``bulge'' surface (light green) lies in between the two, and is entirely in the past of both shocks. The dashed green lines represent families of surfaces with zero quantum expansion in the past and future rightwards null directions respectively. Their intersection determines the location of the bulge surface.}
\label{fig:twoshocks}
\end{figure}

What about if we create a second shock using an operator $a_k \in \mathcal{A}_{R,0}$? This is shown in Figure \ref{fig:twoshocks}. So long as the single shock QES didn't coincide with the right causal surface, it will lie entirely to the left of this second shock, which is parametrically closer to the right causal surface than the left causal surface, and so will be unaffected by it. By identical arguments, there will also generically exist a second QES that lies inside the second shock, but entirely to the right of the first shock, on the right white hole horizon. In this case, the exception occurs when the right QES, in the absence of the left shock, would lie on the left causal surface. In that case, the existence of the left shock may change the location of the QES, or cause it to not exist at all. This is in fact exactly what happens when $a_k$ is unitary (and the sum over $k$ is trivial):  semiclassical unitaries acting within a boundary region $B$ cannot create a new QES for $B$, and so states created from a short wormhole using unitary shocks only contain a single QES, as discussed in Section \ref{sec:scrambleII}.

The formula \eqref{eq:RenyiQES} says that the R\'{e}nyi 2-entropy $S_2(\widehat\Phi)$ should be given by
\begin{align} \label{eq:S2contributions}
e^{-S_2(\widehat\Phi)} = \braket{\widehat\Phi|p(A) \exp[ - \hat A/4G -  h_{\widetilde\Phi,r}]|\widehat\Phi} = \braket{\Phi|p(A) \exp[ - \hat A/4G -  h_{\widetilde\Phi,\ell}]|\widehat\Phi},
\end{align}
where $h_{\widetilde\Phi,r} = - \log \rho_{\widetilde\Phi,r}$ and $ h_{\widetilde\Phi,\ell} = - \log \rho_{\widetilde\Phi,\ell}$ are the modular Hamiltonians in the state $\ket{\widetilde\Phi}$ of the bulk fields that are respectively to the right and left of the \emph{minimal} quantum extremal surface, and $p(A)=\epsilon |g(\epsilon x)|^2$
is the probability distribution for the area.
The minimal QES in this context is just the QES that gives the largest value for \eqref{eq:S2contributions}. We reiterate that the operators $h_{\widetilde\Phi,\ell}$ and $ h_{\widetilde\Phi,r}$ do not really exist because of the Type III nature of the QFT algebras; only the combinations appearing in \eqref{eq:S2contributions} are expected to be well defined.

Recall from Section \ref{sec:curious}, and the more detailed derivation in \eqref{eq:areadif} and \eqref{eq:areaenergy} that the area of the bifurcation surface in a single-shock geometry is
\begin{align}
    \frac{\hat A}{4G}  - \frac{A_0}{4G}= \beta h_L - \beta \hat h_\ell = \beta h_R - \hat h_r,
\end{align}
where $h_\ell$ and $h_r$ are divergent one-sided boost operators that satisfy $\hat h = h_r - h_\ell$. Importantly, we did not need to distinguish here between the areas of the left causal surface, the quantum extremal surface and the right causal surface, all of which are perturbatively close to one another and differ in area only at $O(G^2)$. Since the statement about area differences is gauge-invariant, it is still true even when we describe an $\mathcal{H}_B$ shock in the natural frame for $\mathcal{H}_A$ as in Figure \ref{fig:singleshock}, which causes the separation between the QES and the two causal surfaces to appear large.

If we create a second $\mathcal{H}_A$ shock to the right of the $\mathcal{H}_B$ shock, as in Figure \ref{fig:twoshocks} then backreaction from the second shock will change the area of the right causal surface relative to the other two. But the areas of the left QES and the left causal surface, which lie entirely to the left of the $\mathcal{H}_A$ shock, will still be equal, and will still satisfy
\begin{align}
    \frac{\hat A}{4G}  - \frac{A_0}{4G}= \beta h_L -  \beta h_{\ell} ,
\end{align}
because the right-hand side also only involves physics to the left of the $\mathcal{H}_A$ shock. Here $h_{\ell}$ is still the one-sided boost energy of the bulk modes in the left exterior; since the only excited modes in the left exterior are in $\mathcal{B}_{L,0}$, we morally have $h_{\ell} \in \mathcal{B}_{L,0}$ (except for the fact that it is divergent).

In \eqref{eq:S2contributions}, the divergence in $\hat A/4G$ is supposed to be renormalized by the addition of $h_{\widetilde\Phi,r} = - \log \rho_{\widetilde\Phi,r}$ where $\rho_{\widetilde\Phi,r}$ is the density matrix of $\ket{\widetilde\Phi}$ on all modes to the right of the QES. In the case of the left QES, the algebra of operators to the right of the QES is $\mathcal{B}_{R,0} \otimes \mathcal{B}(\mathcal{H}_A)$. Recall from Section \ref{sec:renyi} that, by analogy with the formula \eqref{goodrel} for Type I and II algebras, the modular operator $\Delta_{\widetilde\Phi|\Psi}^{B|A}$ should be interpreted as a product of the density matrix for $\ket{\widetilde\Phi}$ on $\mathcal{B}_{R,0} \otimes \mathcal{B}(\mathcal{H}_A)$ with the inverse density matrix for $\ket{\Psi_B} \in \mathcal{H}_B$ on $\mathcal{B}_{L,0}$. As in \eqref{eq:rhoonesidedboost}, the latter density matrix should be proportional to $\exp(- \beta \hat h_\ell)$. It follows that we should identify
\begin{align} \label{eq:leftQESmodular}
    \log \Delta_{\widetilde\Phi|\Psi}^{B|A} = - h_{\widetilde\Phi,r} + \beta \hat h_\ell,
\end{align}
and hence
\begin{align} \label{eq:leftQESsgen}
    \frac{\hat A}{4G} +  h_{\widetilde\Phi,r}   - \frac{A_0}{4G}= \beta h_L - \log \Delta_{\widetilde\Phi|\Psi}^{B|A}.
\end{align}
If we substitute \eqref{eq:leftQESsgen} into \eqref{eq:S2contributions}, we find that, when the left QES is minimal,
\begin{align} \label{eq:leftQESrenyi}
  e^{-S_2(\widehat\Phi)} &= e^{-A_0/4G} \braket{\widehat\Phi|\,p(A)\, e^{-\beta h_L} \,\Delta_{\widetilde\Phi|\Psi}^{B|A} \,|\widehat\Phi}
  \\&=e^{-A_0/4G}\int_{-\infty}^\infty dx\, \varepsilon^2\,|g( \varepsilon x)|^4\,e^{-\beta x} \, \braket{\widetilde\Phi| \Delta_{\widetilde\Phi|\Psi}^{B|A}|\widetilde\Phi}.
\end{align}
In the second equality, we used the formulas $h_L = x$ and $p(A) = \varepsilon |g(\varepsilon x)|^2$. Additionally, we used the identification of the subspaces $\mathcal{H}_A^* \otimes \mathcal{H}_B^* \subseteq \mathcal{H}$ and $\mathcal{H}_A^* \otimes \mathcal{H}_B^* \subseteq \mathcal{H}_A \otimes \mathcal{H}_B$ to replace $\ket{\Phi}$ in the definition \eqref{eq:widehatPhiscramble} of $\ket{\widehat\Phi}$ by $\ket{\widetilde\Phi}$.
The formula \eqref{eq:leftQESrenyi} is exactly the contribution to \eqref{eq:renyiformula} that we obtain from the first term in \eqref{eq:threeterms} (up to the usual subtraction of a divergent constant $A_0/4G$). 

An identical argument tells us that the area of the right QES is equal to that of the right causal surface and is given by
\begin{align} \label{eq:rightQESarea1}
    \frac{\hat A}{4G} - \frac{A_0}{4G} = \beta h_R - \beta \hat h_{r},
\end{align}
where $\hat h_{r}$ is the one-sided boost energy of the $\mathcal{A}_{R,0}$ modes in the right exterior. This time, the density matrix of $\ket{\Psi_A} \in \mathcal{H}_A$ (if it existed) would be proportional to $\exp(-\beta \hat h_r)$. We learned from \eqref{eq:deltapsicalc} that the restriction of $h_\Psi$ to $\mathcal{H}_A^* \otimes \mathcal{H}_B^*$ is equal to $h_{\Psi_A} + h_{\Psi_B}$. Here $h_{\Psi_A} = - \log \Delta_{\Psi_A}$ with $\Delta_{\Psi_A}$ the modular operator of $\ket{\Psi_A} \in \mathcal{H}_A$ on the algebra $\mathcal{A}_{R,0}$ (and similar for $h_{\Psi_B}$). We therefore have
\begin{align} \label{eq:betahRtwoshock}
    \beta h_R = \beta h_L + h_{\Psi} = \beta h_L + h_{\Psi_A} + h_{\Psi_B}.
\end{align}
We can formally write $h_{\Psi_A} = \beta \hat h_r - h_{\Psi_A, \ell}$, where the density matrix of $\ket{\Psi_A}$ on $\mathcal{A}_{L,0}$ would be proportional to $\exp(-h_{\Psi_A, \ell})$. Substituting \eqref{eq:betahRtwoshock} into \eqref{eq:rightQESarea1} leads to
\begin{align}
    \frac{\hat A}{4G} - \frac{A_0}{4G} = \beta h_L + h_{\Psi_A} + h_{\Psi_B} - \beta \hat h_r = \beta h_L + h_{\Psi_B} - h_{\Psi_A, \ell}.
\end{align}
As in the case of the left QES, to cancel the divergence in $\hat A/4G$, we need to add $h_{\widetilde\Phi,r} = - \log \rho_{\widetilde\Phi,r}$ where $\rho_{\widetilde\Phi,r}$ is the density matrix of $\ket{\widetilde\Phi}$ on all modes to the right of the QES. In this case, the modes to the right of the QES are just those in $\mathcal{A}_{R,0}$. We can again use the analogy with the formula \eqref{goodrel} for Type I and II algebras, to identify\footnote{Note that $h_{\widetilde\Phi,r}$ here is \emph{not} the same as $h_{\widetilde\Phi,r}$ in \eqref{eq:leftQESmodular}, since we are now considering a different QES.}
\begin{align}
    \log \Delta_{\widetilde\Phi|\Psi}^A = - h_{\widetilde\Phi,r} + h_{\Psi_A, \ell}.
\end{align}
It follows that
\begin{align} \label{eq:rightQESsgen}
    \frac{\hat A}{4G} - \frac{A_0}{4G} +  h_{\widetilde\Phi,r} =\beta h_L + h_{\Psi_B} - \log \Delta_{\widetilde\Phi|\Psi}^A.
\end{align}
Substituting \eqref{eq:rightQESsgen} into \eqref{eq:S2contributions} tells us that
\begin{align}
 e^{-S_2(\widehat\Phi)} &= e^{-A_0/4G} \braket{\widehat\Phi|\,p(A)\, e^{-\beta h_L - h_{\Psi_B}} \,\Delta_{\widetilde\Phi|\Psi}^A \,|\widehat\Phi}
  \\&=e^{-A_0/4G}\int_{-\infty}^\infty dx\, \varepsilon^2\,|g( \varepsilon x)|^4\,e^{-\beta x} \, \braket{\widetilde\Phi| \Delta_{\widetilde\Phi|\Psi}^A \otimes \Delta_{\Psi_B}|\widetilde\Phi}.
\end{align}
This is the same answer that one obtains from substituting the third term in \eqref{eq:threeterms} into \eqref{eq:renyiformula}.

The R\'{e}nyi QES formula \eqref{eq:RenyiQES} is derived using a saddle point approximation. However, to include nonperturbative corrections in a more careful calculation, we should really sum over contributions of all relevant saddles. In a replica trick calculation of the R\'{e}nyi 2-entropy, unlike for higher R\'{e}nyi entropies and for the von Neumann entropy, there is exactly one saddle point for each quantum extremal surface \cite{Penington:2019kki}. It is therefore completely consistent with our expectations that \eqref{eq:renyiformula} involves a sum over saddle point contributions from both the left and right quantum extremal surfaces. 

All that remains is to explain the negative, second term in \eqref{eq:threeterms}. Both the left and right quantum extremal surfaces that we have so far described are local minima of generalized entropy in the spatial directions (on an appropriate Cauchy slice), while being local maxima in time; such surfaces are sometimes called ``throats'' \cite{Brown:2019rox}. It was shown in \cite{Brown:2019rox} that, any time two spacelike-separated throats exist, there will always exist a third QES in between them that can be found by a so-called ``maximinimax procedure.'' Unlike the original two surfaces, this third QES is a local maximum in both space and time. It is therefore referred to as a ``bulge surface.'' 

In our case, the approximate location of the bulge QES can understood heuristically as follows. We assume for simplicity that $\ket{\widehat\Phi}$ is spherically symmetric, although the qualitative conclusions should be true more generally. In the single-shock geometry shown in Figure \ref{fig:singleshock} there is a quantum apparent horizon for the right white hole -- i.e. a family of spherically symmetric surfaces where the quantum expansion in the past, rightwards null direction is zero -- that starts at the quantum extremal surface and approaches the white hole event horizon at past infinity. Moving along this apparent horizon towards past infinity, the generalized entropy monotonically increases. Because the $\mathcal{H}_A$ modes lie parametrically closer to the right causal surface than the left QES, the generalized entropy of this family of surfaces is unaffected by the presence of the $\mathcal{H}_A$ shock until far in the past of the $\mathcal{H}_B$ shock -- and even then the effect on the past, rightwards expansion is minimal because the $\mathcal{H}_A$ modes are left moving. Similarly, there is a family of surfaces with future rightwards expansion zero -- and monotonically increasing generalized entropy -- that begins at the right QES and approaches the left white hole event horizon in the distant past. This family of surfaces is unaffected by the presence of the $\mathcal{H}_B$ shock until far in the past of the $\mathcal{H}_A$ modes, and there the effect on the future rightwards expansion is minimal. The two families intersect at the bulge QES, which lies in the past of both shocks -- in a location where both families of surfaces are under control. Perturbations of this QES towards the future, left or right cause a decrease in generalized entropy due to the presence of the shocks, while perturbations towards the past decrease generalized entropy due to the classical geometry of a Schwarzschild black hole. It is therefore a local maximum in both space and time, as claimed.

What answer would the bulge QES give for the R\'{e}nyi entropy? Suppose we imagine creating the state $\ket{\widehat\Phi}$ by first exciting the $\mathcal{H}_B$ modes and then exciting the $\mathcal{H}_A$ modes. In the natural frame for the $\mathcal{H}_B$ modes, the bulge surface lies very close to the right white hole horizon; it also lies to the past of all the modes in $\mathcal{H}_B$. Before the $\mathcal{H}_A$ modes are not excited, the area of the bulge surface will therefore be equal to the area of the right white hole horizon at past infinity. We therefore have
\begin{align} \notag
\frac{\hat A}{4G} - \frac{A_0}{4G}& = \beta h_R\\ & = \beta h_L + h_{\Psi_B}.\label{eq:bulgearea1}
\end{align}
In the second equality we used \eqref{eq:betahRtwoshock}, and then set $h_{\Psi_A} = 0$ because the $\mathcal{H}_A$ modes have not yet been excited. Exciting the $\mathcal{H}_A$ modes will in general change $h_R$. However, as long as the excitation is dressed to the right boundary, it will not change the area of the bulge surface, and it will also not change $h_L$ or $h_{\Psi_B}$, since all three only involve physics entirely to the left of the $\mathcal{H}_A$ modes. It follows that the second line of \eqref{eq:bulgearea1} is in fact valid for arbitrary states $\ket{\widehat\Phi}$, even ones where the $\mathcal{H}_A$ modes \emph{are} excited and hence the first line of \eqref{eq:bulgearea1} is not true.

We could also have made an equivalent argument where we imagine first exciting the $\mathcal{H}_A$ modes. Before the $\mathcal{H}_B$ modes are excited, the area of the bulge surface will be equal to the area of the left white hole at infinity, giving $(\hat A - A_0)/4G = \beta h_L = \beta h_R - h_{\Psi_A}$. Exciting the $\mathcal{H}_B$ modes using a gravitational dressing to the left boundary leaves $\hat A$, $h_R$ and $h_{\Psi_A}$ unchanged, and we again find
\begin{align} \label{eq:Abulge}
    \frac{\hat A}{4G} - \frac{A_0}{4G} =  \beta h_R - h_{\Psi_A} = \beta h_L + h_{\Psi_B}.
\end{align}
As usual, when applying \eqref{eq:RenyiQES}, we consider the sum of the area operator $\hat A/4G$ with the modular Hamiltonian $h_{\widetilde\Phi,r}$ of all quantum fields to the right of the QES. In this case all modes in $\mathcal{H}_A$ are to the right of the bulge QES, while all modes in $\mathcal{H}_B$ are to the left of the QES. We therefore have $h_{\widetilde\Phi,r} = - \log \tilde\rho_A$ where $\tilde\rho_A$ is the reduced density matrix of $\ket{\widetilde\Phi}$ on $\mathcal{H}_A$. Note that, unlike for the two previous quantum extremal surfaces, $h_{\widetilde\Phi,r}$ is UV-finite because -- within the bulk Hilbert space that we are considering -- the algebra of modes to the right of the QES is the Type I algebra $\mathcal{B}(\mathcal{H}_A)$, rather than a Type III algebra.\footnote{Fortunately (again in contrast to the previous two cases) the formula \eqref{eq:Abulge} is also UV-finite, and so the sum $\hat A/4G + h_{\widetilde\Phi,r}$ does not diverge. The fact that we obtained a UV-finite formula for $\hat A/4G$ may seem confusing, since the divergence in $A/4G$ just comes from the renormalization of $G$. To understand this, first note that we are not including the Rindler modes near the bulk QES, which would usually lead to a divergence in $h_{\widetilde\Phi,r}$, as part of the effective bulk Hilbert space $\mathcal{H}_A^* \otimes \mathcal{H}_B^*$. Consider a simpler situation where we integrate out all the bulk fields (i.e. fix all bulk modes to lie in the Hartle-Hawking state $\ket{\Psi}$). The bare area term $\hat A/4G$ would still be divergent, but we could also define a renormalized area term $\hat A/4G$ that includes a contribution from the bulk entropy of the integrated-out bulk modes. This renormalized area term would be UV-finite, and in fact would satisfy $\hat A/4G - A_0/4G = \beta h_R$. The formula for $\hat A/4G$ in \eqref{eq:Abulge} is similar: it is really a renormalized area obtained after integrating out the Rindler modes near the bulk QES (but not the modes in $\mathcal{H}_A^* \otimes \mathcal{H}_B^*$).} The QES prescription formula \eqref{eq:RenyiQES} applied to the bulge QES therefore gives, when combined with \eqref{eq:bulgearea1},
\begin{align}\notag
    e^{-S_2(\widehat\Phi)} &= e^{-A_0/4G} \braket{\widehat\Phi|\,p(A)\, e^{-\beta h_L - h_{\Psi_B}} \,\tilde\rho_A \,|\widehat\Phi}
    \\&= e^{-A_0/4G}\int_{-\infty}^\infty dx\, \varepsilon^2\,|g( \varepsilon x)|^4\,e^{-\beta x} \, \braket{\widetilde\Phi| \tilde\rho_A \otimes \Delta_{\Psi_B} |\widetilde\Phi}.
\end{align}
This has the same magnitude, although obviously the opposite sign, as the contribution to \eqref{eq:renyiformula} from the negative term in \eqref{eq:threeterms}. A similar result holds for states containing $n$ shocks: in that case there are generically $n$ throats, one inside each shock, which explain the $n$ positive terms in the $n$ shock analogue of \eqref{eq:bigexpand}; each pair of throats has a bulge in between them, and the magnitude of \eqref{eq:RenyiQES} applied to the bulge QESs agrees with the $(n-1)$ negative terms.

Why does a bulge surface give a negative contribution to the R\'{e}nyi entropy, when the throat surfaces gave positive contributions? We leave the full answer to this question to future work, but emphasize that it is a very interesting and important one. In random unitary toy models of the black hole dynamics, this negative term is related to subleading Weingarten contributions that are crucial to showing that the dynamics are indeed exactly unitary.\footnote{A simple toy model of the state $\ket{\Phi}$ is given by a state $\ket{\phi} = \sum_k a_k U b_k \ket{\mathrm{MAX}}$ with $(a_k, b_k)$ trace-zero operators acting on a Hilbert space $\mathcal{H}$ with large dimension $d$, $U$ a Haar random unitary on $\mathcal{H}$ and $\ket{\mathrm{MAX}}$ a purification of the maximally mixed state on $\mathcal{H}$. A standard computation using Weingarten functions shows that the reduced density matrix $\rho_\phi$ on $\mathcal{H}$ satisfies $$\braket{\Tr(\rho_\phi^2)} \approx \frac{1}{d}\left[\tr[a_i^\dagger a_j]\tr[a_k^\dagger a_l]\tr[b_i^\dagger b_j b_k^\dagger b_l] + \left( \tr[a_i^\dagger a_j a_k^\dagger a_l] - \tr[a_i^\dagger a_k] \tr[a_k^\dagger a_l] \right) \tr[b_i^\dagger b_l]\tr[b_j b_k^\dagger]\right],$$ where the expectation value on the left-hand side is over the Haar ensemble and $\tr[a] = 1/d \, \Tr[a]$ is the expectation value of $a$ in the maximally mixed state. There is an obvious correspondence between the right-hand side of this formula and \eqref{eq:bigexpand}, which in turn led to \eqref{eq:threeterms}; the negative piece, analogous to the bulge QES term, comes from a negative subleading Weingarten contribution.} Recently, similar negative terms were successfully obtained in \cite{Stanford:2021bhl} using a very careful analysis of the gravitational path integral -- in the specific case of an out-of-time-order four point function. However no general gravitational account of these negative terms currently exists. It is extremely tempting to conjecture that the desired general explanation is simply that bulge extremal surfaces always contribute with a negative sign (at least in R\'{e}nyi 2-entropy calculations). If true, this could successfully explain a number of other examples where subleading Weingarten terms with negative coefficients play an important role in black hole dynamics; however we leave a detailed discussion of such things to future work. We shall however make one final observation about \eqref{eq:threeterms}: when $a_k$ is unitary (and the sum over $k$ in \eqref{eq:twoshockstate} is trivial with only one value of $k$) the bulge QES and the right QES merge and vanish. This is only consistent with the formula \eqref{eq:threeterms} because the same condition causes the second and third terms of \eqref{eq:threeterms} to cancel -- in fact for unitary $a_k$ both have magnitude $\braket{\widetilde\Phi|\Delta_{\Psi_B}|\widetilde\Phi}$. As a result, \eqref{eq:threeterms} is equal to the first term i.e the contribution from the left QES.  However this only works because the second and third terms contribute with opposite sign. A throat QES and a neighbouring bulge QES can, and frequently do, merge and disappear, whereas two throats or two bulges never can.

\section*{Acknowledgements}
GP is supported by the UC Berkeley Physics Department, the Simons Foundation through the ``It from Qubit'' program, the Department of Energy via the GeoFlow consortium (QuantISED Award DE-SC0019380), and AFOSR award FA9550-22-1-0098.  EW is supported in part by NSF Grant Phy-1911298. VC is supported by a grant from the Simons Foundation (816048, VC). 

\appendix 
\section{1/N corrections and the canonical ensemble \label{app:corrections}}

It was argued in \cite{wittengcp} that the large $N$ algebra for the canonical ensemble is a crossed product once one includes $1/N$ corrections in a formal power series. Because we are working perturbatively around the limit $N \to \infty$, this algebra really valued in the ring $\mathbb{C}[[1/N]]$ of formal power series in $1/N$, rather than being valued in $\mathbb{C}$. As a result, the usual classification of von Neumann algebras does not obviously apply. Nonetheless it is natural to think of this perturbative algebra as a Type II$_\infty$ von Neumann factor. In contrast, in this paper we found the Type II$_\infty$ algebra $\mathcal{A}_R$ in the strict large $N$ limit by switching from the canonical to the microcanonical ensemble. We thereby avoided the need to consider $1/N$ corrections. In this appendix we provide some brief comments about the canonical ensemble construction. Our main observation is that it is hard to define a trace, or, consequently, density matrices, on the canonical algebra, even once $1/N$ corrections are included, because the natural candidate trace that would exist in a crossed product algebra over $\mathbb{C}$ is not well defined as a power series, or even a Laurent series, in $1/N$. Nonetheless it turns that one can define \emph{entropies} for states as a Laurent series in $1/N$ in a fairly satisfactory manner. As we saw in Section \ref{sec:sgen} for the microcanonical ensemble algebra, this entropy ends up being to the generalized entropy of the dual bulk state.

Recall that in Section \ref{sec:canon} the operator $U = (H_R - E_0)/N$ had a large $N$ limit for the canonical ensemble, and that in this limit we also had $U = (H_L - E_0)/N$. Including the operator $U$ along with $\mathcal{A}_{R,0}$ led to a large $N$ algebra $\mathcal{A}_{R,\mathrm{can}}^{(0)} = \mathcal{A}_{R,0} \otimes \mathcal{A}_U$ where $\mathcal{A}_U$ is the algebra of bounded functions of $U$.\footnote{The superscript zero in $\mathcal{A}_{R,\mathrm{can}}^{(0)}$ indicates that we are expanding to zeroth order in a formal $1/N$ power series.}

For most of this section we shall work to 1st order in $1/N$. Formally this means that the algebra $\mathcal{A}_{R,\mathrm{can}}^{(1)}$ is defined over the ring $\mathbb{C}[[1/N]]/<[1/N]^2>$, where $<[1/N]^2> \subseteq \mathbb{C}[[1/N]]$ is the ideal generated by $1/N^2$. Less formally, it means that $1/N^2 = 0$. Since $H_R - H_L = O(1)$ for generic states created by acting on $\ket{\mathrm{TFD}}$ with single-trace operators, we no longer have $(H_L - E_0)/N = (H_R - E_0)/N$ at large $N$ once we include $O(1/N)$ corrections. The operator $U_L = (H_L - E_0)/N$ continues to commute with $\mathcal{A}_{R,0}$. However
\begin{align} \label{eq:perturbativeenergy}
U_R = \frac{H_R - E_0}{N} = U_L + \frac{\hat h}{N}
\end{align}
does not.  The first-order perturbative algebra $\mathcal{A}_{R,\mathrm{can}}^{(1)}$ is generated by $\mathcal{A}_{R,0}$ together with $U_R$.\footnote{There are also perturbative corrections to the algebra $\mathcal{A}_{R,0}$, but these will not be qualitatively important since all the nontrivial operators in $\mathcal{A}_{R,0}$ are already noncentral.} The commutant algebra $\mathcal{A}_{L,\mathrm{can}}^{(1)}$ is generated by $U_L$ together with 
\begin{align}
a_0' + \frac{1}{N} [\hat h, a_0'] \frac{d}{dU_L} ~~~~~~~~~~~~~\forall a_0' \in \mathcal{A}_{L,0}
\end{align}
The center of $\mathcal{A}_{R,\mathrm{can}}^{(1)}$ is trivial since e.g. $U_L \not\in \mathcal{A}_{R,\mathrm{can}}^{(1)}$ while $U_R \not\in \mathcal{A}_{L,\mathrm{can}}^{(1)}$. In fact, if we pretend for a second that $N$ is not perturbatively small, and is instead finite, then $\mathcal{A}_{R,\mathrm{can}}^{(1)}$ becomes exactly the Type II$_\infty$ factor $\mathcal{A}_R$, with $U_R = h_R /N$. Unfortunately, however, if we formally write the trace \eqref{eq:crossedtrace} in terms of the operator $U_L$, we get
\begin{align}
\tr[a] = \int_{-\infty}^\infty dU_L e^{\beta NU_L} \braket{\Psi|a|\psi}~.
\end{align}
In the large $N$ limit, this does not lead to a formal power series in $1/N$, or even a Laurent series, because a positive power of $N$ appears in the exponent.

Clearly a new strategy is needed, which we now describe. Recall that modular flow generates an outer automorphism of a Type III von Neumann factor, but generates an inner automorphism for a Type I or II factor $\mathcal{A}$. This inner automorphism is generated by $\log \rho \in \mathcal{A}$. Since the von Neumann entropy $S$ satisfies $S = -\langle\log \rho \rangle$, one could try to define $S$ as the expectation value of an operator in $\mathcal{A}$ that generates modular flow -- without any reference to density matrices or traces.

Since $\mathcal{A}_{R,0}$ is a Type III$_1$ algebra, modular flow in $\mathcal{A}_{R,\mathrm{can}}^{(0)} = \mathcal{A}_{R,0} \otimes \mathcal{A}_U$ generates an outer automorphism. For modular flow by an $O(1)$ modular time, this statement is unaffected by perturbative $O(1/N)$ corrections and so will continue to be true for $\mathcal{A}_{R,\mathrm{can}}^{(1)}$. 

However we can also consider modular flow by an $O(1/N)$ modular time $s/N$. For a semiclassical state
\begin{align} \label{eq:cqstatedefapp}
    \ket{\widehat\Phi} = \int dU g(U) \ket{\Phi}\ket{U},
\end{align}
we have $\Delta_{\widehat \Phi} = \Delta_{\Phi} + O(1/N)$. Hence
\begin{align}
    \Delta_{\widehat\Phi}^{is/N} a_0 \Delta_{\widehat\Phi}^{-is/N} = -i \frac{s}{N} [h_\Phi, a_0],
\end{align}
and $\Delta_{\widehat\Phi}^{is/N} U_R \Delta_{\widehat\Phi}^{-is/N}= U_R$. The operator $h_\Phi/ N \not\in\mathcal{A}_{R,\mathrm{can}}^{(1)}$. However
\begin{align}
    [ h_\Phi/ N, a] &= [\beta U_L + h_{\Phi|\Psi}/ N, a]\,\,\,\,\,\,\forall a \in \mathcal{A}_R,
    \\ [ h_\Phi/ N, U_R] &= [\beta U_L + h_{\Phi|\Psi}/ N, U_R] = 0
\end{align}
and $(\beta U_L + h_{\Phi|\Psi} /N) = \beta U_R + 1/N(h_{\Phi|\Psi} - h_{\Psi}) \in \mathcal{A}_{R,\mathrm{can}}^{(1)}$. This tells us that perturbative modular flow is indeed an inner automorphism for the algebra $\mathcal{A}_{R,\mathrm{can}}^{(1)}$.

In a Type I or II factor $\mathcal{A}$, it follows directly from \eqref{eq:deltafromdens} that
\begin{align} \label{eq:pertmodflow}
    \frac{d}{ds}\left[\Delta_{\Phi}^{is} a_0 \Delta_{\Phi}^{-is}\right]_{s=0} = i [\log \rho_{\Phi}, a_0]~,
\end{align}
where $\rho_\Phi \in \mathcal{A}$ is the density matrix of $\ket{\Phi}$. It is therefore tempting to try to use \eqref{eq:pertmodflow} as a definition of $\log \rho$. 
More precisely, we could try to define a ``rescaled modular Hamiltonian'' for the state $\ket{\widehat\Psi}$ to be an operator $h^R_{\widehat\Phi} \in \mathcal{A}_{R,\mathrm{can}}^{(1)}$ that satisfies
\begin{align}\label{eq:firstpassent}
    \frac{d}{ds}\left[\Delta_{\Phi}^{is/N} a_0 \Delta_{\Phi}^{-is/N}\right]_{s=0} = -i [h^R_{\widehat\Phi}, a_0]
\end{align}
Here $h^R_{\widehat\Phi}$ is analogous to $-1/N \log \rho_{\widehat\Phi}$ in a Type I or II algebra over $\mathbb{C}$. 
We could then define a Laurent-series valued entropy $$S \in N \left(\mathbb{C}[[1/N]]/<[1/N]^2>\right)$$ by
\begin{align}
    S({\widehat\Phi})_{\mathcal{A}_{R,\mathrm{can}}^{(1)}} = N \braket{\widehat\Phi| h^R_{\widehat\Phi} |\widehat\Phi}.
\end{align}
Comparison with \eqref{eq:firstpassent} suggests that in fact
\begin{align}
    h^R_{\widehat\Phi} =  \beta U + h_{\Phi|\Psi}/N~,
\end{align}
and hence
\begin{align}
S({\widehat\Phi})_{\mathcal{A}_{R,\mathrm{can}}^{(1)}} = N \braket{\beta U} - S_{\mathrm{rel}}(\Phi||\Psi)~.
\end{align}
However this is all far too quick: without affecting \eqref{eq:firstpassent}, we could have added an arbitrary c-number $C_{\widehat\Phi}$ to $(\beta U + h_{\Phi|\Psi} /N)$. We could have also added $f(U_R)/N$ for any function $f$, since such operators are central at first order in $1/N$. 

Moreover, unlike the arbitrary state-independent constant that appears in the standard definition of entropies for Type II$_\infty$ algebras, here we would have to choose an independent constant $C_{\widehat\Phi}$ for each state $\ket{\widehat\Phi}$. As a result, a definition of entropy based purely on \eqref{eq:firstpassent} will be hopelessly ambiguous. In fact we could always consistently choose $C_{\widehat\Phi}$ such that the entropy of every state is zero.

Fortunately we can resolve most of this ambiguity using Connes cocycle flow. Recall that the Connes cocycle $u_{\Phi|\Psi}(s) \in \mathcal{A}$ is defined by
\begin{align} \label{eq:ccdef}
    u_{\Phi|\Psi}(s) = \Delta_{\Phi|\Psi}^{is} \Delta_{\Psi}^{-is} = \Delta_{\Phi}^{is} \Delta_{\Psi|\Phi}^{-is}~.
\end{align}
This satisfies the chain rule
\begin{align} \label{eq:conneschainrule}
    u_{\Phi_1|\Phi_2}(s) u_{\Phi_2|\Phi_3}(s) = u_{\Phi_1|\Phi_3}(s).
\end{align}
It will also be helpful later to note that
\begin{align} \label{eq:usefulequation1}
    u_{\Phi|\Psi}(s) u_{\Phi|\Psi}'(-s) = \Delta_{\Phi}^{is} \Delta_{\Psi}^{-is}
\end{align}
for $u_{\Phi|\Psi}' \in \mathcal{A}'$ because $\Delta_{\Phi|\Psi}^{-1} = \Delta_{\Psi|\Phi}'$. For a Type I or II algebra $\mathcal{A}$, it follows from \eqref{eq:deltafromdens} that
\begin{align}
    -i \left.\frac{d u_{\Phi|\Psi}(s)}{ds}\right|_{s=0} = \log \rho_{\Phi} - \log \rho_{\Psi}.
\end{align}
We can use this to reduce the ambiguity in  $h_{\widehat\Phi}^R$ down to a single, state-independent central operator by requiring
\begin{align} \label{eq:CCconcon}
    \frac{i}{N} \left.\frac{d u_{\widehat\Phi_1|\widehat\Phi_2}(s)}{ds}\right|_{s=0} = \left[h_{\widehat\Phi_1}^R - h_{\widehat\Phi_2}^R\right],
\end{align}
for any pair of states $\ket{\widehat\Phi_1}, \ket{\widehat\Phi_2}$. More precisely, suppose that for one fixed state $\ket{\widehat\Phi_0}$ we find an operator $h_{\widehat\Phi_0}^R \in \mathcal{A}_{R,\mathrm{can}}^{(1)}$ such that
\begin{align} \label{eq:initstate}
    \left[h_{\widehat\Phi_0}^R, a\right] = \left[-\log \Delta_{\widehat\Phi_0}, a \right]
\end{align}
for all operators $a \in \mathcal{A}_{R,\mathrm{can}}^{(1)}$. If so, then $h_{\widehat\Phi_0}^R$ will be unique up to an element of the center of $\mathcal{A}_{R,\mathrm{can}}^{(1)}$. For any other state $\ket{\widehat\Phi}$, consistency with \eqref{eq:CCconcon} then fixes
\begin{align}
    h_{\widehat\Phi}^R = h_{\widehat\Phi_0}^R + 1/N \partial_s u_{\widehat\Phi|\widehat\Phi_0}|_{s=0}.
\end{align}
This automatically satisfies \eqref{eq:firstpassent} thanks to \eqref{eq:usefulequation1}. Finally the chain rule \eqref{eq:conneschainrule} ensures that
\begin{align}
    1/N \partial_s u_{\widehat\Phi_1|\widehat\Phi_2}|_{s=0} = h_{\widehat\Phi_1}^R - h_{\widehat\Phi_2}^R,
\end{align}
and so \eqref{eq:CCconcon} is satisfied for \emph{any} pair of states $\ket{\widehat\Phi_1}, \ket{\widehat{\Phi}_2}$. To summarize: whenever perturbative modular flow is an inner automorphism for any state $\ket{\widehat\Phi_0}$, we can find operators $h_{\widehat\Phi}^R$ for all states $\ket{\widehat\Phi}$ that are consistent with both \eqref{eq:firstpassent} and \eqref{eq:CCconcon}. Moreover these operators will be unique up to a  state-independent element of the center of $\mathcal{A}_{R,\mathrm{can}}^{(1)}$.

Concretely, suppose we choose
\begin{align} \label{eq:hatpsicanon}
    \ket{\widehat\Psi} = \int dU_L f(U_L) \ket{\Psi}\ket{U_L}
\end{align}
for some $f(U_L)$ to have $h_{\widehat\Psi}^R = \beta U_R = \beta U_L + \beta\hat h/N$. At zeroth order in $1/N$, we have $\mathcal{A}_R^{(can),0} = \mathcal{A}_{R,0} \otimes \mathcal{A}_U$. If $\ket{\widehat\Phi}$ is defined as in \eqref{eq:cqstatedefapp}, then $\ket{\widehat\Psi}$ and $\ket{\widehat\Phi}$ are both product states, and hence their relative modular operator is a product of relative modular operators on $\mathcal{A}_{R,0}$ and $\mathcal{A}_U$. Since $\mathcal{A}_U$ is central, it follows directly from \eqref{eq:relmoddef} that the relative Tomita operator $S_{g|f} = g(U_L)\overline{f(U_L)}^{-1} T$ where $T h(U_L) \ket{U_L} = \overline{h(U_L)} \ket{U_L}$ for all $h \in L^2(\mathbb{R})$. Hence $\Delta_{g|f} = |g(U_L)|^2|f(U_L)|^{-2}$. We therefore have
\begin{align}
    \log\Delta_{\widehat\Phi|\widehat\Psi} &= \log \Delta_{\Phi|\Psi} + \log \Delta_{g|f} + O(1/N)
   \\& =\log \Delta_{\Phi|\Psi} + \log |g(U_R)|^2 - \log |f(U_R)|^2 + O(1/N).
\end{align}
Finally we obtain
\begin{align}
    h_{\widehat\Phi}^R = \beta U_L + \frac{1}{N}[h_{\Phi|\Psi} - \log |g(U_R)|^2 + \log |f(U_R)|^2].
\end{align}
Note that we can use the freedom to add a state-independent central operator to $h_{\widehat\Phi}^R$ to replace $1/N \log |f(U_R)|^2$ by $\alpha(U_R)/N$ for any function $\alpha$; we can also add an $O(1)$ constant $S_0$. This leads to a final formula for the entropy
\begin{align} \label{eq:entropycanon}
    S = N \beta \langle U_R \rangle + N S_0 - S_\mathrm{rel}(\Phi||\Psi) - \langle\log |g(U_R)|^2\rangle + \langle \alpha(U_R) \rangle.
\end{align}
If we identify $x = NU$, then this agrees with \eqref{eq:entrform} up to the ambiguity in $S_0$ (which is also present in \eqref{eq:entrform}) and $\alpha(U_R)$ (which is not).

It may seem unsatisfactory that the entropy \eqref{eq:entropycanon} is only defined up to an arbitrary state-independent function $\alpha(U_R)$, whereas in the microcanonical ensemble the only ambiguity was a state-independent constant. However this is a feature of the canonical ensemble in the large $N$ limit rather than a bug in our definition of entropy. Over an $O(1)$ range of energies around $E_0 = O(N^2)$,  the microcanonical entropy $S(E)$ (i.e. the logarithm of the density of states) is approximately linear up to corrections that vanish in the limit $N \to \infty$:
\begin{align}
    S(E) = S(E_0) + \beta (E - E_0) + O(1/N),
\end{align}
where $S(E_0) = O(N^2)$ and $\beta = O(1)$. 

However this is not true over the $O(N)$ range of energies necessary to describe the canonical ensemble. Let $u = E/N^2 = U_R/N$ and let $s(u) = S(N^2 u)/N^2$. In the limit $N\to\infty$, $s(u)$ converges to a finite nonlinear function that, for holographic theories, describes the horizon area of an AdS-Schwarzschild black hole as a function of energy. Over an $O(N)$ range of energies we then have
\begin{align} \label{eq:S(E)O(N)}
    S(E) = N^2 \left[s\left(E_0/N^2 + U_R/N\right)\right] = S(E_0) + N \beta U_R + \frac{1}{2} U_R^2 \frac{d^2 s}{du^2} + O(1/N).
\end{align}
Here we have used the fact that $ds/du = dS/dE = \beta$. In other words there is an $O(1)$ contribution to the entropy that is controlled by $d^2s/du^2$. In terms of more standard thermodynamic quantities we have
\begin{align}
    \frac{d^2 s}{d u ^2} = \frac{d \beta}{d u} = - \frac{N^2}{T^2} \frac{d T}{d E} = - \frac{N^2}{T^2 C_\mathrm{BH}}
\end{align}
where $C_\mathrm{BH} = dE/dT = O(N^2)$ is the heat capacity of the black hole. However, to first order in $1/N$, the large $N$ algebra $\mathcal{A}_{R,\mathrm{can}}^{(1)}$ is unaffected by the variation in the temperature as a function of $E$. Since it does not know the correct value of the heat capacity $C_\mathrm{BH}$, the entropy of $\mathcal{A}_{R,\mathrm{can}}^{(1)}$ cannot correctly fix the last term in \eqref{eq:S(E)O(N)}. The freedom to choose $\alpha(U_R)$ in \eqref{eq:entropycanon} captures this ambiguity. 

To resolve the ambiguous term, we need to go to second order in $1/N$ and consider the algebra $\mathcal{A}_{R,\mathrm{can}}^{(2)}$. For our purposes the important change at $O(1/N^2)$ is that $\hat h / N = U_R - U_L$ is no longer linearly related to $h_\Psi = - \log \Delta_\Psi$. Instead we have
\begin{align}\label{eq:1/N2URcorrect}
    U_R = U_L + \frac{1}{\beta N} h_\Psi - \frac{d^2 s}{d u^2} \frac{U_L }{\beta^2 N^2} h_\Psi,
\end{align}
where the third term comes from the $O(1/N)$ linear variation
\begin{align}
    T = T_0 + \frac{N U_L}{C_\mathrm{BH}} + O(1/N^2)
\end{align}
in the effective temperature $T$ of the microcanonical ensemble as a function of $U_L$.

To fix $\alpha$, it is sufficient to consider the single state $\ket{\widehat\Psi}$ defined in \eqref{eq:hatpsicanon}. In this case, we can compute $\log \Delta_{\widehat\Psi}$ to first order in $1/N$ via the general formula \eqref{eq:cqmodular} for the entropy of states of the form \eqref{eq:hatpsicanon}. We have
\begin{align}\label{eq:mod1/N^2}
    \log \Delta_{\widehat\Psi} = \log \Delta_\Psi + \log |f(U_R)|^2 - \log |f(U)|^2 = - h_\Psi + \frac{h_\Psi}{\beta N} \frac{d}{dU}[ \log |f(U)|^2],
\end{align}
where in the second equality we expanded $\log |f(U_R)|^2$ to first order in $1/N$ using $U_R = U + h_{\Psi}/\beta N + O(1/N^2)$. The formula \eqref{eq:mod1/N^2} is consistent with the rescaled modular Hamiltonian
\begin{align}
    h_{\widehat\Psi}^R = \beta U_R
\end{align} 
for the second-order algebra $\mathcal{A}_{R,\mathrm{can}}^{(2)}$ if and only if the $O(1/N)$ term in \eqref{eq:mod1/N^2} generates the same action as the $O(1/N^2)$ term in $\beta U_R$ that we get from \eqref{eq:1/N2URcorrect}. This requires\footnote{In principle it would be consistent for the two sides of \eqref{eq:matching1/N^2} to differ by a nontrivial element of the commutant algebra $\mathcal{A}_{L,\mathrm{can}}^{(2)}$; this never happens because both sides are proportional to $h_\Psi$.}
\begin{align} \label{eq:matching1/N^2}
    \frac{1}{\beta N^2} \frac{d}{dU}[ \log |f(U)|^2]\, h_\Psi = - \frac{d^2 s}{d u^2} \frac{U }{\beta N^2} h_\Psi, 
\end{align}
and hence
\begin{align}
    |f(U)|^2 \propto \exp( -\frac{1}{2} \frac{d^2 s}{du^2} U^2) \propto \exp(S(E) - \beta E).
\end{align}
This is exactly the energy distribution of the thermofield double state $\ket{\mathrm{TFD}}$, which indeed has modular Hamiltonian $(\beta H_R + \mathrm{const})$ at finite $N$. We conclude that the consistency of \eqref{eq:entropycanon} at $O(1/N^2)$ requires 
\begin{align}
\alpha(U_R) =  (d^2 s/du^2) U_R^2/2 + \mathrm{const}.
\end{align}

\section{Other conserved charges} \label{app:otherscharges}

In Section \ref{sec:canon} we briefly mentioned the existence of other central single-trace operators (beyond the operator $U$) associated to additional symmetries of the boundary theory beyond time translations. We then proceeeded to ignore those operators throughout the main part of the paper. We now rectify that omission. These operators were previously studied in Section 4 of \cite{wittengcp}, and we closely follow the discussion there. Our new contribution is to identify the existence of two distinct large $N$ limits associated to the fixed chemical potential ensemble and a finite charge ensemble. This distinction is closely analogous to the distinction between the large $N$ algebras for the canonical and microcanonical ensembles described in Section \ref{sec:alg}.

The microcanonical and canonical ensembles both break the full symmetry group of the boundary theory down to the group $\mathbb{R}_t$ of time translations, along with a compact group $G$ of symmetries that commute with $\mathbb{R}_t$. This includes rotations and also additional internal symmetries of the theory that are dual to gauge symmetries in the bulk. The group $G$ is generated by a set of $\mathrm{dim}\, G$ charges $Q^i$, where, in both ensembles, fluctuations in the charges $Q^i$ grow as $O(N)$ in the large $N$ limit. It follows that the operators $U^i = Q^i/N$ have a sensible large $N$ limit. Since the action of $G$ on $\mathcal{A}_{R,0}$ has a finite large $N$ limit, we have $[Q^i, a_0] = O(1)$ and hence $[U^i,a_0] \to 0$. The operators $U^i$ are therefore central at large $N$. The resulting large $N$ Hilbert space is $\mathcal{H} \otimes L^2(\mathfrak{g})$ where $\mathcal{H}$ was the large $N$ Hilbert space before we added the operators $U^i$ and $\mathfrak{g}$ is the Lie algebra of $G$. The large $N$ right boundary algebra is $\mathcal{A}_R \otimes \mathcal{A}_{\mathfrak{g}}$ where $\mathcal{A}_{\mathfrak{g}}$ is the algebra of bounded functions on $\mathfrak{g}$.\footnote{For concreteness, we are working in the microcanonical ensemble here, as we shall do for the rest of this appendix. In the canonical ensemble $\mathcal{A}_R$ should be replaced by $\mathcal{A}_{R,\mathrm{can}}^{(0)} \cong \mathcal{A}_{R,0} \otimes \mathcal{A}_U$.} The subalgebra $\mathcal{A}_{\mathfrak{g}}$ is central at leading order in $N$. 

From here one could consider perturbative corrections in a formal $1/N$ power series as was done in \cite{wittengcp} or, for the operator $U$, in Appendix \ref{app:corrections}. However one can obtain similar results to those in \cite{wittengcp} in the strict large $N$ limit by switching from a fixed potential ensemble to an ensemble over finite charge states. To do so, we replace the state $|\widetilde{\text{TFD}}\rangle$ in \eqref{eq:microtfd} by
\begin{align} \label{eq:doubletildeTFD}
    \ket{\widetilde{\widetilde{\text{TFD}}}} = \sum_i e^{-S(E_i, R_i)/2} f(E_i - E_0) g(R_i) \ket{i}_L \ket{i}_R.
\end{align}
Here $R_i$ is the irreducible representation of $G$ that contains the state $\ket{i}$, and $g$ is a map from the space of irreducible representations of $G$ to the complex numbers such that
\begin{align}
    \sum_R |g(R)|^2 = 1.
\end{align}
Finally $e^{S(E,R)}$ is the density of states of energy $E$ in the representation $R$. The charges $Q^i$ have a finite large $N$ limit when acting on states created from \eqref{eq:doubletildeTFD} by acting with single-trace operators. It follows that we can use the state \eqref{eq:doubletildeTFD} to construct a large $N$ algebra generated by noncentral single-trace operators, the renormalized Hamiltonian $h_R$, and the charges $Q^i$.

To understand the structure of the resulting large $N$ Hilbert space and algebra, it is helpful to first study the bulk side of the AdS/CFT duality. The global symmetry group of the boundary theory, which corresponds to the asymptotic symmetry group of the bulk theory, is $G_L \times G_R$ where $G_L$ and $G_R$ act on the left and right boundaries respectively. However the eternal black hole solution is only invariant under the diagonal action $G_D$. At the classical level, this leads to a moduli space of solutions $G_{\mathcal{M}} \cong (G_L \times G_R)/ G_D$. This space is isomorphic to the group $G$. As a simple example, if $G = SO(D-1)$ is the group of rotations of the asymptotic boundary, $G_{\mathcal{M}}$ describes the relative orientations of the two boundaries. Just like the timeshift, this is a physical degree of freedom in the bulk theory. The usual eternal black hole corresponds $1 \in G_{\mathcal{M}}$ where the two boundaries have the same orientation. 

We can quantize small fluctuations of the background fields (along with the timeshift mode) for any $g \in G_{\mathcal{M}}$ to obtain a Hilbert space $\mathcal{H}_{g}$ that will be isomorphic to $\mathcal{H}$. These Hilbert spaces form a bundle $\mathcal{V}$ over $G_{\mathcal{M}}$ with fibre $\mathcal{H}$. The full quantum Hilbert space $\widetilde{\mathcal{H}}$ is the space of sections of the bundle $\mathcal{V}$. If we choose a left-invariant trivialization of $\mathcal{V}$, then we can identify
\begin{align} \label{eq:hilbertotherconserved}
    \widetilde{\mathcal{H}} \cong \mathcal{H} \otimes L^2(G_{\mathcal{M}})
\end{align}
such that $G_L$ acts solely on $L^2(G_{\mathcal{M}})$ and bulk QFT observables dressed to the right boundary act solely on $\mathcal{H}$. The full right boundary algebra $\widetilde{\mathcal{A}}_R$ is then the crossed product of the Type II algebra $\mathcal{A}_{R}$ described in Section \ref{sec:micro} by the action $G_D$. This algebra is generated by $\mathcal{A}_{R}$ together with the group $G_R$, which acts by right multiplication on $L^2(G_{\mathcal{M}})$ and by the action of $G_D$ on $\mathcal{H}_0$; see \cite{wittengcp} for a detailed definition of a crossed product by a non-Abelian group.

This is all exactly analogous to the crossed product by the modular group in the microcanonical ensemble, except that the group of time translations has been replaced by $G$. The only novel features are that $G$ is a) non-Abelian and b) compact. The latter property means that the crossed product does not affect the type of von Neumann algebra.

To relate this to our original boundary construction, we will need a few facts about the Hilbert space $L^2(G)$. For any compact Lie group $G$, the Peter-Weyl theorem says that
\begin{align}
    L^2(G) \cong \oplus_R \,\mathcal{H}_R \otimes \mathcal{H}_R^*,
\end{align}
where the direct sum is over all irreducible unitary representations $R$ of the group $G$. The action of $G$ on $L^2(G)$ by right multiplication acts on $\mathcal{H}_R$ as the representation $R$, while left multiplication acts on $\mathcal{H}_R^*$ as the representation $R^*$. The von Neumann algebra generated by $G_R$ is equivalently generated by bounded operators acting on $\mathcal{H}_R$ together with bounded functions of the representation $R$; the latter are central.

We now return to the boundary description. The holographic duality tells us that the boundary operators $Q_i$ should generate the action $G_R$ of the bulk theory, while $\mathcal{A}_{R,0}$ is dual to the bulk QFT algebra $\mathcal{A}_{r,0}$ and $h_R$ is dual to the renormalized right ADM mass as before. If we can identify the large $N$ limit of the state $\ket{\widetilde{\widetilde{\text{TFD}}}}$ with an element of the semiclassical bulk Hilbert space $\widetilde{H}$, then we can use the duality to identify $\widetilde{H}$ and $\widetilde{A}_R$ with the large $N$ boundary Hilbert space and algebra constructed from $\ket{\widetilde{\widetilde{\text{TFD}}}}$ using $\mathcal{A}_{R,0}$, $h_R$ and $Q^i$. $\ket{\widetilde{\widetilde{\text{TFD}}}}$ is annihilated by $q^i = Q^i_R - Q^i_L$. It is therefore invariant under the diagonal action $G_D$ of the group $G$. For any representation $R$, there exists a unique canonical maximally entangled state $\ket{\omega_R} \in \mathcal{H}_R \otimes \mathcal{H}_R^*$ that is invariant under $G_D$. It follows that we should identify $\ket{\widetilde{\widetilde{\text{TFD}}}}$ with the state
\begin{align} \label{eq:doubletildebulk}
\ket{\widetilde{\widetilde{\text{TFD}}}} \sim \sum_R \int_{-\infty}^\infty dx\, f(x) g(R) \ket{\Psi} \ket{x} \ket{\omega_R}.
\end{align}
It is easy to check that this reproduces all the correct large $N$ correlation functions. The state \eqref{eq:doubletildebulk} is cyclic for the von Neumann algebra  $\widetilde{\mathcal{A}}_R$, so the completion of the vector space spanned by acting with right boundary operators on $\ket{\widetilde{\widetilde{\text{TFD}}}}$ at large $N$ is indeed isomorphic to the bulk Hilbert space \eqref{eq:hilbertotherconserved}.

\section{Crossed-product isomorphism is trace preserving \label{app:isotrace}}
In this appendix we prove the claim \eqref{eq:isotrace}. As discussed in Section \ref{sec:micro}, the only freedom in the trace is an overall multiplicative constant. Therefore, it suffices to show that the constant is 1 for some element of $\mathcal{A}_r$. In particular, consider operators of the form
\begin{align}
    \widehat{a} = \int_{-\infty}^{\infty}ds ~a(s) e^{is (\beta x + h_\Psi)},
\end{align}
where we take $a(s)$ to be holomorphic in the strip $0 \leq \text{Im} \,s \leq 1$. First, we compute
\begin{align}
    \text{Tr}\widehat{a} &= \int_{-\infty}^{\infty}dxds ~ e^{\beta x(is+1)}\langle \Psi|a(s)|\Psi\rangle  \label{eq:tracecontour}
    \\ &= \int_{-\infty-i}^{\infty-i}ds \int_{-\infty}^{\infty}dx ~ e^{is \beta x}\langle \Psi|a(s+i)|\Psi\rangle 
    \\ &= 2\pi\langle \Psi|a(i)|\Psi\rangle.
\end{align}
Using \eqref{eq:isoproof}, along with $[u'_{\Phi|\Psi}(p), a(s)] = 0$, we get 
\begin{align}
    u'_{\Phi|\Psi}(p/\beta)\,\widehat{a}\,u'_{\Phi|\Psi}(p/\beta)^{\dagger} = \int_{-\infty}^{\infty}ds ~ a(s) \Delta_{\Psi|\Phi}^{-is} e^{is \beta x} = \int_{-\infty}^{\infty}ds ~ b(s) e^{is \beta x + h_\Phi},
\end{align}
where 
\begin{align}
    b(s) = a(s) u_{\Psi|\Phi}(-s).
\end{align}
We can do the same calculation as in \eqref{eq:tracecontour}, but with $\text{Tr}_{\Phi}$, which yields   
\begin{align}
    \text{Tr}_{\Phi}\left( u'_{\Phi|\Psi}(p)\widehat{a}u'_{\Phi|\Psi}(p)^{\dagger} \right) &= 2\pi \langle \Phi|b(i)|\Phi\rangle 
    \\ &= 2\pi \langle \Phi|a(i)\Delta_{\Psi|\Phi}|\Phi\rangle 
    \\ &= 2\pi \langle \Psi|a(i)|\Psi\rangle. 
\end{align}
In getting to the last line we used \eqref{eq:Deltacond}. Thus we have the desired result.   

\bibliographystyle{JHEP}
\bibliography{all}

\end{document}